\documentclass[useAMS,usenatbib]{mn2e}

\usepackage{times}
\usepackage{graphicx}
\usepackage{subfigure}

\newcommand{\gsim}{\mathrel{\hbox{\rlap{\lower.55ex \hbox {$\sim$}}
                   \kern-.3em \raise.4ex \hbox{$>$}}}}
\newcommand{\lsim}{\mathrel{\hbox{\rlap{\lower.55ex \hbox {$\sim$}}
                   \kern-.3em \raise.4ex \hbox{$<$}}}}

\title[Stellar, brown dwarf, and multiple star properties]{Stellar, brown dwarf and multiple star properties from a radiation hydrodynamical simulation of star cluster formation}
\author[M.R. Bate]{Matthew R. Bate\thanks{E-mail:
mbate@astro.ex.ac.uk}\\ School of Physics, University of Exeter, Stocker
Road, Exeter EX4 4QL \\ Monash Centre for Astrophysics, School of Mathematical Sciences, Monash University, Clayton, Vic 3168, Australia}

\date{Accepted by MNRAS}
\begin{document}
\maketitle
\begin{abstract}
We report the statistical properties of stars, brown dwarfs and multiple systems obtained from the largest radiation hydrodynamical simulation of star cluster formation to date that resolves masses down to the opacity limit for fragmentation (a few Jupiter masses).   The initial conditions are identical to those of previous barotropic calculations published by Bate, but this time the calculation is performed using a realistic equation of state and radiation hydrodynamics.  The calculation uses sink particles to model 183 stars and brown dwarfs, including 28 binaries and 12 higher-order multiple systems, the properties of which are compared the results from observational surveys. 

We find that the radiation hydrodynamical/sink particle simulation reproduces many observed stellar properties very well.  In particular, whereas using a barotropic equation of state produces more brown dwarfs than stars, the inclusion of radiative feedback results in a stellar mass function and a ratio of brown dwarfs to stars in good agreement with observations of Galactic star-forming regions.  In addition, many of the other statistical properties of the stars and brown dwarfs are in reasonable agreement with observations, including multiplicity as a function of primary mass, the frequency of very-low-mass binaries, and general trends for the mass ratio and separation distributions of binaries.  We also examine the velocity dispersion of the stars, the distributions of disc truncation radii due to dynamical interactions, and coplanarity of orbits and sink particle spins in multiple systems.  Overall, the calculation produces a cluster of stars whose statistical properties are difficult to distinguish from observed systems, implying that gravity, hydrodynamics, and radiative feedback are the primary ingredients for determining the origin of the statistical properties of low-mass stars.
\end{abstract}

\section{Introduction}
\label{introduction}

Understanding the origin of the statistical properties of stellar systems is the fundamental goal of a complete theory of star formation.  Much attention has been paid to the origin of the stellar initial mass function (IMF), and there are many models that have been proposed for its origin \citep*[see the review of][]{BonLarZin2007}.  However, the statistical properties of stellar systems include much more than just the IMF.  A non-exhaustive list also includes the star formation rate and efficiency, the structure and kinematics of stellar groups and clusters, the properties of multiple stellar systems, jets, and protoplanetary discs, and the rotation rates of stars.  A complete model must be able to explain the origin of all the statistical properties of stellar systems, and how these depend on variations in environment and initial conditions.  While simplified analytic or semi-analytic models may be useful for understanding the role that different processes play in the origin of some stellar properties, numerical simulations are almost certainly necessary to help us understand the full complexity of the star formation process.

To investigate the origin of the statistical properties of stellar systems through numerical simulations of star formation is difficult because it is necessary to use sufficient resolution to ensure that the processes involved are accurately modelled while simultaneously producing a large number of stars from which statistical properties can be derived.  One approach is to perform a large number of hydrodynamical calculations of star formation in small molecular cloud cores (e.g. \citealt*{DelClaBat2004}; \citealt{Delgadoetal2004}; \citealt*{GooWhiWar2004c, GooWhiWar2004a, GooWhiWar2004b, GooWhiWar2006, StaHubWhi2007, StaWhi2009}).  Each calculation may only produce a few stars, but from the large ensemble of simulations the statistical properties can be studied.  However, such calculations begin with an arbitrary population of dense cores for their initial conditions, which may or may not be a good representation of real dense cores.  They also neglect evolution of the cores due to external processes such as growth of the cores by accretion, interactions with the turbulent environment in which they are embedded, and interactions between cores and protostellar systems which may be particularly important in dense star-forming regions.

An alternative is to perform single large-scale hydrodynamical calculations of molecular clouds that each produce large numbers of stars.  In these calculations, dense cores may form and evolve self-consistently from hydrodynamical flows on larger scales, and interactions between dense cores and protostellar systems can occur naturally.  Such calculations can be divided into two types: those that resolve small ($\lsim 5$~AU) scales to capture the opacity limit for fragmentation \citep{LowLyn1976,Rees1976}, and those that do not.  

Hydrodynamical calculations that do not resolve small scales miss some of the star formation and most binaries and discs.  They usually seek to investigate the origin of the IMF and/or other large-scale properties such as the star formation rate.  Early simulations of this type (\citealt*{KleBurBat1998}; \citealt{KleBur2000, Bonnelletal2001a, Klessen2001, KleBur2001, BonBat2002}; \citealt*{BonBatVin2003}) used isothermal equations of state and produced large numbers of stars, but used sink particles \citep*{BatBonPri1995} with radii of hundreds of AU.  More recent calculations have the investigated effects of additional physical processes on the origin of the IMF such as radiative transfer (\citealt{Offneretal2009}; \citealt*{UrbMarEva2010, KruKleMcK2011}), but these calculations also employ sink particles with accretion radii of greater than $100$~AU.

The first hydrodynamical calculation to resolve the opacity limit for fragmentation and begin to probe the statistical properties of stars and brown dwarfs used a barotropic equation of state and sink particles with accretion radii of 5~AU, thus resolving many discs and binary and multiple systems \citep*{BatBonBro2002a, BatBonBro2002b, BatBonBro2003}.  This calculation was followed by other similar calculations that probed the dependence of stellar properties on the mean thermal Jeans mass in the molecular cloud, the thermal behaviour of the gas, and the initial turbulent motions \citep{BatBon2005, Bate2005, Bate2009c}.  For example, these calculations showed that, when using a barotropic equation of state, the characteristic stellar mass depends primarily on the initial mean thermal Jeans mass in the cloud and not, for example, the initial turbulent power spectrum.  These calculations were followed by those of other groups that also modelled the formation of stellar groups while simultaneously resolving discs and binaries (\citealt{Lietal2004}; \citealt*{OffKleMcK2008}).  Most recently, calculations  that resolve these small scales and produce stellar groups have also begun to include the effects of additional physical processes such as magnetic fields \citep{PriBat2008}, radiative transfer \citep{Bate2009b}, and both of these combined \citep{PriBat2009}.  Using radiative transfer is found to dramatically decrease the amount of fragmentation, increase the characteristic stellar mass, decrease the proportion of brown dwarfs \citep{Bate2009b,Offneretal2009}, and weaken the dependence of the characteristic mass of the IMF on the initial Jeans mass \citep{Bate2009b}.  The latter effect may help to explain why the IMF is not observed to be strongly dependent on initial conditions, at least in our Galaxy \citep*{BasCovMey2010}. Stronger magnetic fields are found to decrease the star formation rate \citep{PriBat2008,PriBat2009}.  However, in each of these calculations, only a few dozen stars and brown dwarfs were produced, making it difficult to compare statistical properties with observations in any detail.

Currently, the only published hydrodynamical calculations that resolve the opacity limit for fragmentation and produce large numbers of stars, brown dwarfs ($>100$) are those of \cite{Bate2009a}.  Two calculations were performed of 500-M$_\odot$ molecular clouds, one using sink particle accretion radii of 5~AU, and an identical calculation using accretion radii of 0.5~AU that was not followed as far.  The calculations used a barotropic equation of state to model the opacity limit for fragmentation.  The former calculation produced more than 1250 stars and brown dwarfs, including well over 100 multiple systems, that for the first time allowed a detailed comparison of a wide range of stellar properties with observations.  It was found that many of the stellar properties were in good agreement with observed properties.  For example, multiplicity was found to be a strongly increasing function of primary mass, the median separation of multiple systems was found to decrease with primary mass, the mass ratios of very-low-mass (VLM) binaries (primary masses $<0.1$~M$_\odot$) were found to favour near equal-masses, and the relative orbital orientations of triple systems were found to be somewhat aligned.  The good agreement with the observed properties of multiple stellar systems implies that such properties may originate primarily from dissipative $N$-body dynamics, and that other physical processes such as radiative transfer and magnetic fields may play less of a role.  However, the calculations produced far too many brown dwarfs relative to stars compared with a typical Galactic IMF.

In this paper, we repeat the 500-M$_\odot$ calculations of \cite{Bate2009a}, but we use a realistic equation of state and include radiative transfer as implemented in the smaller calculations of \cite{Bate2009b} and \cite{PriBat2009}.  Our aim is to investigate the effect of the realistic equation of state and radiative feedback on the star formation process in more detail than was possible with the earlier smaller calculations.  In particular, we wish to determine whether the inclusion of radiative feedback can produce a more realistic IMF than that obtained by \cite{Bate2009a}, but retain the good agreement that was found when comparing the statistical properties of multiple stellar systems with observations.

\section{Computational method}
\label{method}

The calculations presented here were performed 
using a three-dimensional smoothed particle
hydrodynamics (SPH) code based on the original 
version of \citeauthor{Benz1990} 
(\citeyear{Benz1990}; \citealt{Benzetal1990}), but substantially
modified as described in \citet{BatBonPri1995},
\citet*{WhiBatMon2005}, \citet{WhiBat2006},
\cite{PriBat2007}, and 
parallelised using both OpenMP and MPI.

Gravitational forces between particles and a particle's 
nearest neighbours are calculated using a binary tree.  
The smoothing lengths of particles are variable in 
time and space, set iteratively such that the smoothing
length of each particle 
$h = 1.2 (m/\rho)^{1/3}$ where $m$ and $\rho$ are the 
SPH particle's mass and density, respectively
\cite[see][for further details]{PriMon2007}.  The SPH equations are 
integrated using a second-order Runge-Kutta-Fehlberg 
integrator with individual time steps for each particle
\citep{BatBonPri1995}.
To reduce numerical shear viscosity, we use the
\cite{MorMon1997} artificial viscosity
with $\alpha_{\rm_v}$ varying between 0.1 and 1 while $\beta_{\rm v}=2 \alpha_{\rm v}$
\citep[see also][]{PriMon2005}.

\subsection{Equation of state and radiative transfer}

The calculations presented in this paper were performed using radiation hydrodynamics
with an ideal gas equation of state for the gas pressure
$p= \rho T_{\rm g} \cal{R}/\mu$, where 
$T_{\rm g}$ is the gas temperature, $\mu$ is the mean molecular weight of the gas,
and $\cal{R}$ is the gas constant.  
The thermal evolution takes into account the translational,
rotational, and vibrational degrees of freedom of molecular hydrogen 
(assuming a 3:1 mix of ortho- and para-hydrogen; see
\citealt{Boleyetal2007}).  It also includes molecular
hydrogen dissociation, and the ionisations of hydrogen and helium.  
The hydrogen and helium mass fractions are $X=0.70$ and 
$Y=0.28$, respectively.
The contribution of metals to the equation of state is neglected.

For this composition, the mean molecular weight of the gas is initially
$\mu = 2.38$.  The original calculation of \cite{Bate2009a} was performed using a
barotropic equation of state which took the mean molecular weight 
of the gas to be $\mu = 2.46$ and the initial temperature to be 10~K.
To keep the same initial conditions (i.e.\ the same initial thermal energy
of the gas), we set the initial temperature of the radiation hydrodynamical
calculation to be 10.3~K.

Two temperature (gas and radiation) radiative transfer in the flux-limited
diffusion approximation is implemented as described by \citet{WhiBatMon2005}
and \citet{WhiBat2006}, except that the standard explicit SPH contributions 
to the gas energy equation due to the work and artificial viscosity are used 
when solving the (semi-)implicit energy equations to provide better 
energy conservation.  Energy is generated when work is done on the gas or 
radiation fields, radiation is transported via flux-limited diffusion and energy 
is transferred between the gas and radiation fields depending on their 
relative temperatures, and the gas density and opacity.  The gas and dust
temperatures are assumed to be the same.  Taking solar metallicity 
gas, the opacity is set to be the maximum of 
the interstellar grain opacity tables of \citet{PolMcKChr1985} and, at higher 
temperatures when the dust has been destroyed, the gas opacity
tables of \citet{Alexander1975} (the IVa King model)  \citep[see][for further details]{WhiBat2006}.

The cloud has a free boundary.  To provide a boundary condition for the
radiative transfer we use the same method as \cite{Bate2009b}.  All particles
with densities less than $10^{-21}$~g~cm$^{-3}$ have their gas and radiation 
temperatures set to the initial values of 10.3 K. This gas is two orders of magnitude 
less dense that the initial cloud (see Section \ref{initialconditions}) and, thus, these boundary particles 
surround the region of interest in which the star cluster forms.


\subsection{Sink particles}
\label{sinks}

Using the above realistic equation of state and radiation hydrodynamics
means that as the gas collapses, each of the phases of
protostar formation are captured \citep{Larson1969}.  The initial collapse of
a dense core proceeds almost isothermally, until the compressional
rate heating of the gas exceeds the rate at which the gas can cool.  At this
point the collapse stalls, and a pressure supported fragment forms which
\citeauthor{Larson1969} termed the `first hydrostatic core'.  The typical
initial size and mass of this object is $\approx 5$~AU in radius and 
a few Jupiter masses (M$_{\rm J}$).  This first core accretes gas from the infalling envelope and
if it is rotating rapidly it may undergo rotational dynamical instabilities to form 
a disc \citep{Bate1998}.  Eventually, due to mass accretion \citep{Larson1969}, dynamical
instability \citep{Bate1998}, and/or cooling \citep{Tomidaetal2010b}
the central temperature exceeds $\approx 2000$~K and molecular hydrogen
begins to dissociate, absorbing thermal energy and resulting in a second
collapse \citep{Larson1969} within the first core/disc.  This collapse is
halted when the dissociation is complete and the `second' or `stellar core'
forms \citep{Larson1969}.  This object initially has a radius of $\approx 2$~R$_\odot$
and a mass of $\approx 1-2$~M$_{\rm J}$.  Subsequently it accretes to higher
masses from the surrounding first core/disc and envelope.

The calculations presented here include the physics necessary to follow 
each of these stages of protostar formation and evolution.  Indeed, the
same code has been used to study the formation and evolution of 
first cores, pre-stellar discs, and stellar cores \citep{Bate2010,Bate2011}.
However, as the collapse proceeds, the timesteps required to obey the
Courant-Friedrichs-Levy (CFL) criterion become smaller and smaller. 
Because we wish to evolve the large scales over timescales of
$\sim 10^5$ years, we cannot afford to follow the small scales 
(e.g.\ the stellar cores themselves).

Instead, we follow the evolution of each protostar through the first core
phase and into the second collapse (which begins at densities of
$\sim 10^{-7}$~g~cm$^{-3}$), but we insert a sink particle 
\citep{BatBonPri1995} when the density exceeds
$10^{-5}$~g~cm$^{-3}$.  The timesteps required to follow this evolution
get very short, but the duration of the second collapse phase is quite brief
and the use of individual particle timesteps mean that the calculation 
does not get slowed down for long.   This density is 
just two orders of magnitude before the
stellar core begins to form (density $\sim 10^{-3}$~g~cm$^{-3}$),
and a considerable improvement over previous similar barotropic 
calculations.  For example, \cite{BatBonBro2003} and \cite{Bate2009a}
inserted sink particles well before the onset of second collapse
at densities of $10^{-11}$ and $10^{-10}$~g~cm$^{-3}$, respectively.
At these densities, sink particles might have been inserted before two fragments 
merged or before a fragment was disrupted.  However, the time
taken for a protostar to evolve from $10^{-5}$~g~cm$^{-3}$ to the formation
of a stellar core is much less than a year (the free-fall time at this density is 
only one week!), so there is a no chance of protostellar
fragments merging or begin disrupted between sink particle insertion and
stellar core formation.

In the main calculation discussed in this paper, a sink particle is formed by 
replacing the SPH gas particles contained within $r_{\rm acc}=0.5$ AU 
of the densest gas particle in region undergoing second collapse 
by a point mass with the same mass and momentum.  Any gas that 
later falls within this radius is accreted by the point mass 
if it is bound and its specific angular momentum is less than 
that required to form a circular orbit at radius $r_{\rm acc}$ 
from the sink particle.  Thus, gaseous discs around sink 
particles can only be resolved if they have radii $\gsim 1$ AU.
Sink particles interact with the gas only via gravity and accretion.
There is no gravitational softening between sink particles.
The angular momentum accreted by a sink particle is recorded
but plays no further role in the calculation.

Since all sink particles are created within pressure-supported 
fragments, their initial masses are several M$_{\rm J}$, 
as given by the opacity limit for fragmentation \citep{LowLyn1976,Rees1976}.  
Subsequently, they may accrete large amounts of material 
to become higher-mass brown dwarfs ($\lsim 75$ M$_{\rm J}$) or 
stars ($\gsim 75$ M$_{\rm J}$), but {\it all} the stars and brown
dwarfs begin as these low-mass pressure-supported fragments.

Sink particles are permitted to merge in the calculation if they
passed within 0.01 AU of each other (i.e., $\approx 2$~R$_\odot$).
However, no mergers occurred during the calculation.

\begin{figure*}
\centering
    \includegraphics[width=5.8cm]{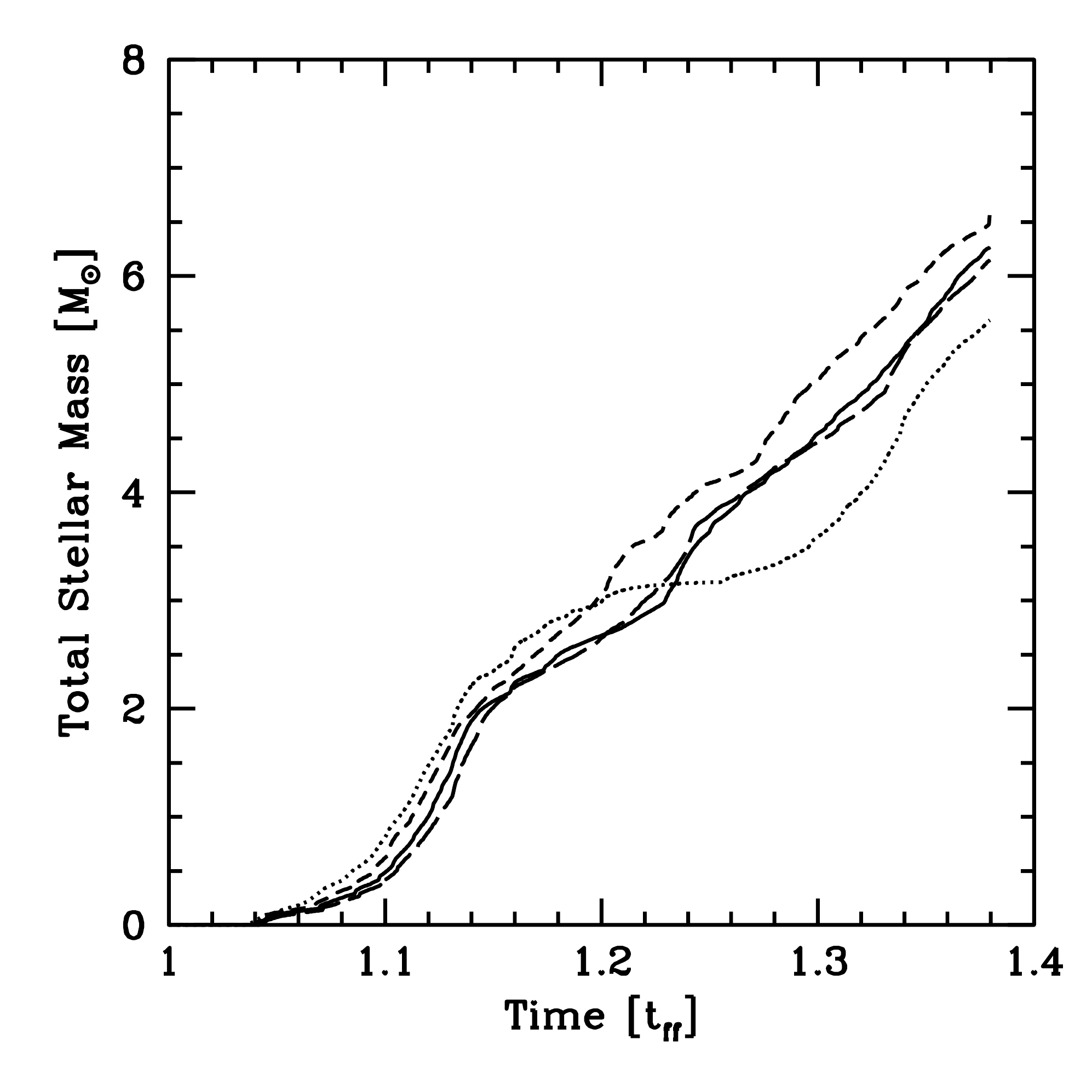}
    \includegraphics[width=5.8cm]{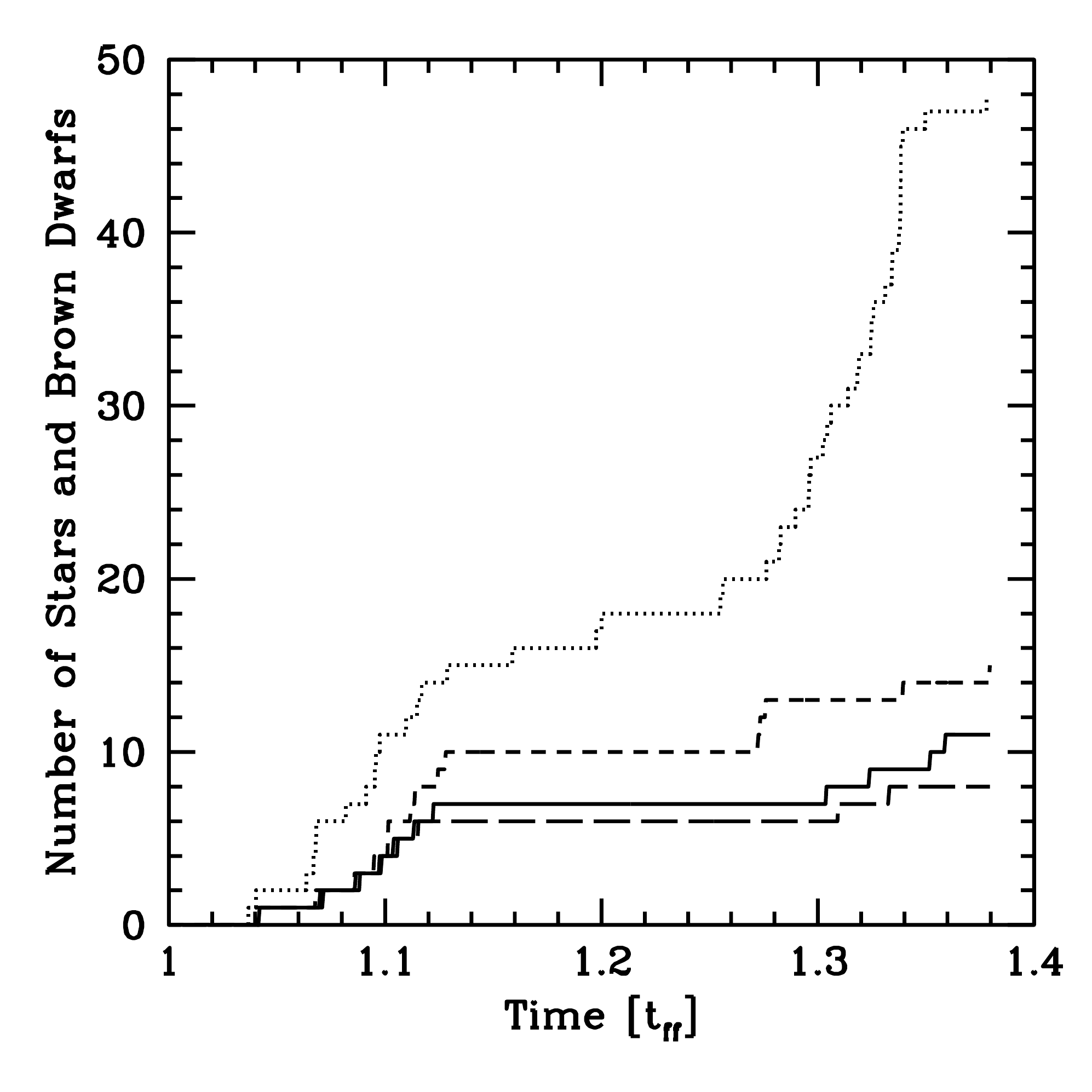}
    \includegraphics[width=5.8cm]{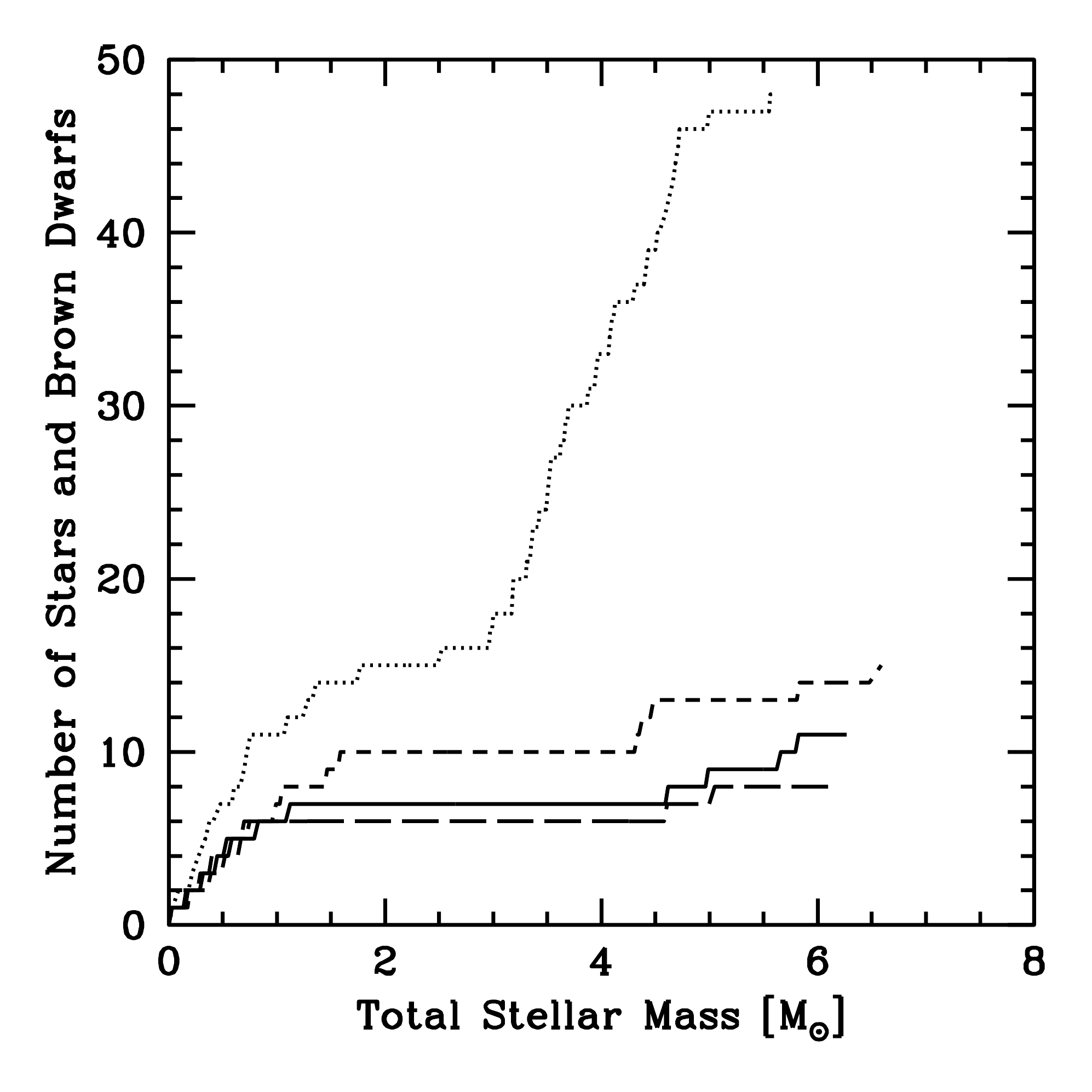}
\caption{The dependence of the star formation rate on the sink particle accretion radius, $r_{\rm acc}$.  For previously published barotropic and radiation hydrodynamical calculations of star formation in a turbulence 50-M$_\odot$ cloud we plot: the total stellar mass (i.e. the mass contained in sink particles) versus time (left panel), the number of stars and brown dwarfs (i.e. the number of sink particles) versus time (centre panel), and the number of stars and brown dwarfs versus the total stellar mass (right panel).  The different line types are for: a barotropic calculation using $r_{\rm acc}=5$~AU (dotted line; Bate et al. 2003); and radiation hydrodynamical calculations with $r_{\rm acc}=5$~AU (short-dashed line; Bate 2009c), $r_{\rm acc}=0.5$~AU (solid line; Bate 2009c), and $r_{\rm acc}=0.05$~AU (long-dashed line; performed for this paper). Time is measured from the beginning of the calculations in terms of the free-fall time ($1.90\times 10^5$~yr). It can be seen that the main change in the star formation is captured when changing from a barotropic equation of state to a radiation hydrodynamical calculation (both with $r_{\rm acc}=5$~AU).  Reducing the accretion radius used for the radiation hydrodynamical calculations has a smaller effect each time the accretion radius is reduced.  }
\label{convergence}
\end{figure*}

\subsection{Limitations of the sink particle approximation}
\label{limitations}

The benefits and potential problems associated with introducing 
sink particles into barotropic star formation calculations performed using 
SPH have been 
discussed by \cite{BatBonPri1995}, \cite{BatBonBro2003} and 
\cite{Bate2009a}.  Some of these problems are avoided
in the calculation presented here.  As mentioned above, it is no
longer possible that a fragment which has been replaced sink particle
might have merged or been disrupted before stellar core formation if
it had not been replaced by a sink particle.
Another problem, found by \cite{Bate2009a}, was that the eccentricities of
binary stellar systems with separations $1-20$~AU were too high 
if sink particles with radii $r_{\rm acc}=5$~AU were used due to
the absence of small-scale dissipation, but that 
this was corrected if the radius was reduced to 0.5~AU.
Therefore, for the calculation presented here we use $r_{\rm acc}=0.5$~AU.
However, other problems with using sink particles remain.
For example, using $r_{\rm acc}=0.5$~AU, dissipative interactions between protostars
on length-scales $\lsim 1$~AU are still neglected.  Similarly, discs smaller than 
$\approx 1$~AU in radius cannot be resolved.

When using radiation hydrodynamics, there is a new problem --- how to handle
radiative feedback.  If each stellar core itself was
resolved \citep[e.g][]{WhiBat2006,Bate2010,Bate2011}, the full radiative feedback
from the protostars would be naturally modelled.  However, introducing sink
particles, means that the evolution inside the sink particle radius is neglected.
In the simplest form where a sink particle consists only of a point mass
and a hole, there is no radiative feedback
from the stellar core and the inner part of an accretion disc on the rest of the calculation.
This is the case for the calculations presented by \cite{Bate2009b} and also for
the calculations presented in this paper.

However, this means that the luminosity of the material inside the sink particle radius
is omitted.  There are three main sources of luminosity that may be inside the sink radius:
the intrinsic luminosity of the protostar (or protostars), the luminosity generated by accretion onto the
protostar(s), and the luminosity of any other matter (e.g. an accretion disc).  If the
protostar accretes from a steady-state accretion disc, the accretion luminosity onto the protostar
will dominate the luminosity of the disc itself.  However, trying to determine the intrinsic luminosity of
the protostar(s) and accretion luminosity is far from easy and we discuss the many problems below.

\subsubsection{Intrinsic protostellar luminosity}

For a young low- or intermediate-mass ($\lsim 3$~M$_\odot$) protostar that is accreting at a 
significant rate ($>10^{-6}$~M$_\odot$~yr$^{-1}$), the accretion luminosity dominates and
the intrinsic luminosity of the stellar object is negligible \citep[e.g.][]{Offneretal2009,HosOmu2009}.  For example, the accretion
luminosity 
\begin{equation}
\label{eq_acclum}
L_{\rm acc} \approx  \frac{G M_* \dot{M}_*}{R_*}.
\end{equation}
of a star of mass $M_*=1$~M$_\odot$ with a radius of 
$R_* =2~{\rm R}_\odot$ \citep[e.g.][]{HosOmu2009} accreting
at $\dot{M}_* = 1\times 10^{-6}~{\rm M}_\odot$~yr$^{-1}$ is $\approx 15$~L$_\odot$ whereas the luminosity of the
stellar object itself is $\approx$ 1~L$_\odot$ \citep{HosOmu2009}.  As will be seen in Section
\ref{origin_imf}, the typical accretion rates in the calculation presented here are an order of
magnitude higher than this ($\approx 10^{-5}$~M$_\odot$~yr$^{-1}$).  Therefore, for the vast 
majority of the stars and brown dwarfs produced, the accretion luminosity dominates over 
stellar luminosity unless the accretion rate drops to very low levels.  Very low accretion 
rates are usually obtained only after a star has had a dynamical interaction and
ejected from the dense star-forming cores, where upon its radiative feedback is no longer important
for the subsequent star formation.  Indeed, we define a star or brown dwarf whose accretion rate
drops below  $10^{-7}$~M$_\odot$~yr$^{-1}$ to have stopped accreting.

For stars with masses $\gsim 3$~M$_\odot$, whether accretion luminosity or stellar luminosity dominates is more complex.  \cite{HosOmu2009} find that for accretion rates of $1 \times 10^{-5}$~M$_\odot$~yr$^{-1}$, accretion luminosity dominates for $\lsim 3$~M$_\odot$, but with accretion rates of $1 \times 10^{-3}$~M$_\odot$~yr$^{-1}$, accretion luminosity dominates for $\lsim 9$~M$_\odot$.
There are only two stars with masses $>3$~M$_\odot$ produced by the main calculation presented here:
a 3.84~M$_\odot$ star and a 3.38~M$_\odot$ star.  However, these two stars have average accretion rates $>5 \times 10^{-5}$~M$_\odot$~yr$^{-1}$, so the accretion luminosity is expected to dominate over stellar luminosity.

We conclude that for low- and intermediate-mass star formation like that considered in
this paper, the intrinsic stellar luminosity can be confidently neglected.

\subsubsection{Accretion luminosity and sink particle accretion radii}

However, the accretion luminosity is a different issue.  Because accretion luminosity scales
$\propto 1/R$ (equation \ref{eq_acclum}), excluding the radiation coming from inside a sink particle
accretion radius means that the luminosity of an accreting stellar object is potentially underestimated by a factor of 
$\approx R_*/r_{\rm acc}$.  Taking a protostellar radius of $R_* \approx 3-5$~R$_\odot$
(typical for protostars of mass $0.1-2$~M$_\odot$ accreting at a rate of $10^{-5}$~M$_\odot$~yr$^{-1}$; \citealt{HosOmu2009}), 
this is a factor of $\approx 200-300$ when using an accretion
radius of $r_{\rm acc}=5$~AU and a factor of $\approx 20-30$ when using $r_{\rm acc}=0.5$~AU. 

\cite{Bate2009b} performed radiation hydrodynamical calculations of star formation that were
similar to the calculation presented in this paper, but for smaller 50-M$_\odot$ clouds.  He performed
calculations using sink particle accretion radii of 5~AU and 0.5~AU (and, therefore, different 
fractions of the accretion luminosity) to determine the effect on the fragmentation.  The initial conditions
for these calculations were identical to the original barotropic calculation of \cite{BatBonBro2003}
which used $r_{\rm acc}=5$~AU.
To investigate this issue further for this paper, we performed another calculation, identical
to those of \cite{Bate2009b}, but using accretion
radii of just 0.05~AU (i.e.\ only $\approx 2-3$ times larger than a rapidly accreting low-mass star).
In Fig.~\ref{convergence}, we plot the total stellar mass (i.e. the mass in sink particles) and the 
number of sink particles as functions of time.  
We also plot the number of sink particles versus the total stellar mass.
It is clear that the main effect of including radiative feedback is captured when going from 
a barotropic equation of state 
to radiation hydrodynamics and $r_{\rm acc}=5$~AU which reduces the number of objects formed by more than a factor of 3.  Decreasing the accretion radii from 
5~AU to 0.5~AU to 0.05~AU has a progressively smaller and smaller effect.  Although at 
$t=1.38$~$t_{\rm ff}$ the calculation with $r_{\rm acc}=0.05$~AU had formed 11 objects while the 
calculation with $r_{\rm acc}=0.5$~AU had formed 8 objects, the time
evolution of the total stellar mass is almost identical for the two smallest accretion radii and until 
$t=1.35$~t$_{\rm ff}$ (before the onset of the second burst of star formation; \citealt{BatBonBro2003}) the number of stars only differs by one.  

Why should increasing the accretion luminosity by a factor of 100 have such little effect compared to
switching from a barotropic to radiation hydrodynamical calculation?  There are
three main reasons.  First, the although the barotropic equation of state models the evolution
of the {\it maximum} temperature/density during the collapse to form a protostar reasonably well,
the temperature surrounding the object {\it even before} stellar core formation may be underestimated
by up to an order of magnitude, particularly in a protostellar disc \citep{WhiBat2006}.  Thus,
using radiation hydrodynamics rather than a barotropic equation of state dramatically decreases
the propensity for massive discs to fragment even without stellar feedback.  
In \cite{BatBonBro2003}, three-quarters of the
brown dwarfs formed by disc fragmentation \citep{BatBonBro2002a}, so this has a particularly 
important effect on the formation of low-mass objects.  Second, in the envelope surrounding
a protostar, the gas temperature depends on luminosity roughly as $L_*^{1/4}$.  Therefore,
a large change in the luminosity does not actually translate into a large change in temperature.
The Jeans mass scales with temperature as $T_{\rm g}^{3/2}$, so this translates into a change in the
Jeans mass of $L_*^{3/8}$.  Therefore, changing the accretion radii from 5~AU, to 0.5~AU to 0.05~AU, only decreases
the Jeans mass near an existing protostar by factors of $\approx 8$, 3.4, and 1.4 compared to 
that expected from including the full accretion luminosity.  Finally, 
protostars that form near each other usually form in a short period of time.  Since when a
new stellar core first forms most of the mass is still in the disc \citep{Bate1998,Bate2010,MacMat2011}, the
the accretion luminosity of the stellar core does not exceed the luminosity of the accreting first core/disc
for some time.  If the nearby fragmentation occurs before the stellar luminosity becomes
significant \citep[e.g.][]{Bate2011}, the luminosity of the nearby protostar will be a good 
approximation despite the use of a sink particle. 

\subsubsection{Problems with estimating sink particle luminosity}

Some studies have elected to try to include radiative feedback from inside the sink particle
radius \citep[e.g.][]{Offneretal2009, UrbMarEva2010, KruKleMcK2011}.   
These investigations employ sink particles with accretion
radii in excess of 100~AU, so if nothing was done the radiative feedback
from within this radius would be underestimated by a factor $\gsim 10^4$.
However, such attempts to include the missing radiative feedback rely on many 
approximations and assumptions.  Problems include: deciding on the mass distribution 
within the accretion radius (for example, how much mass is in a disc versus a stellar
object, or whether there is a single or multiple stellar system inside the accretion radius); 
deciding how much energy comes out in other forms of feedback (e.g. kinetic feedback
from jets, outflows, and winds); deciding how much energy is advected into the stellar object
rather than radiated away; and deciding whether accretion is continuous or occurs in bursts.
This list is not exhaustive, but it gives an indication of the difficulties associated with trying to accurately
estimate the level of radiative feedback coming from inside the sink particle accretion radius.
We now briefly discuss each of these problems and conclude that, generally, the calculations
of \cite{Offneretal2009}, \cite{UrbMarEva2010}, and \cite{KruKleMcK2011} likely {\it overestimate} the
effects of radiative feedback.

The simplest assumption is that all of the mass that enters a sink particle is immediately 
incorporated into a stellar object \cite[e.g.][]{Offneretal2009,UrbMarEva2010, KruKleMcK2011}.
However, when using sink particles with sizes $>100$~AU a considerable
fraction of the sink particle mass will still be in a protostellar disc.  In fact, early in the
star formation process, the vast majority of the mass will be in a disc rather than
the stellar object.  \cite{Bate1998} showed that the first cores formed from the collapse
of rotating molecular cloud cores can in fact be pre-stellar discs (e.g.\ 50~AU in radius) that may last 
thousands of years before a stellar core forms.  Soon after stellar core
formation, the disc mass can exceed the stellar mass by up to two orders of magnitude
\citep{Bate1998,Bate2010,MacInuMat2010,MacMat2011, Bate2011}.  Thus, allocating all of the mass of a sink particle
to the stellar object can overestimate the level of radiative feedback soon after star formation
by up to two orders of magnitude.  

Before stellar core formation, this overestimate can be
even worse.  During the thousands of years between first core/pre-stellar disc formation and the
formation of a stellar core, the
typical luminosity of a first core/disc ranges from $\approx 0.004-0.1$~L$_\odot$ 
depending on its rotation rate \citep{SaiTom2006, SaiTom2011}.  However, the mass
within 100~AU can be substantial (e.g.\ up to $0.2$~M$_\odot$; \citealt{Bate2011}).  
Assuming that all of this mass is
in a stellar core with a radius $\approx 3$~R$_\odot$ and accreting at a rate of $1\times 10^{-5}$
solar masses per year (a typical accretion rate of a young object) gives an accretion
luminosity of 20~L$_\odot$ (i.e.\ $200-5000$ times greater than the true value).  This is particularly 
important because if fragmentation occurs near to an existing protostar it often occurs 
shortly after the first protostar formed.

Later in the star formation process, it may perhaps be assumed that the mass of the disc 
is less than that of the stellar object (i.e.\ the stellar mass should be wrong by less than a factor
of two; \citealt{KraMat2006,KraMatKru2008,Kratteretal2010}).  However, even at this stage significant uncertainties remain.  Another potentially big problem
is that of multiple systems.  For a low-mass sink particle, whether a sink particle contains a
single object or a binary is not very important because, as mentioned above, the intrinsic
stellar luminosity is usually negligible compared to the accretion luminosity and, for a given accretion
rate, all low-mass protostars are expected to have similar radii.  However, for more massive
protostars the intrinsic stellar luminosity is expected to dominate over the accretion luminosity.
For zero-age-main-sequence stars with masses between $4-9$~M$_\odot$ (depending on accretion rate) 
and 20~M$_\odot$ the intrinsic stellar luminosity scales as $\propto M_*^{7/2}$.  Furthermore, observations show that massive stars preferentially have massive companions \citep[e.g.][]{KobFry2007}.
Therefore, if a high-mass sink particle actually contains an equal-mass binary rather than a single object,
its luminosity will be overestimated by a factor of $2^{7/2-1} \approx 5.7$ or, for an equal-mass 
triple system, a factor of 16.  This problem does not occur in the simulations presented in this paper
because they resolve the opacity limit for fragmentation and, thus, are expected to capture the formation of all stars and brown dwarfs.  Furthermore, no massive protostars are formed.   However, if sink particles with large accretion radii are used in calculations that do not resolve the opacity limit for fragmentation, then some massive sink particles may contain a multiple systems.

Another problem with estimating the radiative feedback is determining the fraction of 
gravitational energy which is liberated as radiation rather than other forms of energy or
retained within the accretion radius.  In reality, some fraction of the energy will exit the
sink particle accretion radius as kinetic energy of jets, outflows, and winds.  \cite{OffMcK2011}
estimate that 1/4 of the gravitational energy may exit in kinetic form rather than as radiation for
low-mass stars.  For high-mass stars, the fraction is unknown.

During the process of mass accretion onto a stellar object from a circumstellar disc, some 
of the energy will be advected into the star rather than all of it being radiated away. This 
efficiency factor is not at all well understood.  It is usually assumed that this fraction is very small
\citep*[e.g.][]{BarChaGal2009}, however if it is substantial this can lead to very different evolutionary paths of
the stellar object \citep*[e.g.][]{HarZhuCal2011}.  \cite*{HosOffKru2011} performed pre-main-sequence stellar evolution calculations with either `hot' or `cold' accretion and found that the protostellar radius could be factors of $2-4$ larger with `hot' accretion.  This translates into an accretion luminosity that is $2-4$ times lower than that predicted by `cold' accretion models.

The final problem we discuss here is one that was identified more than two decades ago from 
observations of protostars.  \cite{Kenyonetal1990} found that low-mass protostars are under luminous by 
two orders of magnitude when compared to simple estimates of protostellar accretion.  The survey of \cite{Evansetal2009} recently confirmed this discrepancy, which is know as the ``protostellar luminosity problem".   Several solutions or partial solutions have been proposed to solve the problem \citep[see][]{OffMcK2011}.  As mentioned above,
some of the gravitational potential energy may be released in non-radiative forms (e.g. kinetic
energy) and some may be advected into the stellar object leading to larger radii than usually assumed.
Another possibility, first raised in the discovery paper \citep{Kenyonetal1990} is that protostars do not
accrete steadily, but that most of their mass is accreted in short bursts of accretion.  During this
time, the accretion luminosity would be very high, but most of the time they would be in a low
luminosity state.  Studies of such bursty accretion have become quite popular recently \citep[e.g.][]{VorBas2005,ZhuHarGam2009,BarChaGal2009}.
In terms of the effects of radiative feedback on star formation, if protostars spent the vast most of their
time in a low-luminosity state, this would be similar to reducing the radiative feedback from the
protostar to that emitted during the low luminosity state and ignoring the brief periods of high 
luminosity.  Recently, this argument has been used by \cite*{StaWhiHub2011} to argue that
those calculations that include continuous radiative feedback from sink particles may overestimate
the effects of radiative feedback and, therefore, underestimate the amount of disc fragmentation
and the formation of low-mass objects.

\subsubsection{Summary for sink particle luminosity}

In summary, there is no easy way to accurately include the radiative feedback from a sink particle
on the rest of a radiation hydrodynamical calculation.  It is often assumed that including 
radiative feedback from inside the sink particle radius will
be more accurate than excluding it \citep{Offneretal2009,UrbMarEva2010,KruKleMcK2011}.  
However, as the examples above show, this is far from certain.  
Indeed, as implemented in the literature, radiative feedback from sink particles
almost certainly {\it overestimates} the effects of radiative feedback.  Usually this overestimate
is expected to be at the level of factors of $\approx 2-4$, but at the earliest stages of protostar formation the
overestimate may be as much as 3 orders
of magnitude.  This early radiative feedback would eventually be emitted by the source, but using a simple prescription it is emitted too early and may affect fragmentation at early times. Thus, all that can really be 
said at the present time is that the actual affect of radiative feedback probably lies somewhere 
between the results obtained by excluding radiative feedback and the results that 
attempt to include radiative feedback.   

The choice made in this paper is to neglect radiative feedback coming from inside the 
sink particle radius, but to use as small an accretion radius as is computationally feasible.  
This model is elegant in that the results depend on a single parameter --- the sink particle 
accretion radius.    The effect of the missing radiation on the degree of fragmentation 
is tested by using smaller calculations with different accretion radii (Fig.~\ref{convergence}).  
By decreasing the accretion radius by an {\it order of magnitude} (from 0.5 to 0.05~AU) we find 
that using sink particles with accretion radii of 0.5~AU {\it may} 
overestimate the degree of fragmentation by up to a factor of $\approx 11/8 = 1.4$.   
To put this in context, \cite{KruKleMcK2011}, who attempt to include sink particle luminosity, 
reduce their accretion radius by only a factor of {\it two} (from 400 AU to 200 AU) and produce 
70\% more stars.  

Naively, the calculations discussed in the remainder of this paper using sink particles with accretion radii of 0.5~AU underestimate the accretion luminosity by factors of $20-30$ (taking protostellar accretion radii of $3-5$~M$_\odot$; \citealt{HosOmu2009}).
However, it has been pointed out (P. Andr\'e, private communication) that 
because protostars are {\it observed} to be under-luminous by 1--2 orders of magnitude
\citep{Kenyonetal1990,Evansetal2009},
neglecting the radiative feedback from inside sink particles with accretion radii 
of $\approx 0.5$~AU might be more accurate than including the `missing' luminosity.
If this is the case, then the level of radiative feedback in calculation discussed 
in the rest of this paper may be close to reality.

\begin{table*}
\begin{tabular}{lccccccccccc}\hline
Calculation & Initial Gas & Initial  & Accretion & Gravity & Time & No. Stars & No. Brown  & Mass of Stars \&  & Mean  & Median \\
& Mass  & Radius & Radii & Softening & & Formed & Dwarfs Formed & Brown Dwarfs & Mass & Mass  \\
 & M$_\odot$ & pc  & AU & AU & $t_{\rm ff}$ & & & M$_\odot$ & M$_\odot$ & M$_\odot$ \\ \hline
B2009a Main & 500 & 0.404 & 5 & 4 & 1.04 & $\geq$102 & $\leq$119 & 32.6 & 0.15 &  0.06 \\ 
& & &  & & 1.20 & $\geq 235$& $\leq 355$& 92.1 & 0.16 & 0.05  \\ 
& & &  & & 1.50 & $\geq$459 & $\leq$795 & 191 & 0.15 & 0.06 \\  
B2009a Rerun & 500 & 0.404 &  0.5 & 0 & 1.04 & $\geq$94 & $\leq$164 & 32.0 & 0.12 & 0.05\\  
Radiation Hydro & 500 & 0.404 & 0.5 & 0 & 1.04 & $\geq 50$ & $\leq 19$ & 28.4 & 0.41 & 0.24 \\  
& & &  & & 1.20 & $\geq 147$& $\leq 36$& 88.2 & 0.48 & 0.21 \\ \hline  
\end{tabular}
\caption{\label{table1} The parameters and overall statistical results for the two calculations of Bate (2009a) and the calculation presented here.  The initial conditions for all calculations were identical.  The Bate (2009a) calculations used a barotropic equation of state and the main calculation used sink particles with $r_{\rm acc}=5$~AU and gravitational softening inside 4~AU, while the rerun calculation used $r_{\rm acc}=0.5$~AU and no gravitational softening.  The calculation presented here was identical to the rerun calculation, except that it used a realistic equation of state with radiation hydrodynamics.  The calculations were run for different numbers of initial cloud free-fall times due to computational limitations.  Brown dwarfs are defined as having final masses less than 0.075 M$_\odot$.  The numbers of stars (brown dwarfs) are lower (upper) limits because some brown dwarfs were still accreting when the calculations were stopped.  Including radiative feedback decreases the number of objects formed at the same time by a factor of $\approx 3.2-3.7$, and increases the mean and median masses of the objects by factors of $\approx 3$ and $\approx 4$, respectively.  The amount of gas converted into stars only decreases by $4-11$\% compared to the barotropic calculations at the same times.}
\end{table*}

\subsection{Initial conditions}
\label{initialconditions}

The initial conditions for the calculation presented in this paper are identical to those 
of \cite{Bate2009a}.  We followed the collapse 
of an initially uniform-density molecular cloud containing 500 M$_\odot$ of molecular gas.
The cloud's radius was 0.404 pc (83300 AU) giving an initial density of 
$1.2\times 10^{-19}$~g~cm$^{-3}$.  At the initial temperature of 10.3 K, the mean 
thermal Jeans mass was 1 M$_\odot$ (i.e., the cloud contained 500 thermal Jeans masses).  
Although the cloud was uniform in density, we imposed an initial 
supersonic `turbulent' velocity field in the same manner
as \citet*{OstStoGam2001} and \cite{BatBonBro2003}.  
We generated a divergence-free random Gaussian velocity field with 
a power spectrum $P(k) \propto k^{-4}$, where $k$ is the wavenumber.  
In three dimensions, this results in a
velocity dispersion that varies with distance, $\lambda$, 
as $\sigma(\lambda) \propto \lambda^{1/2}$ in agreement with the 
observed Larson scaling relations for molecular clouds 
\citep{Larson1981}.
The velocity field was generated on a $128^3$ uniform grid and the
velocities of the particles were interpolated from the grid.  
The velocity field was normalised so that the kinetic energy 
of the turbulence was equal to the magnitude of the gravitational potential 
energy of the cloud.
Thus, the initial root-mean-square (rms) Mach number of the turbulence 
was ${\cal M}=13.7$.
The initial free-fall time of the cloud was $t_{\rm ff}=6.0\times 10^{12}$~s or 
$1.90\times 10^5$ years.

Since the initial conditions for the calculation are identical to those of
\cite{Bate2009a} and including radiative transfer does not alter the 
temperature of the gas significantly until just before the first protostar forms,
the early evolution of the cloud was not repeated.  Instead, the radiation
hydrodynamical calculation was begun from a dump file from the 
original calculation just before the maximum density exceeded $10^{-15}$~g~cm$^{-3}$.
This method was also used for the radiation hydrodynamical calculations
presented by \cite{Bate2009b}, which were radiation hydrodynamical 
versions of the earlier barotropic calculations published by \cite{BatBonBro2003}
and \cite{BatBon2005}.

\subsection{Resolution}
\label{resolution}

The local Jeans mass must be resolved throughout the calculation 
to model fragmentation correctly (\citealt{BatBur1997, Trueloveetal1997, Whitworth1998, Bossetal2000}; \citealt*{HubGooWhi2006}).  
The original barotropic calculation of \cite{Bate2009a} used $3.5 \times 10^7$ particles to model the 
500-M$_\odot$ cloud and resolve the Jeans mass down to its minimum 
value of $\approx 0.0011$ M$_\odot$ (1.1 M$_{\rm J}$) at the maximum 
density during the isothermal phase of the collapse, 
$\rho_{\rm crit} = 10^{-13}$ g~cm$^{-3}$.  Using radiation hydrodynamics results in
temperatures at a given density no less than those given by the original 
barotropic equation of state \citep[e.g.][]{WhiBat2006} and, thus, the Jeans mass is also resolved in
the calculations presented here.

The calculation was performed on the University of Exeter 
Supercomputer, an SGI Altix ICE 8200.  Using 256 compute cores,
it required 6.3 million CPU hours (i.e. 34 months of wall time).

\begin{figure*}
\centering \vspace{-0.0cm}
    \includegraphics[width=16.2cm]{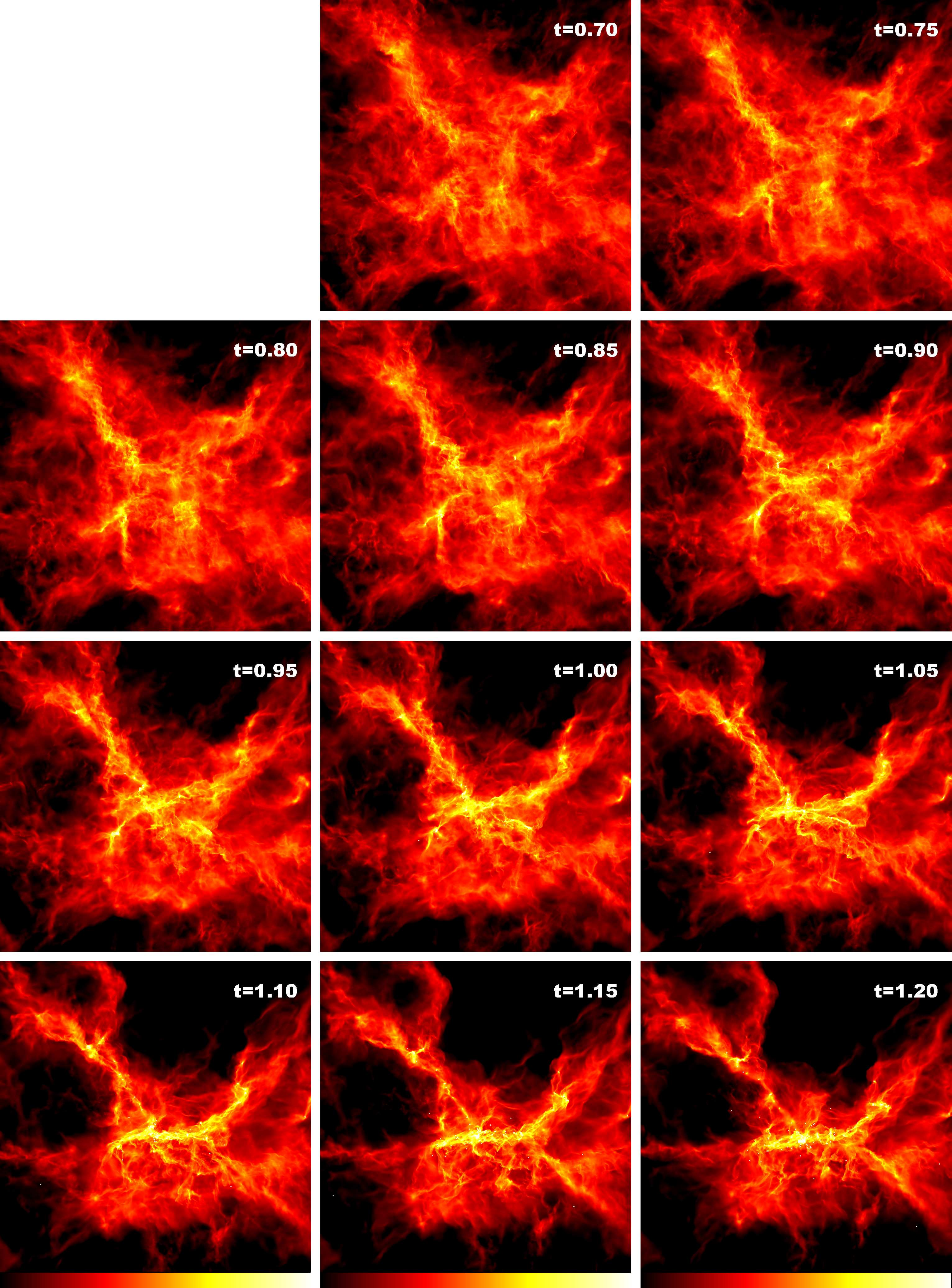}
\caption{The global evolution of column density in the radiation hydrodynamical calculation.  Shocks lead to the dissipation of the turbulent energy that initially supports the cloud, allowing parts of the cloud to collapse.  Star formation begins at $t=0.727~t_{\rm ff}$ in a collapsing dense core.  By $t=1.10t_{\rm ff}$ the cloud has produced six main sub-clusters, and by the end of the calculation these sub-clusters started to merge and dissolve. Each panel is 0.6 pc (123,500 AU) across.  Time is given in units of the initial free-fall time, $t_{\rm ff}=1.90\times 10^5$ yr.  The panels show the logarithm of column density, $N$, through the cloud, with the scale covering $-1.4<\log N<1.0$ with $N$ measured in g~cm$^{-2}$. White dots represent the stars and brown dwarfs.}
\label{global_density}
\end{figure*}

\begin{figure*}
\centering \vspace{-0.0cm}
    \includegraphics[width=16.2cm]{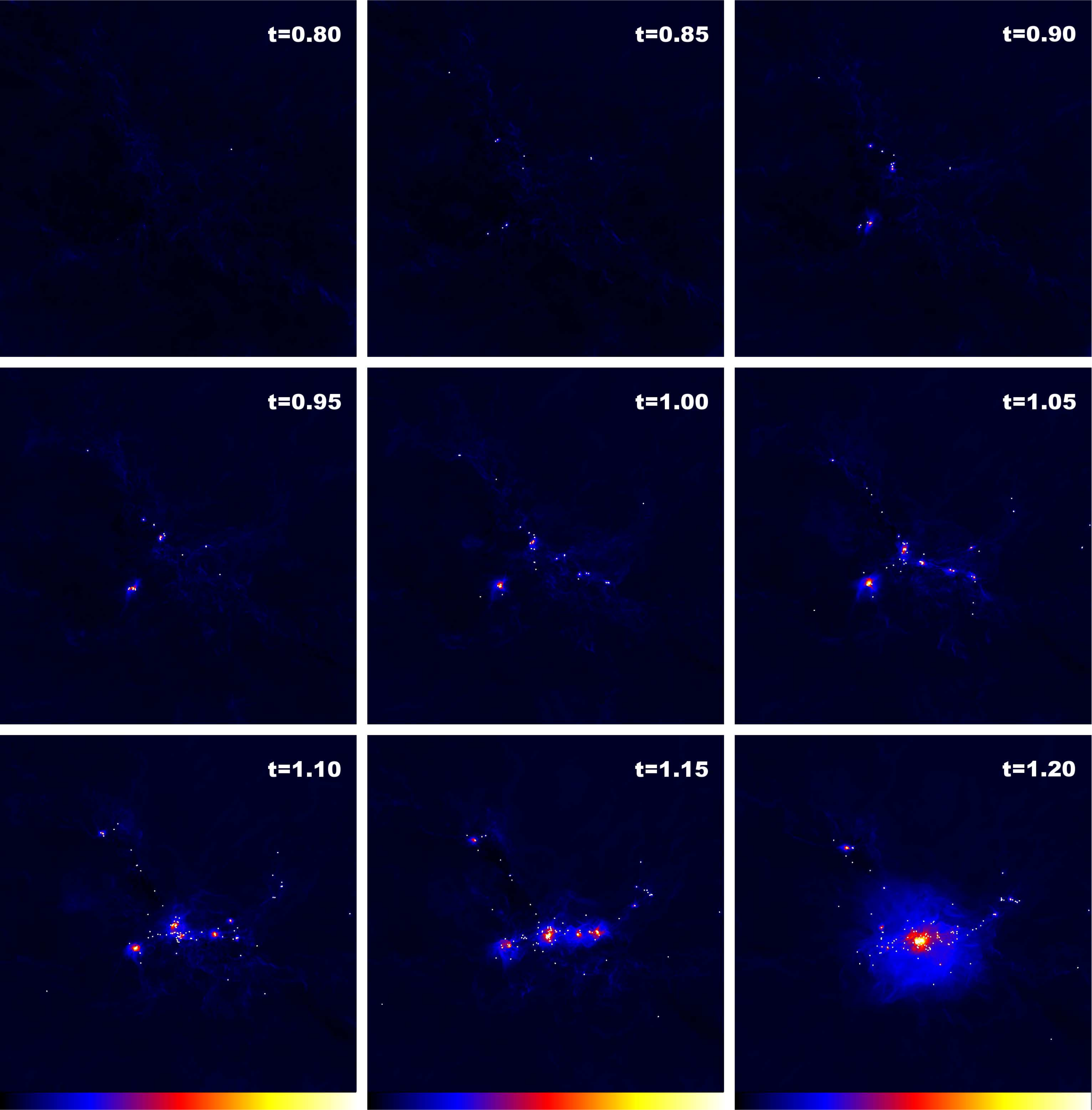}
\caption{The global evolution of gas temperature in the radiation hydrodynamical calculation.  Initially, the gas is almost isothermal on large-scales, but as groups of stars begin to form they locally heat the gas.  Soon after $t=1.15t_{\rm ff}$, the merger of two stellar groups at the centre of the cluster and increased accretion rates onto the stars heats the inner 0.2~pc of the cluster.
Each panel is 0.6 pc (123,500 AU) across.  Time is given in units of the initial free-fall time, $t_{\rm ff}=1.90\times 10^5$ yr.  The panels show the logarithm of mass weighted gas temperature, $T_{\rm g}$, through the cloud, with the scale covering $9-50$~K.  White dots represent the stars and brown dwarfs.}
\label{global_temp}
\end{figure*}

\section{Results}
\label{results}

The calculation presented here is a radiation hydrodynamical version of the barotropic calculations presented by \cite{Bate2009a}.  \citeauthor{Bate2009a} presented the results from two calculations that differed only in the value of the sink particle accretion radius used and the gravitational softening between sink particles.  The main calculation with $r_{\rm acc}=5$~AU produced 1254 stars and brown dwarfs in $1.50 t_{\rm ff}$ (285,350 years) and the rerun calculation used $r_{\rm acc}=0.5$~AU and produced 258 objects in $1.038 t_{\rm ff}$ (197,460 years).  See Table \ref{table1} for a summary of the statistics, including the numbers of stars and brown dwarfs produced by the two calculations, the total mass that had been converted to stars and brown dwarfs, and the mean and median stellar masses.  \cite{Bate2009a} compared a large number of statistical properties of the stars and brown dwarfs formed in the calculations with observations, finding that many were in good agreement with observations (see the Introduction).  In this paper, we use the same analysis methods as those used by \cite{Bate2009a}, and we compare and contrast the results both with the results from the earlier barotropic calculations and with the results of observational surveys.

\begin{figure*}
\centering
    \includegraphics[width=5.8cm]{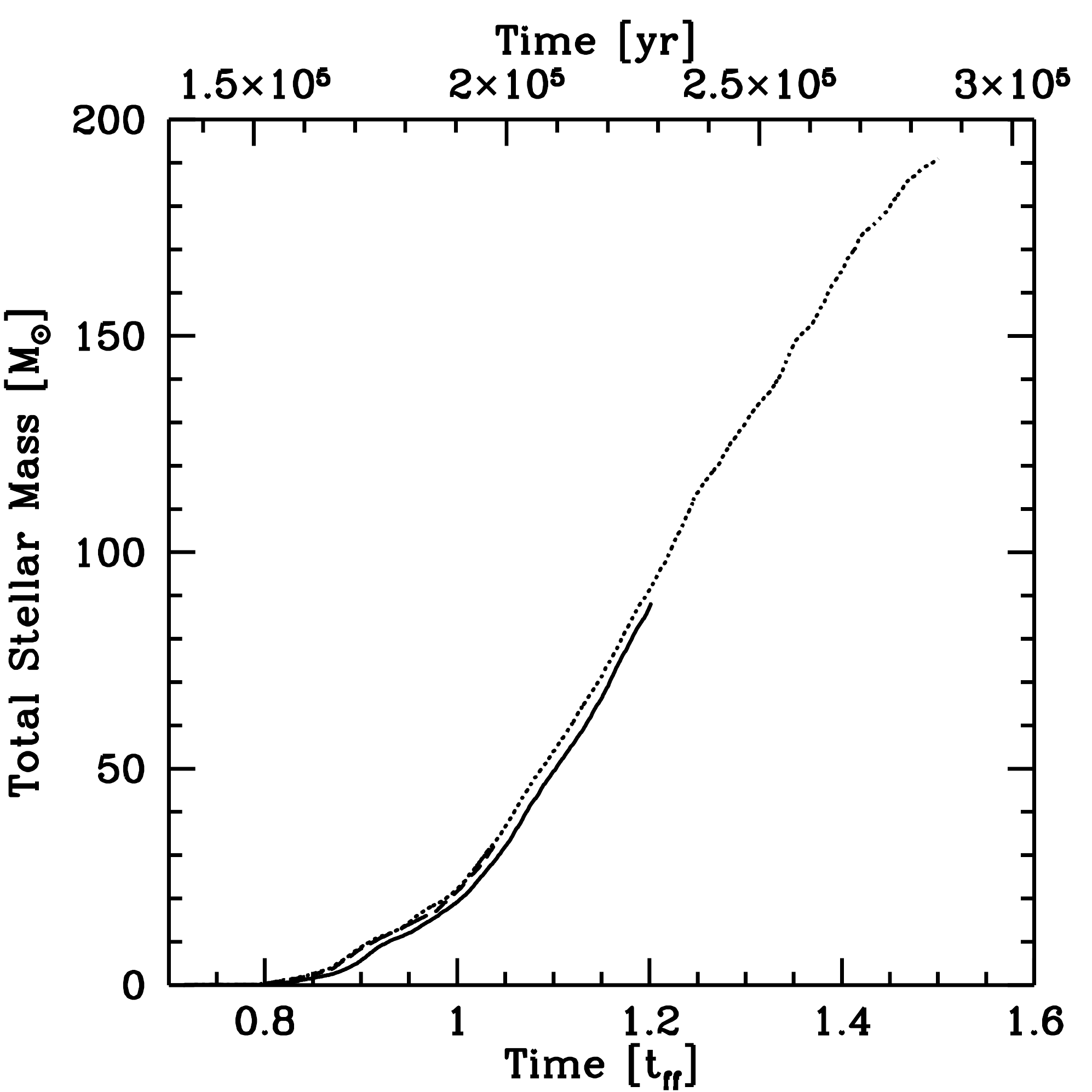}
    \includegraphics[width=5.8cm]{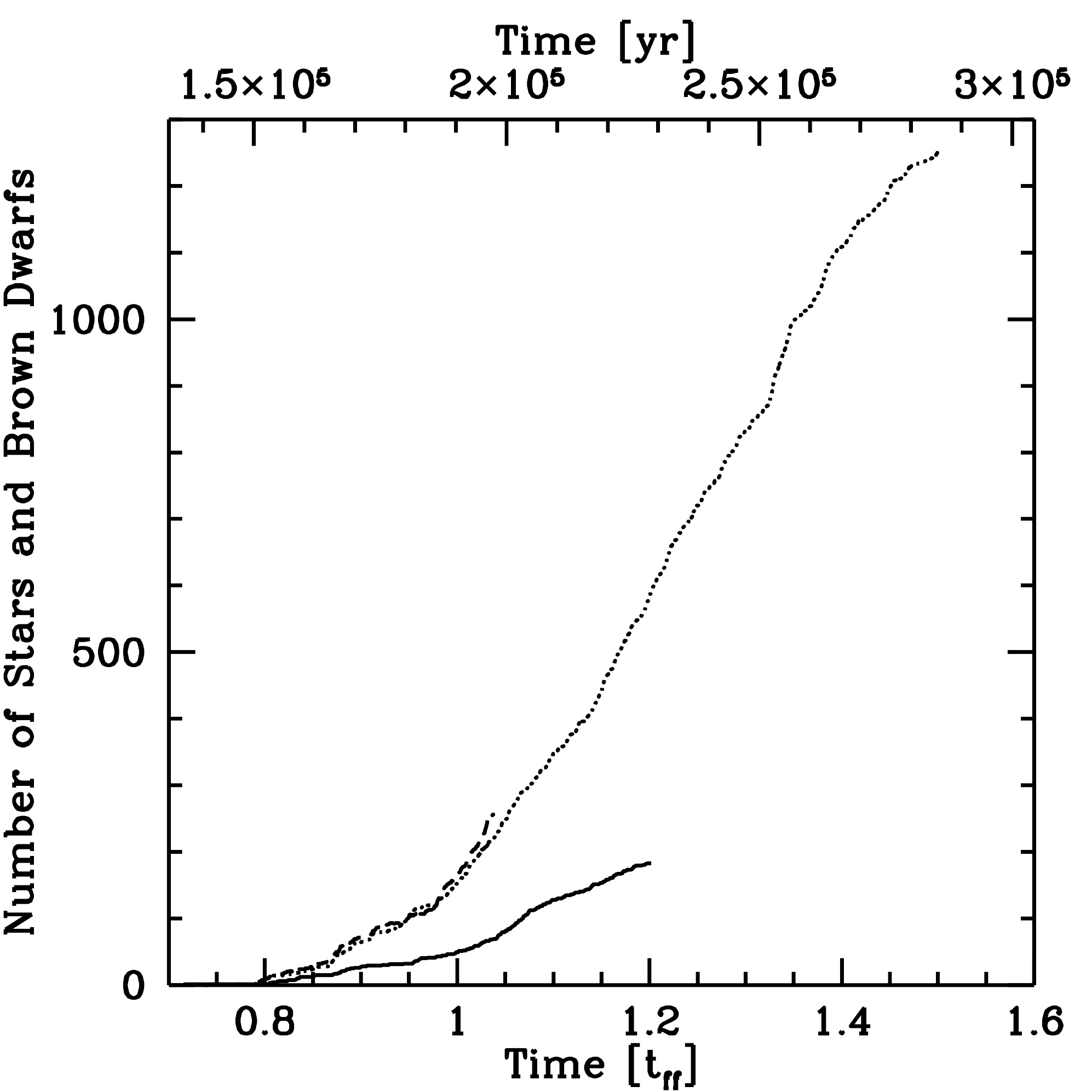}
    \includegraphics[width=5.8cm]{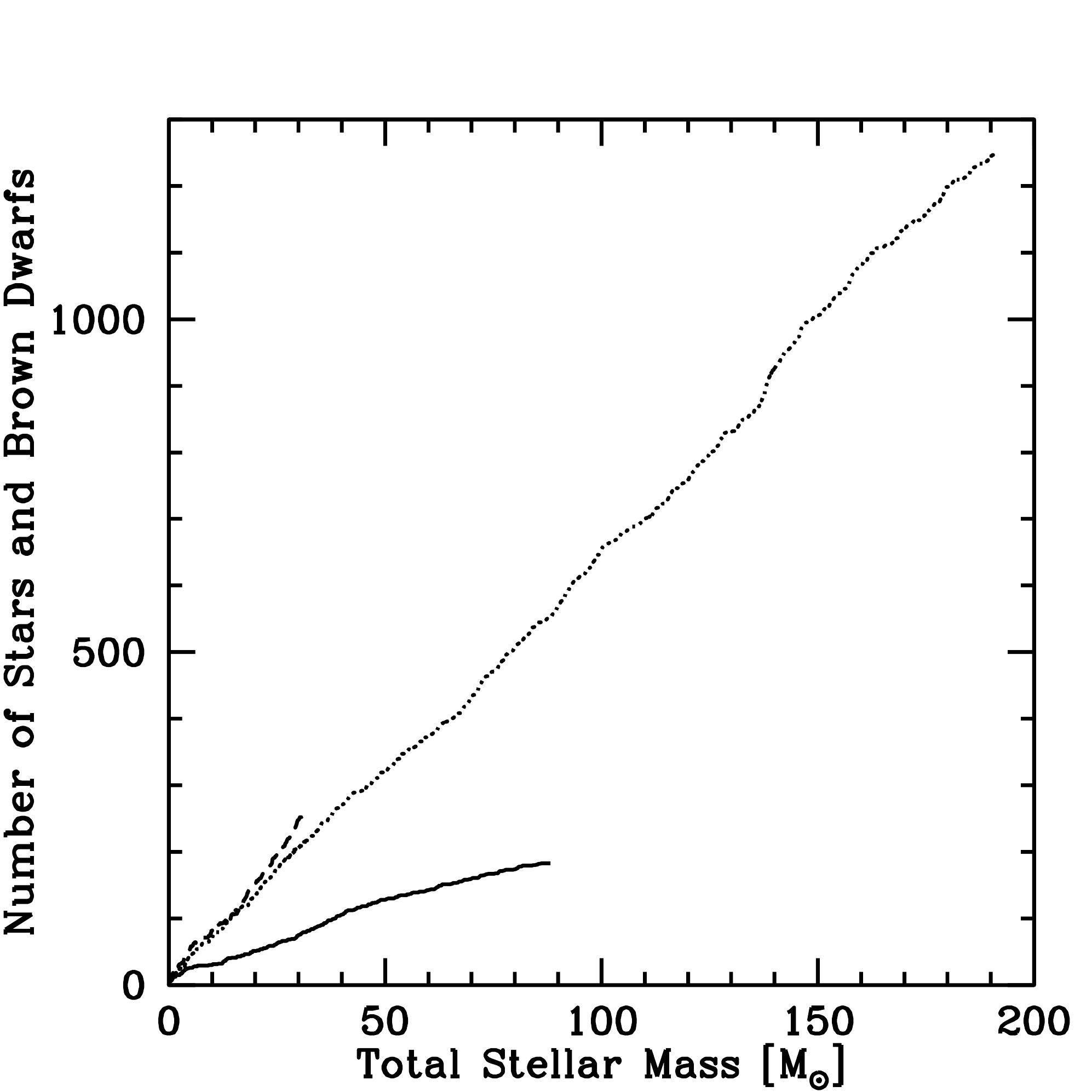}
\caption{The star formation rate obtained using a barotropic equation of state or radiation hydrodynamics for the 500-M$_\odot$ turbulent molecular cloud.  We plot: the total stellar mass (i.e. the mass contained in sink particles) versus time (left panel), the number of stars and brown dwarfs (i.e. the number of sink particles) versus time (centre panel), and the number of stars and brown dwarfs versus the total stellar mass (right panel).  The different line types are for: the main barotropic calculation using $r_{\rm acc}=5$~AU (dotted line; Bate 2009a); the rerun barotropic calculation using $r_{\rm acc}=0.5$~AU (dashed line; Bate 2009a); and the radiation hydrodynamical calculation with $r_{\rm acc}=0.5$~AU (solid line). Time is measured from the beginning of the calculation in terms of the free-fall time of the initial cloud (bottom) or years (top). The rate at which mass is converted into stars is almost unaltered by the introduction of radiative feedback, but the number of stars and brown dwarfs is decreased by a factor of three.  }
\label{massnumber}
\end{figure*}

\subsection{The evolution of the star-forming cloud}

As mentioned in Section \ref{initialconditions}, the radiation hydrodynamical calculation was not re-run from the initial conditions, but was started from the last dump file of the original barotropic calculation before the density exceeded $10^{-15}$ g~cm$^{-3}$.  Before this point the initial `turbulent' velocity field had generated density structure in the gas, some of which was collected into dense cores which had begun to collapse.  Those readers interested in this early phase should refer to \cite{Bate2009a} for figures and further details.

The radiation hydrodynamical calculation was started from $t=0.700~t_{\rm ff}$. In the barotropic calculation, the first sink particle was inserted at $t=0.715~t_{\rm ff}$, some $2.9\times 10^3$ years later.  Using radiation hydrodynamics, the first sink particle is inserted at $t=0.727~t_{\rm ff}$.  The slightly later time ($2.2\times 10^3$ yrs) is primarily because in the original calculation sink particles were inserted when the density exceeded $10^{-10}$ g~cm$^{-3}$ (when the fragment was in the `first hydrostatic core' stage of evolution) whereas with the radiation hydrodynamics we do not insert sink particles until halfway through the second collapse phase at a density of $10^{-5}$ g~cm$^{-3}$ (see Section \ref{sinks} for further details).  The first core phase typically lasts several thousand years, depending on the amount of rotation \citep{Bate1998, SaiTom2006, SaiTomMat2008, Bate2010, MacInuMat2010, MacMat2011, Bate2011} and accretion rate \citep{Tomidaetal2010b}.

In the panels of Figs.~\ref{global_density} and \ref{global_temp}, we present snapshots of the global evolution of the calculation.  Fig.~\ref{global_density} displays the column density (using a red-yellow-white colour scale), while Fig.~\ref{global_temp} displays the mass-weighted temperature in the cloud (using a blue-red-yellow-white colour scale).  The column-density snapshots may be compared with the figures in \cite{Bate2009a} that show the barotropic evolution.  Animations of both the main barotropic calculation and the radiation hydrodynamical calculation can be downloaded from http://www.astro.ex.ac.uk/people/mbate/~. 

\cite{Bate2009b} found that barotropic and radiation hydrodynamical calculations diverge quickly on small scales.  In particular, it was found that many massive circumstellar discs that quickly fragmented in barotropic calculations did not fragment, or took much longer to fragment, when using radiation hydrodynamics.  The accretion luminosity released as gas falls onto a disc and then spirals in towards the central protostar is often sufficient to heat a disc and prevent its fragmentation.  An individual low-mass protostar can also produce substantial heating of the surrounding cloud out to thousands of AU (depending on the protostar's accretion rate) which can inhibit the fragmentation of nearby collapsing filaments.  \cite{Bate2009b} found that the temperature field around rapidly accreting protostars could vary significantly on timescales of hundreds to thousands of years due to variations in the accretion rates of the protostars and their discs, particularly when protostars undergo dynamical interactions that perturb their discs.  The same effects are found in the calculation presented here, though with an order of magnitude more objects it is impossible for us to compare and contrast the barotropic and radiation hydrodynamical evolutions of individual objects as was done by \cite{Bate2009b}.  The temperature variability is best seen in an animation. 

The large-scale influence of the radiative feedback from the young protostars can be seen in Fig.~\ref{global_temp}.  Each panel in this figure measure 0.6 pc (123,500 AU) across. As in the calculations of \cite{BonBatVin2003} and \cite{Bate2009a}, the star formation in the cloud occurs in small groups, often formed within larger-scale filaments.  Initially, each group contains only a few low-mass objects and the heating of the surrounding gas is limited to their immediate vicinity (a few thousand AU).  However, as the stellar groups grow in number and some of the stars grow to greater masses, the heating can be seen to extend to larger and larger scales.  At $t \approx 1.15~t_{\rm ff}$, the merger of several stellar groups occurs near the centre of the cloud and the protostellar accretion rates also increase.  This produces a burst of radiation that heats the centre of the cloud out to distances of $\approx 0.2$~pc ($\approx 80,000$ AU).  Several bursts between this time and the end of the calculation ($t=1.20~t_{\rm ff}$) continuously heat the centre of the cloud.

\cite{Bate2009a} followed the main barotropic calculation to $1.50~t_{\rm ff}$ (285\,350 yr) at which time 38\% of the gas had been converted into 1254 stars and brown dwarfs.  Unfortunately, due to the extra computational expense of resolving the gas near sink particles to 0.5~AU and the implicit radiative transfer we are only able to follow the radiation hydrodynamical case to $1.20~t_{\rm ff}$ (228\,280 yr) which is $9.0 \times 10^4$ years after the star formation began.  At this stage 88.2 M$_\odot$ of gas (18\%) has been converted into 183 stars and brown dwarfs.  Table \ref{table1} gives the numbers of stars and brown dwarfs and their mean and median masses for the radiation hydrodynamical calculation, and for the barotropic calculations.  The information is given at the end points of each of the calculations and, where possible, at the same times to allow direct comparison between the different calculations.

As was found by \cite{Bate2009b} for smaller 50-M$_\odot$ clouds  (see also Fig.~\ref{convergence}), the radiative feedback dramatically reduces the number of objects formed.  Comparing the main barotropic calculation and the radiation hydrodynamical calculation at $t=1.20~t_{\rm ff}$, the former had produced 590 objects while the latter has only produced 183 (less than 1/3).  However, this reduction in the number of objects is not the same for all stellar masses.  The main barotropic calculation produced 235 stars and 355 brown dwarfs in the same time that the radiation hydrodynamical calculation produced 147 stars and 36 brown dwarfs.  Thus, the number of stars has only been reduced to 63\% of the barotropic value, but the number of brown dwarfs has been slashed to just 10\%!  This change in the stellar mass distribution is also reflected in the mean and median masses (Table \ref{table1}) with the mean mass increasing by a factor of three from 0.16~M$_\odot$ to 0.48~M$_\odot$ and the median mass increasing by a factor of 4.2 from 0.05~M$_\odot$ to 0.21~M$_\odot$  (measured at $1.20~t_{\rm ff}$).  The maximum stellar mass is 3.84~M$_\odot$, whereas the main barotropic calculation had produced a star of mass 3.13~M$_\odot$ at the same time and went on to produce a star of 5.4~M$_\odot$ by the end of the calculation.  We investigate the change in the distribution of stellar masses further in the next section.  Before that, in Fig.~\ref{massnumber} we examine the star formation rate in terms of mass and number of stars and brown dwarfs.  In the left panel, we plot the total stellar mass as a function of time for the barotropic calculations of \cite{Bate2009a} and the radiation hydrodynamical calculation.  It can be seen that in terms of stellar mass, there is a slow star formation rate of $\approx 5\times 10^{-4}$~M$_\odot$~yr$^{-1}$ from $\approx 0.8-1.0~t_{\rm ff}$ followed by an increase to $\approx 2\times 10^{-3}$~M$_\odot$~yr$^{-1}$ after $\approx 1.0~t_{\rm ff}$.  The star formation rate is quite constant after this transition and, despite the dramatic effect of the radiative feedback in heating the cloud (Fig.~\ref{global_temp}), there is no evidence of a decreasing rate near the end of the radiation hydrodynamical calculation.  In the main barotropic calculation, there is a hint of a decrease after $1.40~t_{\rm ff}$.   This is not surprising since after this point more than a third of the gas has been converted to stars and some of the remaining gas is drifting off to large distances.  

Comparing the barotropic and radiation hydrodynamical calculations, the star formation rate in terms of M$_\odot$~yr$^{-1}$ is almost entirely unaffected by the inclusion of radiative feedback. At the end of the radiation hydrodynamical calculation, 88.2~M$_\odot$ of gas has been converted to stars while, at the same time, 92.1~M$_\odot$ of gas had been converted to stars in the main barotropic calculation (a difference of only 4\%).  \cite{Bate2009b} also found that radiative feedback had little effect on the rate at which gas was converted to stars --- with one type of initial condition the rate decreased by 4\%, while in the other it increased by 15\%.  Similarly, \cite{KruKleMcK2011} recently modelled star formation in a 1000~M$_\odot$  cloud and found very similar star formation rates in terms of M$_\odot$~yr$^{-1}$ with and without radiative feedback.

We note that a general problem with hydrodynamical models of star formation in bound molecular clouds (whether they include radiative feedback or not) is that the star formation rate is much quicker than observed.  The observed star formation efficiency per free-fall time is $3-6$\% \citep{Evansetal2009}, whereas the peak rate in the calculation presented here is 76\% (i.e. an order of magnitude greater).  The solution(s) to this problem may be that star formation occurs in globally unbound molecular clouds \citep{ClaBon2004,Clarketal2005}, or that magnetic support \citep{PriBat2008,PriBat2009}, kinetic feedback \citep{MatMcK2000,KruMatMcK2006,NakLi2007} or a combination \citep[e.g.][]{Wangetal2010} reduce the star formation rate.   Investigating these effects is beyond the scope of this paper, but they certainly warrant future investigation.

Rather than altering the rate at which gas is converted into stars, the effect of radiative feedback is to convert mass into {\it fewer} stars and brown dwarfs by inhibiting fragmentation of the gas.  The reduction in the rate of production of new protostars is clear from the centre and right panels of Fig.~\ref{massnumber}.  Throughout the evolution, the radiation hydrodynamical calculation consistently produces new objects at about 1/3 the rate of the barotropic calculations.  However, as with the rate at which mass is converted into stars, there is no evidence that the radiative heating of the central regions of the cloud is reducing the rate at which new stars are being formed.  This is in contrast to the results obtained by \cite{KruKleMcK2011}, who found that the initial rate of protostar formation was similar with and without radiative feedback, but that as the calculation progressed the rate of protostar formation dropped off much faster with radiative feedback than without.  Part of this difference between the results here and those of \citeauthor{KruKleMcK2011} is certainly due to the different initial conditions.  The initial conditions of \citeauthor{KruKleMcK2011} are more than seven times denser (twice the mass within a cloud radius of 0.26~pc rather than 0.4~pc) and they are centrally condensed rather than uniform in density.  This produces qualitatively different protostar formation even {\it without} radiative feedback: \citeauthor{KruKleMcK2011} find a monotonically decreasing protostar formation rate as a function of mass without feedback, whereas we obtain a constant rate (dotted line, right panel of Fig.~\ref{massnumber}).  The centrally-condensed initial conditions of \citeauthor{KruKleMcK2011} also favour the formation of massive stars \citep{Girichidisetal2011}, while uniform initial conditions result in the formation of many sub-clusters \citep{BonBatVin2003} which only later merge into a single cluster \citep{Bate2009a}.  The higher density cloud of \citeauthor{KruKleMcK2011} gives a star formation rate of $2.4 \times 10^{-2}$~M$_\odot$~yr$^{-1}$, more than an order of magnitude greater than obtained here.  The combination of a higher star formation rate and a bias towards massive star formation mean that when radiative feedback is included it has a much greater effect.  Finally, as discussed in Section \ref{limitations},  \citeauthor{KruKleMcK2011} include radiative feedback from sink particles which may be over-estimated in contrast to our feedback which may be under-estimated.

If the radiation hydrodynamical calculation presented here was followed further then, as shown by \cite{Bate2009a}, the eventual result would be that most of the small groups and sub-clusters would merge into a single cluster surrounded by a halo of ejected objects.  Unfortunately, the calculation cannot be followed that far due to computational limitations.  However, it is already clear by comparing the main barotropic calculation with the radiation hydrodynamical calculation at the same time, that the number of ejected objects is substantially lower with radiative feedback.  This is because of the reduction in small-scale fragmentation.  The ejected objects in the barotropic calculations primarily come from small dense multiple systems.  With radiative heating, each multiple system or stellar group contains fewer objects so their are fewer dynamical interactions and ejections.


\begin{figure}
\centering
    \includegraphics[width=8.4cm]{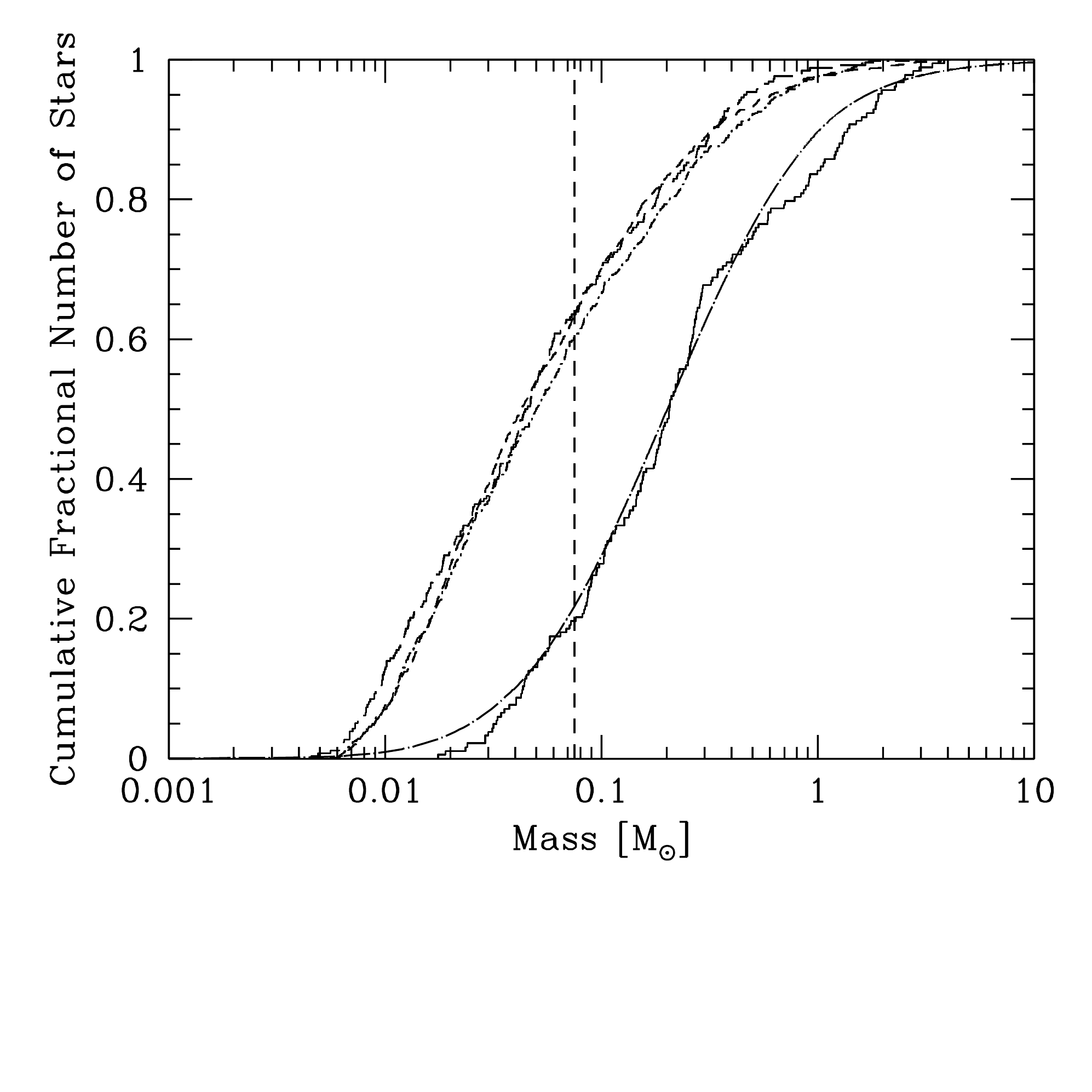}\vspace{-1.5cm}
\caption{The cumulative IMF produced at the end ($t=1.20~t_{\rm ff}$) of the radiation hydrodynamical calculation (solid line), compared to the IMF obtained from the main barotropic calculation by Bate (2009a) at $t=1.20~t_{\rm ff}$ (long-dashed line) and $t=1.50~t_{\rm ff}$ (dotted line), and the rerun barotropic calculation at $t=1.04~t_{\rm ff}$ (short-dashed line).  We also plot using the dot-long-dashed line, the cumulative IMF from the parameterisation of the observed IMF by Chabrier (2005).  The vertical dashed line marks the stellar/brown dwarf boundary.  The form of the barotropic IMF does not change substantially from 1.04 to 1.50~${\rm ff}$ (Kolmogorov-Smirnov tests show the three distributions to be indistinguishable), but introducing radiative feedback substantially increases the median stellar mass and changes the IMF from being dominated by brown dwarfs to being dominated by stars.  A Kolmogorov-Smirnov test gives less than 1 chance in $10^{19}$ that the barotropic and radiation hydrodynamic IMFs at $t=1.20~t_{\rm ff}$ are drawn from the same underlying distribution.  On the other hand, a Kolmogorov-Smirnov test gives a 50.5 percent probability that the radiation hydrodynamic IMF could have been drawn from Chabrier's fit to the observed IMF (i.e. the two mass functions are indistinguishable).}
\label{cumimf_baro}
\end{figure}

\subsection{The initial mass function}
\label{imf}

In past barotropic calculations of the formation of stellar groups and clusters, the number of brown dwarfs produced has consistently been greater than the number of stars, in disagreement with observations of Galactic star-forming regions \citep{BatBonBro2003,BatBon2005,Bate2005,Bate2009a,Bate2009c}.  The radiation hydrodynamical calculations of \cite{Bate2009b} showed that radiative feedback provides a potential solution to this over production of low-mass objects.  Although the clouds studied were an order of magnitude less massive than the calculation presented here, \cite{Bate2009b} found that the number of objects formed when including radiative feedback was reduced by a factor of $\approx 3.8$ compared to the barotropic calculation of \cite{BatBonBro2003} and the mean and median masses of the objects increased by factors of $\approx 4$.   The effects of radiative feedback found here are very similar: a reduction in the number of objects by a factor of $590/183=3.2$ and increases of the mean and median masses by factors of $\approx 3$ and $\approx 4$, respectively (Table \ref{table1}).  However, with an order of magnitude more objects we are able to examine the change in the mass function in more detail.  

\begin{figure}
\centering
    \includegraphics[width=8.4cm]{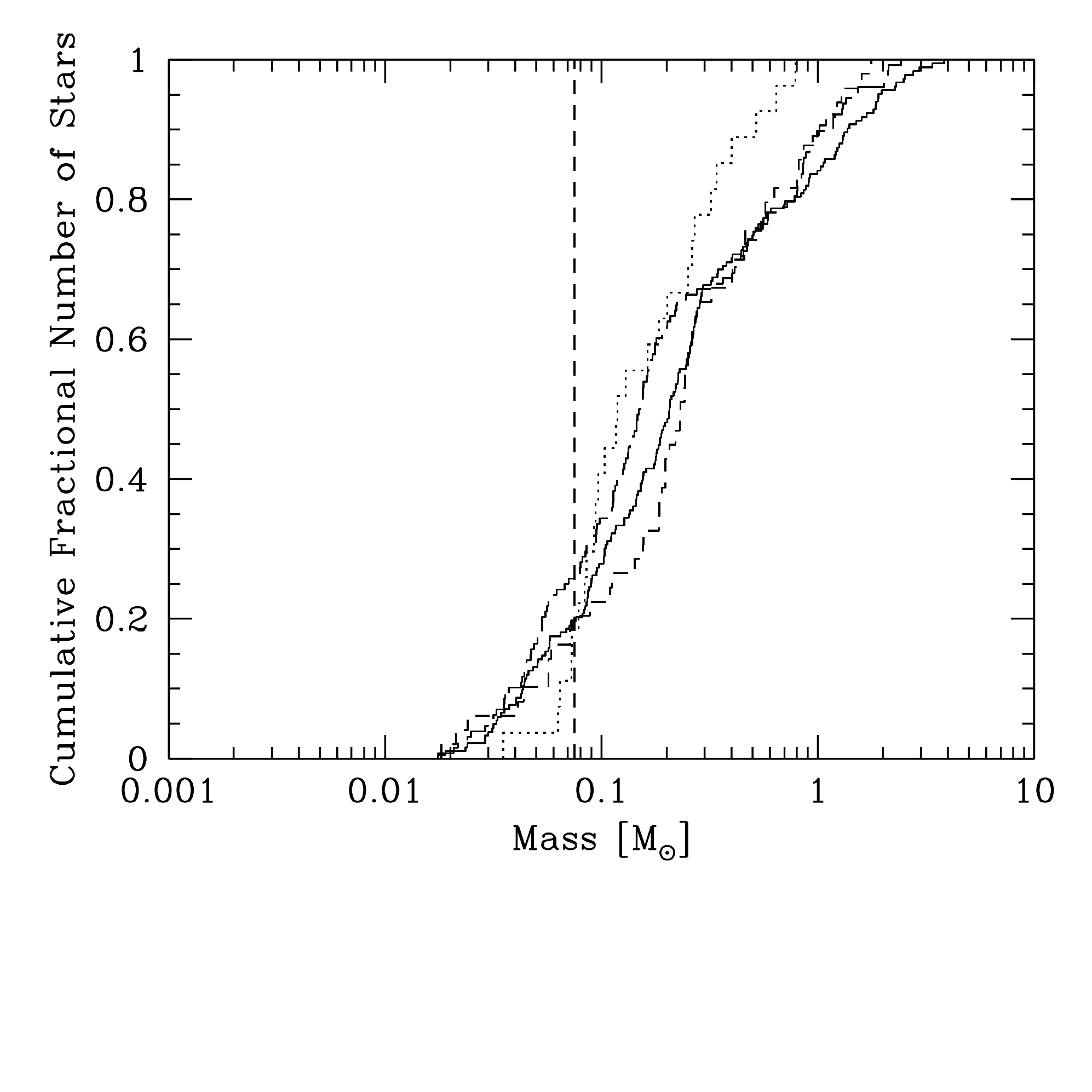}\vspace{-1.5cm}
\caption{The cumulative stellar mass distributions produced at various times during the radiation hydrodynamical calculation.  The times are $t=0.90~t_{\rm ff}$ (dotted line), $t=1.00~t_{\rm ff}$ (short-dashed line), $t=1.10~t_{\rm ff}$ (long-dashed line), and $t=1.20~t_{\rm ff}$ (solid line).  The vertical dashed line marks the stellar/brown dwarf boundary.  The form of the stellar mass distribution does not change significantly during the radiation hydrodynamical calculation, though as more stars are formed the maximum stellar mass increases.}
\label{cumimf_RT}
\end{figure}

\begin{figure*}
\centering
    \includegraphics[width=8.4cm]{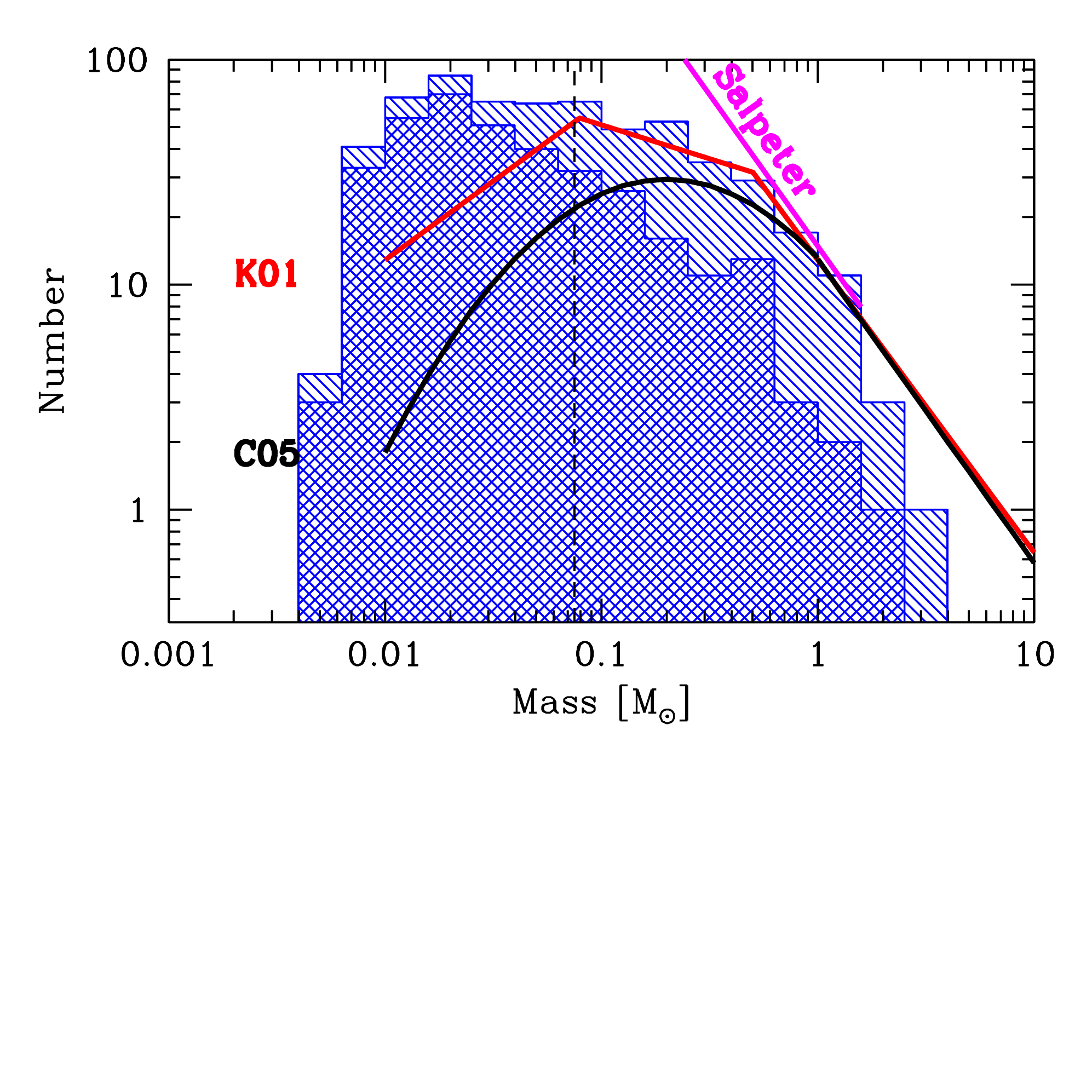}
    \includegraphics[width=8.4cm]{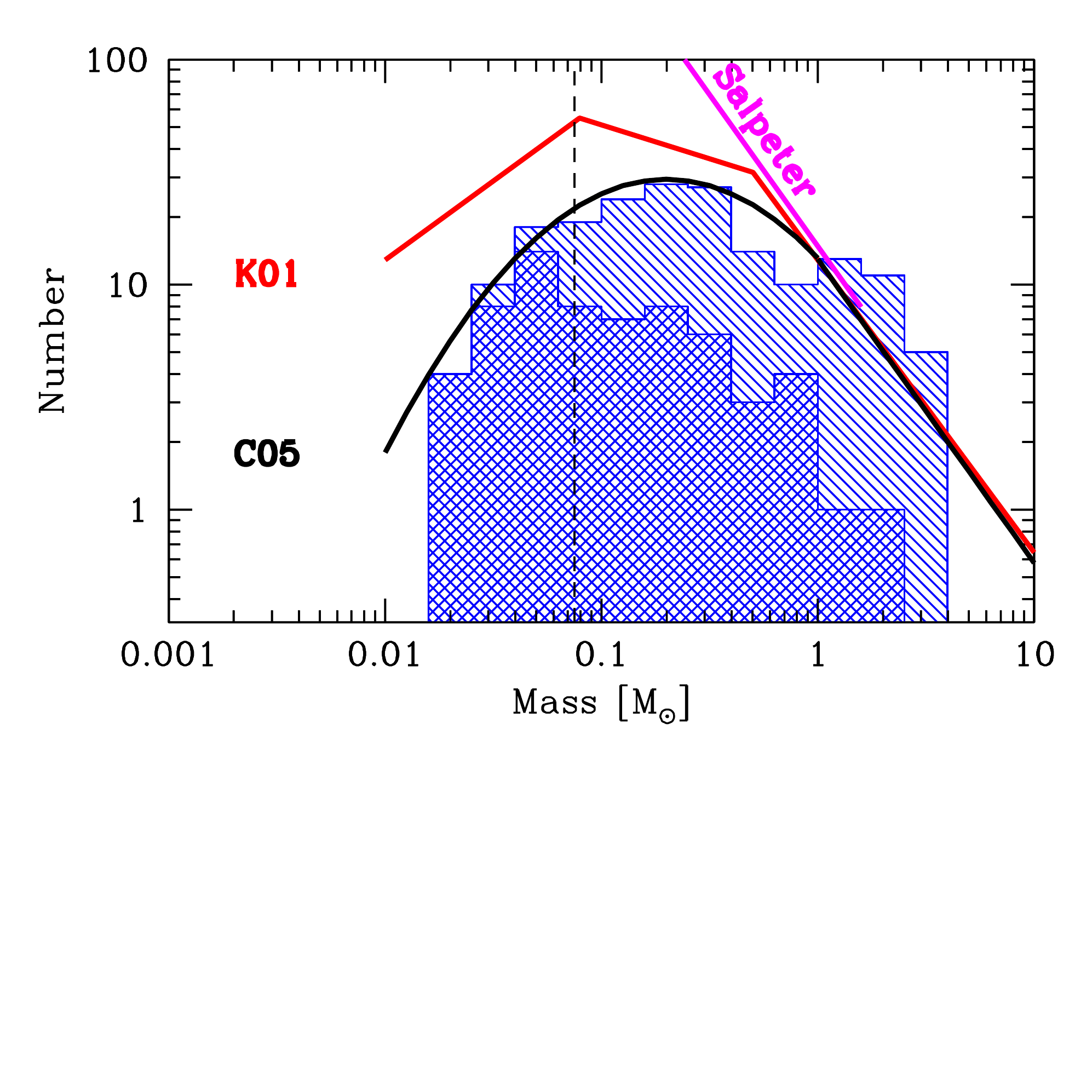}\vspace{-2.8cm}
\caption{Histograms giving the initial mass function of the 590 stars and brown dwarfs at $t=1.20t_{\rm ff}$ from the main barotropic calculation of Bate (2009a) (left), and the 183 objects formed at the same time in the radiation hydrodynamical calculation (right).  The double hatched histograms are used to denote those objects that have stopped accreting, while those objects that are still accreting are plotted using single hatching.  The radiation hydrodynamical calculation produces far fewer brown dwarfs and low-mass stars and more stars with masses $\gsim 1.5$~M$_\odot$ and is in good agreement with the Chabrier (2005) fit to the observed IMF for individual objects.  Two other parameterisations of the IMF are also plotted: Salpeter (1955) and Kroupa (2001).}
\label{imfcomp}
\end{figure*}

In Fig.~\ref{cumimf_baro}, we compare the cumulative IMF at the end of the radiation hydrodynamical calculation (solid line) with the IMFs of the main and rerun barotropic calculations \citep{Bate2009a}.  With radiative feedback there is a clear increase of the median stellar mass and a huge decrease in the fraction of brown dwarfs.  Thus, in agreement with earlier works \citep{Bate2009b, Offneretal2009, UrbMarEva2010, KruKleMcK2011}, we find radiative feedback is crucial for determining the IMF even for low-mass star formation.

Note that, in fact, the calculation produces a protostellar mass function (PMF) rather than an IMF \citep{FleSta1994a, FleSta1994b,McKOff2010} because some of the objects are still accreting when the calculation is stopped.  In this paper, we refer to the mass function as an `IMF' because we compare it to the observed IMF since the PMF cannot yet be determined observationally.  However, it should be noted that how a PMF transforms into the IMF depends on the accretion history of the protostars and how the star formation process is terminated.   One issue that can be studied from the calculation presented here is whether the distribution of stellar masses evolves in form during the formation of the stellar cluster or whether the mass distribution is `fully-formed' so that no matter when the distribution is examined it is always consistent with being drawn from a constant underlying mass function.  From Fig.~\ref{cumimf_baro}, the median stellar mass and the overall shape of the distribution obtained using a barotropic equation of state does not change with time, except that the maximum mass increases with time.  This is in contrast to the isothermal calculation of \cite{KruKleMcK2011}, who find that the median mass increases by a factor of two during their calculation {\it without} radiative feedback.  This difference is probably due to the very different initial conditions, but the fact their sink particles are 800 times larger than those used here so that they cannot capture small-scale fragmentation may also play a role.  

In Fig.~\ref{cumimf_RT}, we plot the cumulative stellar mass distributions at four different times during the radiation hydrodynamical calculation.  Performing Kolmogorov-Smirnov tests comparing the final distribution to each of the early distributions shows that each intermediate distribution is consistent with being drawn randomly from the same distribution as the final distribution.  Of course the intermediate and final distributions are not statistically independent, but the test still shows that the stellar mass distribution keeps the same form as it evolves during the formation of the cluster.  This also means that in stopping the calculation at $t=1.20~t_{\rm ff}$ we do not seem to have stopped at a special point in the evolution, at least as far as the mass function is concerned.  The only thing that changes is that as the number of stars increases with time, so the maximum stellar mass increases.  Again, this is in contrast to \cite{KruKleMcK2011}, who find that with radiative feedback their median stellar mass increases by almost a factor of 4 from 0.55~M$_\odot$ when 10\% of the cloud's mass has been converted to stars to 2~M$_\odot$ when 50\% the stars contain 50\% of the total mass.  As mentioned above, this difference is probably due to a combination of the denser, centrally-condensed initial conditions, the radiative feedback from sink particles, and the use of much larger sink particles.

The differential form of the IMF at the end of the radiation hydrodynamical calculation is shown in the right-hand panel of Fig.~\ref{imfcomp} and is compared with the parameterisations of the observed IMF given by \citet{Chabrier2005}, \citet{Kroupa2001}, and \citet{Salpeter1955}.  In the left-hand panel of the figure, we provide the IMF from the main barotropic calculation of \cite{Bate2009a} at the same time for comparison.  
In agreement with the smaller radiation hydrodynamical calculations of \cite{Bate2009b}, the introduction of the radiative feedback has clearly addressed the problem of the over-production of brown dwarfs and low-mass stars that occurs when using a barotropic equation of state \citep{Bate2009a}.   In fact, comparing the histogram of objects with the parameterisation of the observed IMF by \cite{Chabrier2005}, the agreement is almost too good to be true.  The cumulative mass distribution from the calculation (solid line) is compared with that of \cite{Chabrier2005} (dot-long-dashed line) in Fig.~\ref{cumimf_baro}.  A Kolmogorov-Smirnov test gives a 50.5 percent probability that the radiation hydrodynamical IMF could have been drawn from \citeauthor{Chabrier2005}'s fit to the observed IMF (i.e. the two mass functions are indistinguishable).

However, despite more than a decade of observational work, the form of the IMF in the sub-stellar regime is still quite uncertain.  Although it is now generally accepted that the number of stars is larger than the number of brown dwarfs in Galactic star-forming regions \citep{Chabrier2003, Greissletal2007, Andersenetal2008}, considerable uncertainty remains.  Rather than trying to determine the exact form or slope of the substellar IMF, a popular method is to compare the number of brown dwarfs in an observed region to the number of stars with masses less than that of the Sun. \citet{Andersenetal2008} find that the ratio of stars with masses $0.08-1.0$ M$_\odot$ to brown dwarfs with masses $0.03-0.08$ M$_\odot$ is $N(0.08-1.0)/N(0.03-0.08)\approx 5\pm 2$.  By combining the results of two radiation hydrodynamical calculations of star formation in 50-M$_\odot$ molecular clouds, \cite{Bate2009b} found a ratio of stars to brown dwarfs of $25/5 \approx 5$.  This number is in agreement with observational constraints, but the statistical uncertainty is large.  Here we obtain a ratio of $N(0.08-1.0)/N(0.03-0.08) = 117/31 \approx 3.8$.  Eight of the 31 low-mass objects and 84 of the 117 stars were still accreting when the calculation was stopped, so there is some uncertainty in this figure due to unknown future evolution.  But the value is well within observational uncertainties.  For the main barotropic calculation of \cite{Bate2009a}, this ratio was much lower: $212/146 \approx 1.45$ at $t=1.20~t_{\rm ff}$ and $408/326 = 1.25$ at $t=1.50~t_{\rm ff}$.

Below 0.03~M$_\odot$, the IMF is very poorly constrained, both observationally and from the calculation presented here.  The radiation hydrodynamical calculation produced 6 objects with masses in this range, with a minimum mass of 17.6 M$_{\rm J}$.  All of these objects had stopped accreting when the calculation was stopped.  This is very different to the main barotropic calculation.  At the same time ($1.20~t_{\rm ff}$), the barotropic calculation had produced 217 brown dwarfs (40 still accreting) with masses less than 30~M$_{\rm J}$ with a minimum mass of only 5.6~M$_{\rm J}$.  Even discounting objects that were still accreting, the inclusion of radiative feedback has cut the production of these very-low-mass brown dwarfs by a factor of $\approx 30$ and significantly increased the minimum mass.  It is interesting to note that the minimum mass is substantially higher than the opacity limit for fragmentation \citep{LowLyn1976,Rees1976,Silk1977a,Silk1977b, BoyWhi2005}.  This is because the opacity limit provides a minimum mass, but generally objects will accrete from their surroundings to greater masses.  Perhaps more importantly, the minimum mass is also greater than the estimated masses of some objects observed in star-forming regions \citep{ZapateroOsorioetal2000,Kirkpatricketal2001,ZapateroOsorioetal2002, Kirkpatricketal2006,Lodieuetal2008,Luhmanetal2008,Bihainetal2009,Weightsetal2009, Burgessetal2009,Luhmanetal2009a,Quanzetal2010}.  While an exact cut off is difficult to determine from either numerical simulations or observations, the results of the radiation hydrodynamical calculation presented here do imply that brown dwarfs with masses $\lsim 15$~M$_{\rm J}$ should be very rare.

\begin{figure}
\centering
    \includegraphics[width=8.4cm]{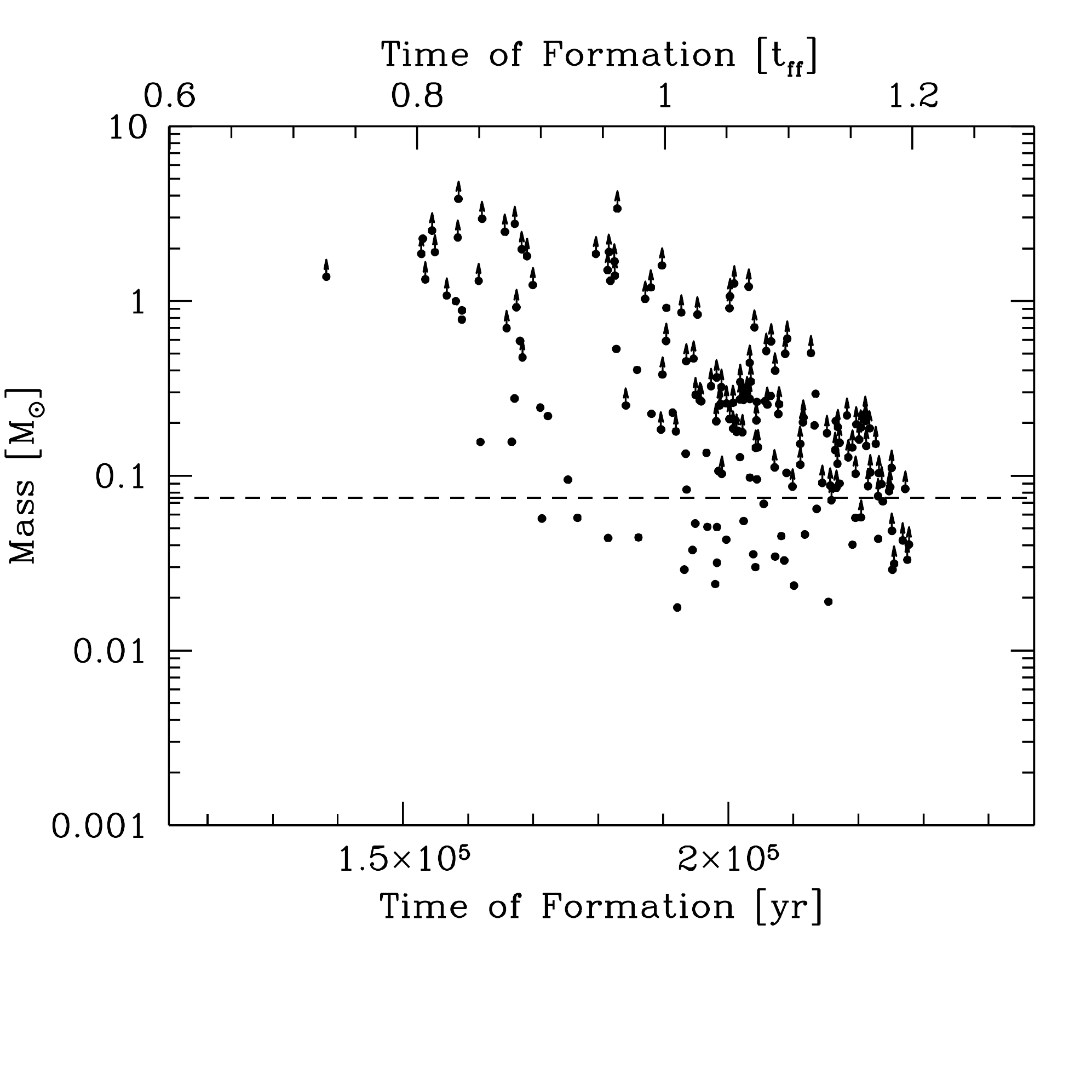}\vspace{-1cm}
\caption{Time of formation and mass of each star and brown dwarf at the end of the radiation hydrodynamical calculation.  It is clear that the objects that are the most massive at the end of the calculation are actually some of the first sink particles to form.  Objects that are still accreting significantly at the end of the calculation are represented with vertical arrows.  The horizontal dashed line marks the star/brown dwarf boundary.  Time is measured from the beginning of the calculation in terms of the free-fall time of the initial cloud (top) or years (bottom).}
\label{massform}
\end{figure}

\begin{figure}
\centering\vspace{-0.2cm}
    \includegraphics[width=8.4cm]{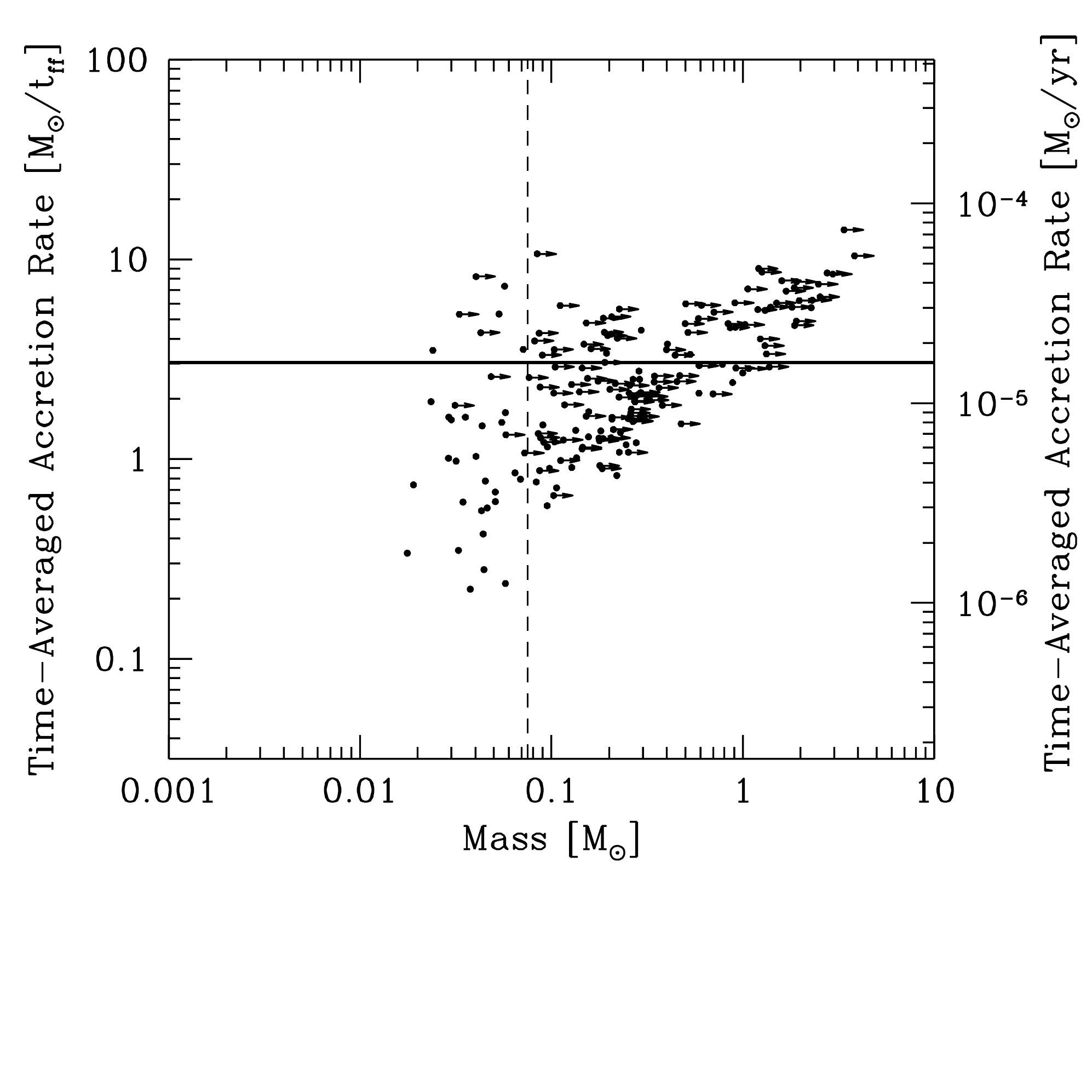}\vspace{-1.5cm}
\caption{The time-averaged accretion rates of the objects formed in the radiation hydrodynamical calculation versus their final masses.  The accretion rates are calculated as the final mass of an object divided by the time between its formation and the termination of its accretion or the end of the calculation.  Objects that are still accreting significantly at the end of the calculation are represented with horizontal arrows.  There is no dependence of mean accretion rate on final mass for objects with less than $\sim 0.5$ M$_\odot$ (just a  dispersion).  However, there is a low-accretion rate region of exclusion for the most massive objects since only objects with mean accretion rates greater than their mass divided by their age can reach these high masses during the calculation.  The horizontal solid line gives the mean of the accretion rates: $1.54\times 10^{-5}$ M$_\odot$~yr$^{-1}$. The accretion rates are given in M$_\odot/t_{\rm ff}$ on the left-hand axes and M$_\odot$~yr$^{-1}$ on the right-hand axes. The vertical dashed line marks the star/brown dwarf boundary.
}
\label{accrate}
\end{figure}

\subsubsection{The origin of the initial mass function}
\label{origin_imf}

\cite{BatBon2005} analysed two barotropic star cluster formation simulations that began with clouds of different densities to determine the origin of the IMF in those calculations \citep[see also][]{Bate2005}.  They found that the IMF resulted from competition between accretion and `ejection'.  There was no significant dependence of the mean accretion rate of an object on its final mass.  Rather, there was a roughly linear correlation between an object's final mass and the time between its formation and the termination of its accretion.  Furthermore, the accretion on to an object was usually terminated by a dynamical interaction between the object and another system.  Note that such an interaction does not necessarily require that the object is ejected from the cluster.  Many times this is the case, but moving an object into a lower density part of the cloud (e.g. out of its natal core) or substantially increasing the object's speed without it becoming unbound can also dramatically reduce its accretion rate (c.f., the Bondi-Hoyle accretion formula $\dot{M} \propto \rho/(c_{\rm s}^2 + v^2)^{3/2}$, where $v$ is the velocity of the object relative to the gas). Thus, \citeauthor{BatBon2005} found that objects formed with very low masses (a few Jupiter masses) and accreted to higher masses until their accretion was terminated, usually, by a dynamical encounter.  This combination of competitive accretion and stochastic dynamical interactions produced the mass distributions, and \cite{BatBon2005} presented a simple semi-analytic model which could describe the numerical results in which the characteristic stellar mass was given by the product of the typical accretion rate and the typical time between a object forming and having a dynamical interaction that terminated its accretion.  \cite{Bate2009a} found the IMF in their larger barotropic calculations also originated in this manner.  They found the mean accretion rate of a low-mass star did not depend on its final mass, but that objects that accreted for longer ended up with greater masses and that protostellar accretion was usually terminated by dynamical interactions.  Here we analyse the radiation hydrodynamical calculation using the same methods.

\begin{figure}
\centering\vspace{-0.2cm}
    \includegraphics[width=8.4cm]{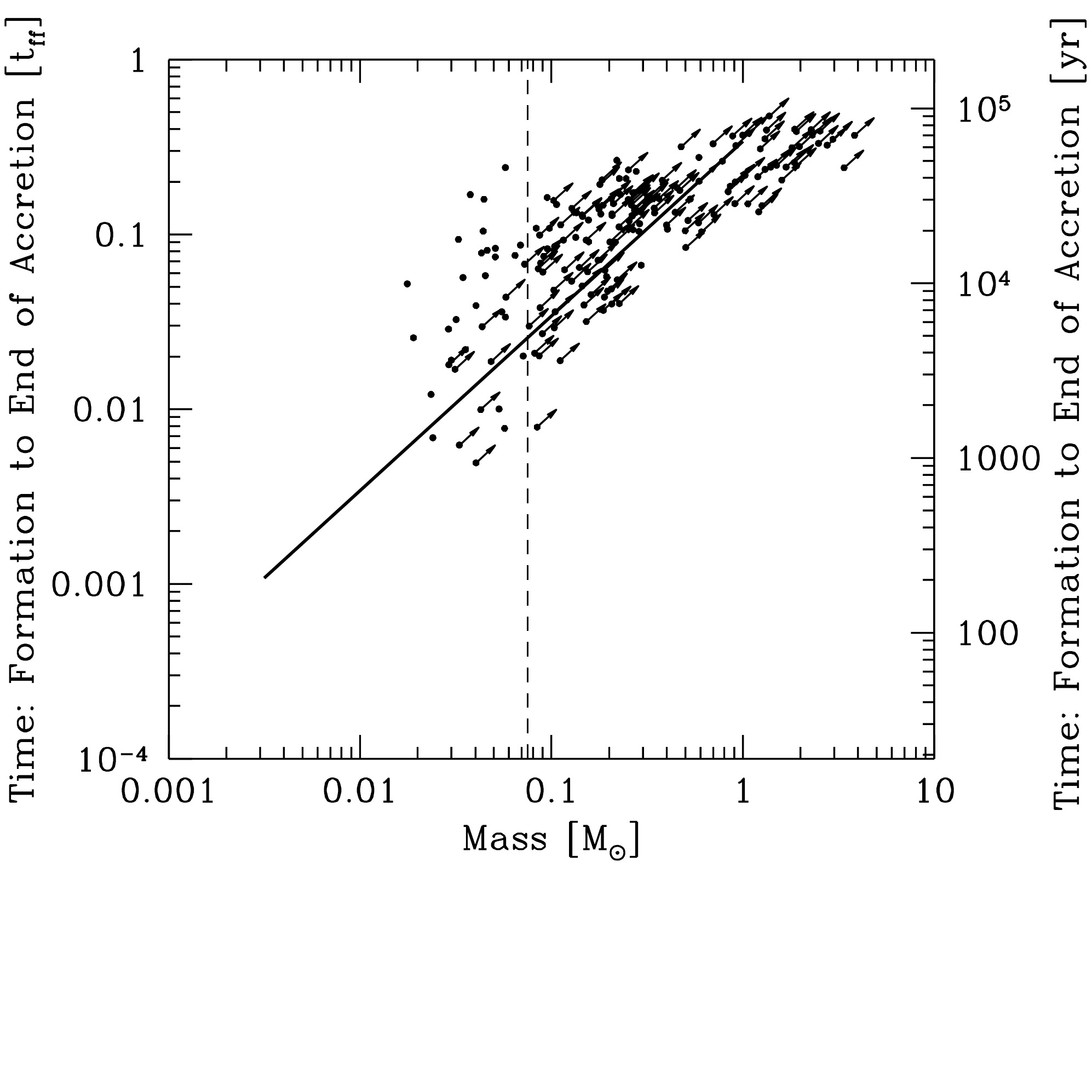}\vspace{-1.5cm}
\caption{The time between the formation of each object and the termination of its accretion or the end of the radiation hydrodynamical calculation versus its final mass.  Objects that are still accreting significantly at the end of the calculation are represented with arrows.  As in past barotropic calculations, there is a clear linear correlation between the time an object spends accreting and its final mass.  The solid line gives the curve that the objects would lie on if each object accreted at the mean of the time-averaged accretion rates. The accretion times are given in units of the $t_{\rm ff}$ on the left-hand axes and years on the right-hand axes. The vertical dashed line marks the star/brown dwarf boundary.
}
\label{acctime}
\end{figure}

In Fig.~\ref{massform}, we plot the final mass of an object versus the time at which it formed (i.e. the time of insertion of a sink particle).  It is clear that the most massive stars at the end of the calculation were some of the first to begin forming.  During the calculation, as other lower-mass stars have formed and some have had their accretion terminated, these stars have continued to grow to higher and higher masses.  \cite{Maschbergeretal2010} have argued that such a cluster formation process naturally produces a relation between cluster mass and maximum stellar mass similar to that which is observed \citep{WeiKro2006,WeiKroBon2010}, although others argue that the observations are also consistent with random sampling from a universal IMF  (\citealt{Lambetal2010}; \citealt*{FumdaSKru2011}).

\begin{figure}
\centering\vspace{-0.2cm}
    \includegraphics[width=8.4cm]{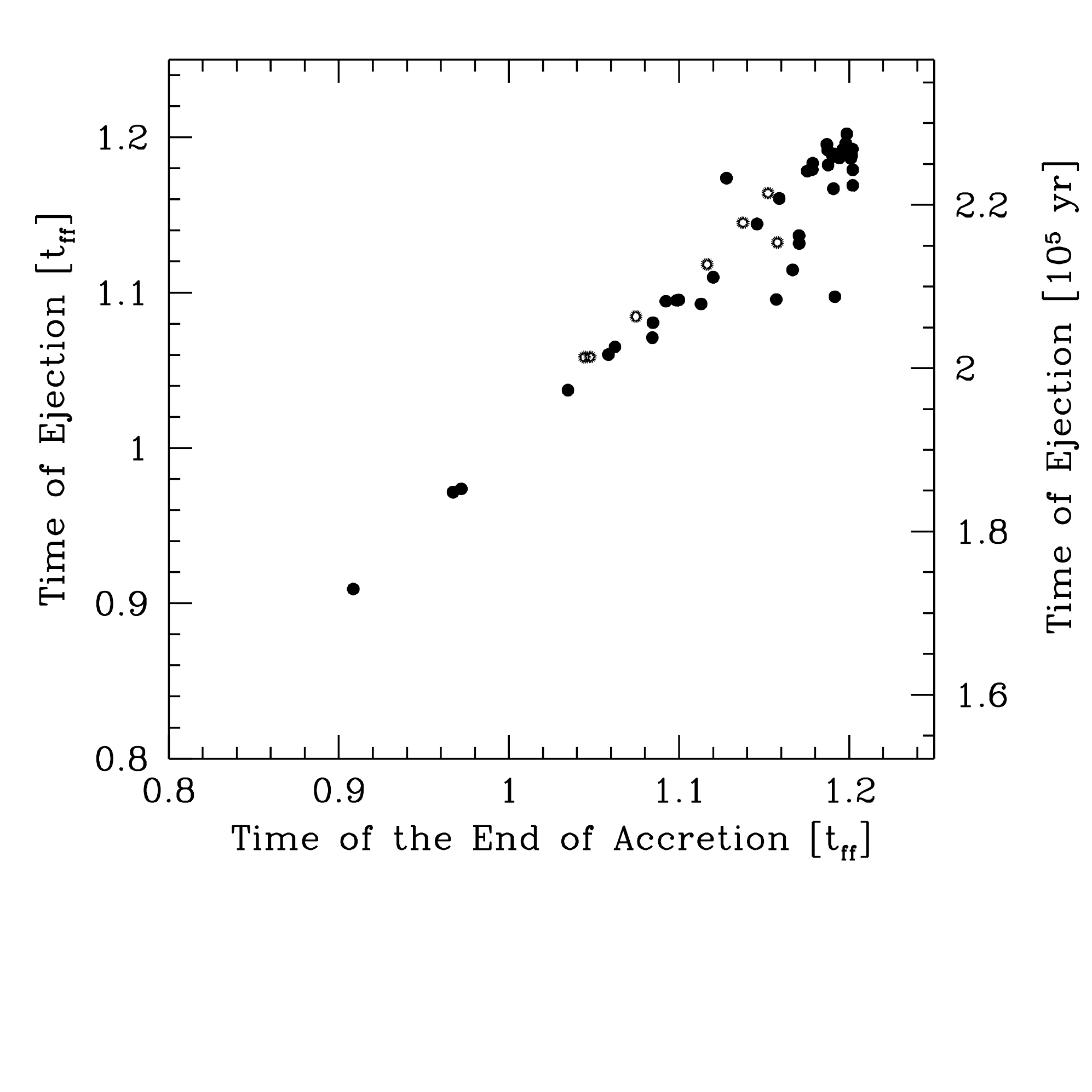}\vspace{-1.5cm}
\caption{For each single object that has stopped accreting by the end of the main calculation, we plot the time of the ejection of the object from a multiple system versus the time at which its accretion is  terminated.  As in past barotropic calculations, these times are correlated, showing that the termination of accretion on to an object is usually associated with dynamical ejection of the object. Open circles give those objects where multiple `ejections' are detected by the ejection detection algorithm and, hence, the ejection time is ambiguous (see the main text).  Binaries have been excluded in the plot because it is difficult to determine when a binary has been ejected.
}
\label{ejtime_vs_acctime}
\end{figure}

In Figs.~\ref{accrate}, \ref{acctime}, and \ref{ejtime_vs_acctime}, we plot similar figures to those found in \cite{BatBon2005} and \cite{Bate2005, Bate2009a, Bate2009c}.  These figures display the same trends as found in the barotopic calculations.  Fig.~\ref{accrate} gives the time-averaged accretion rates of all the objects formed in the radiation hydrodynamical calculation versus the object's final mass.  The time-averaged accretion rate is the object's final mass divided by the time between its formation (i.e. the insertion of a sink particle) and the end of its accretion (defined as the last time its accretion rate drops below 10$^{-7}$ M$_\odot$~yr$^{-1}$) or the end of the calculation.  As in the barotropic calculations, there is no dependence of the time-averaged accretion rate on an object's final mass, except that objects need to accrete at a rate at least as quickly as their final mass divided by their age (i.e., the lower right potion of Fig.~\ref{accrate} cannot have any objects lying in it).  This means that the most massive stars have higher time-averaged accretion rates than the bulk of the stars and VLM objects.  But, on the other hand, if the calculation were continued longer, objects that accrete with lower time-averaged accretion rates could also reach high masses.   Note that these results should not be used to infer that the typical accretion rate remains independent of mass above 3~M$_\odot$.  The calculation presented here does not produce any high-mass protostars, but other studies have reported that the accretion rates of protostars with masses $\gsim 3$~M$_\odot$ do increase with mass \citep[e.g.,][]{UrbMarEva2010,KruKleMcK2011}. 

The mean of the accretion rates is $1.5 \times 10^{-5}$ M$_\odot$~yr$^{-1}$, which is within a factor of $\approx 2$ of the mean accretion rates of the barotropic calculations in all of the above papers.  Thus, the mean accretion rate does not depend significantly on cloud density \citep{BatBon2005}, the equation of state of high-density gas \citep{Bate2005}, the total mass of the gas cloud \citep{Bate2009a}, or on whether the calculation is performed using a barotropic equation of state or radiation hydrodynamics.  It only depends on the mean temperature of the initial cloud \citep{Bate2005}, in that it scales roughly as $T_{\rm g}^{3/2}$ (or, equivalently, $c_{\rm s}^3/G$, where is the mean sound speed on large-scales; \citealt{Shu1977}).  The dispersion in the accretion rates is about 0.37 dex, also similar to the previous barotropic simulations.  Thus, rather than the final mass of a star depending on its average accretion rate, the primary determinant of the final mass of a star or brown dwarf is the period over which it accretes.  Fig.~\ref{acctime} very clearly shows the linear relation (with some dispersion) between the period of time over which an object accretes and its final mass.  This means that the higher characteristic stellar mass produced when radiative feedback is included is due to an increase in the average time over which an object accretes.

\begin{figure}
    \includegraphics[width=8.4cm]{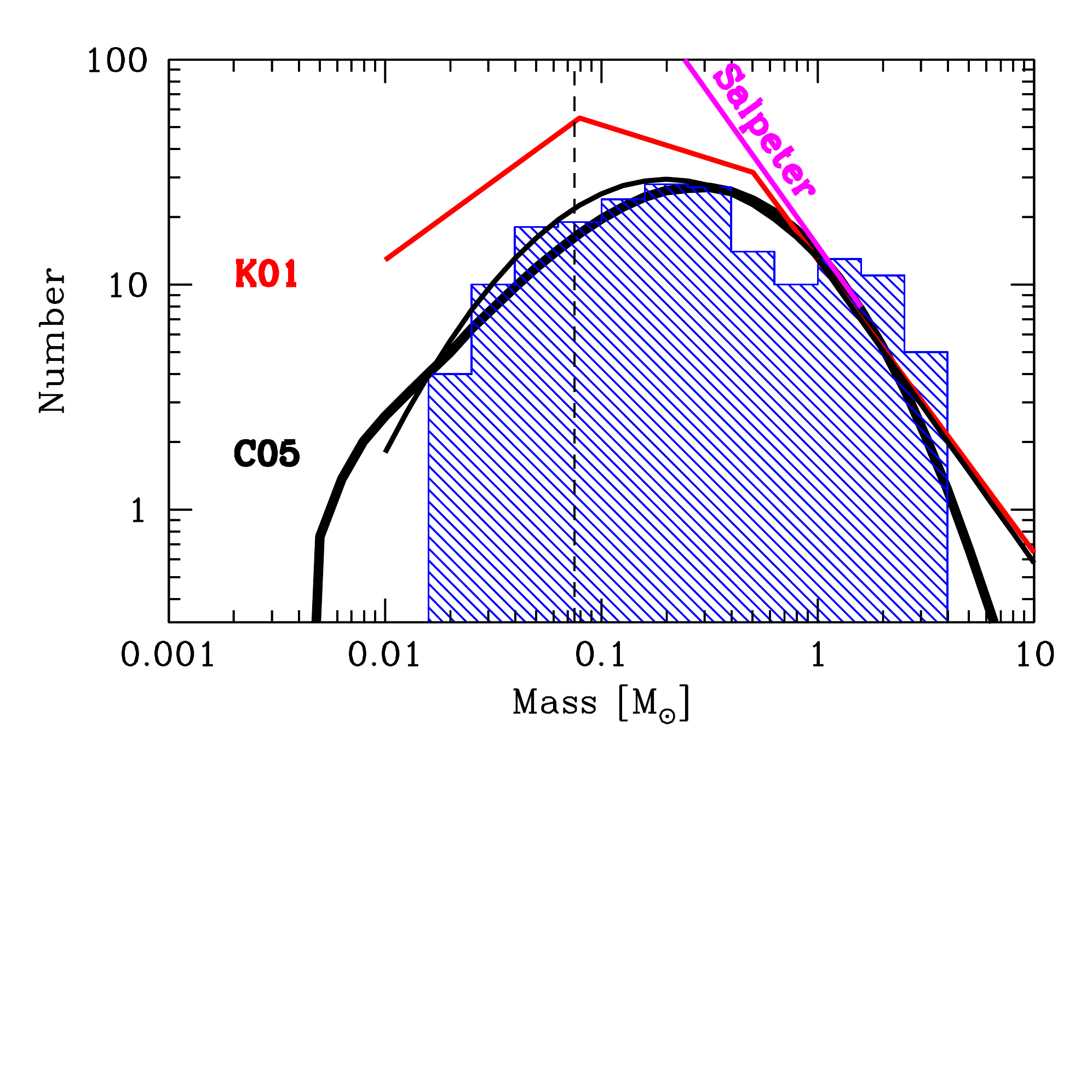}\vspace{-3cm}
\caption{\label{imffit} The initial mass functions produced by the radiation hydrodynamical calculation (histogram) and a comparison with the fit using the simple accretion/ejection IMF model (thick curve) of \citet{BatBon2005}.  Statistically, the hydrodynamical and the model IMFs are indistinguishable (a Kolmogorov-Smirnov test gives a 19 percent probability that the hydrodynamical IMF could have been drawn from the model IMF).  Also shown are the \citet{Salpeter1955} slope (solid straight line), and the \citet{Kroupa2001} (solid broken line) and \citet{Chabrier2005} (thin curve) mass functions.  The vertical dashed line is the stellar-substellar boundary.} 
\end{figure}

In Fig.~\ref{ejtime_vs_acctime}, for each object that has stopped accreting by the end of the main calculation (excluding the components of binaries), we plot the time at which the object undergoes a dynamical ejection versus the time that its accretion is terminated.  The strong correlation shows that accretion is usually terminated by a dynamical encounter with other objects, as seen in the barotropic calculations. We define the time of ejection of an object as the last time the magnitude of its acceleration drops below 2000 km~s$^{-1}$~Myr$^{-1}$ \citep{Bate2009a} or the end of the calculation. The acceleration criterion is based on the fact that once an object is ejected from a stellar multiple system, sub-cluster, or cluster through a dynamical encounter, its acceleration will drop to a low value. We exclude binaries because they have large accelerations throughout the calculation which frequently results in false detections of ejections.  Also, in Fig.~\ref{ejtime_vs_acctime}, we use two different symbols (filled circles and open circles).  For the former we are confident of the ejection time.  However, for those objects denoted by the open circles, we find that at least two `ejections' more than 2000 years apart have occurred.  These are usually objects that have had a close dynamical encounter with a multiple system that has put them into long-period orbits rather than ejecting them.  In these cases, we chose the `ejection' time closest to the accretion termination time but we use an open symbol to denote our uncertainty in whether or not we have identified the best time for the dynamical encounter.

We find that, excluding binaries, for 40 objects out of 47 (85\%) the accretion termination time and the ejection time are within 2000 years of each other.  If we also exclude those objects for which we are uncertain in our identifications of the ejection times as described above, we find 33 objects out of 40 (83\%) are consistent with ejection terminating their accretion.  These are probably lower limits in the sense that it is difficult to determine in an automated way the time at which an ejection occurs and an erroneous value is much more likely to differ from the accretion termination time by more than 2000 years than coincide with it.  In any case, it is clear that for the majority of objects their accretion is terminated by dynamical encounters with other stellar systems.

In Fig.~\ref{imffit}, we compare the IMF obtained from the radiation hydrodynamical calculation with the semi-analytic accretion/ejection IMF model of \cite{BatBon2005} using parameters determined from the radiation hydrodynamical calculation \citep[see also][]{Bate2009c}.  The parameters are: the mean accretion rate and its dispersion (given above), period of time over which stars form (i.e. 90,000 yrs), the characteristic ejection time, and the minimum stellar mass.  The characteristic ejection time, $\tau_{\rm eject}=62,400$~yr, is chosen such that the mean number of objects that have finished accreting over the time period equals that from the radiation hydrodynamical calculation (64 objects).  The minimum stellar mass primarily determines the minimum mass cut-off to the IMF, rather than the shape of the rest of the IMF.  For the semi-analytic IMF in Fig.~\ref{imffit} we choose 5 Jupiter-masses, but 10-15 Jupiter-masses result in similarly good fits.  A Kolmogorov-Smirnov test comparing the semi-analytic IMF to the IMF obtained from the radiation hydrodynamical calculation shows that the latter is consistent with being randomly drawn from the former (probability 19\%). 

In conclusion, the origin of the IMF in the radiation hydrodynamical calculation is the same as in the past barotropic calculations: the IMF originates from competition between accretion and dynamical encounters.  Objects end up with low masses if their accretion is terminated (by a dynamical encounter) soon after they form.  Objects end up with high masses by accreting for an extended period.  The reason the characteristic stellar mass is larger when radiative feedback is included is that objects typically accrete for longer before their accretion is terminated.  This is because the radiative feedback increases the typical distance between objects \citep{Bate2009b}, and so dynamical interactions take longer to occur.

\begin{figure}
\centering
    \includegraphics[width=8.4cm]{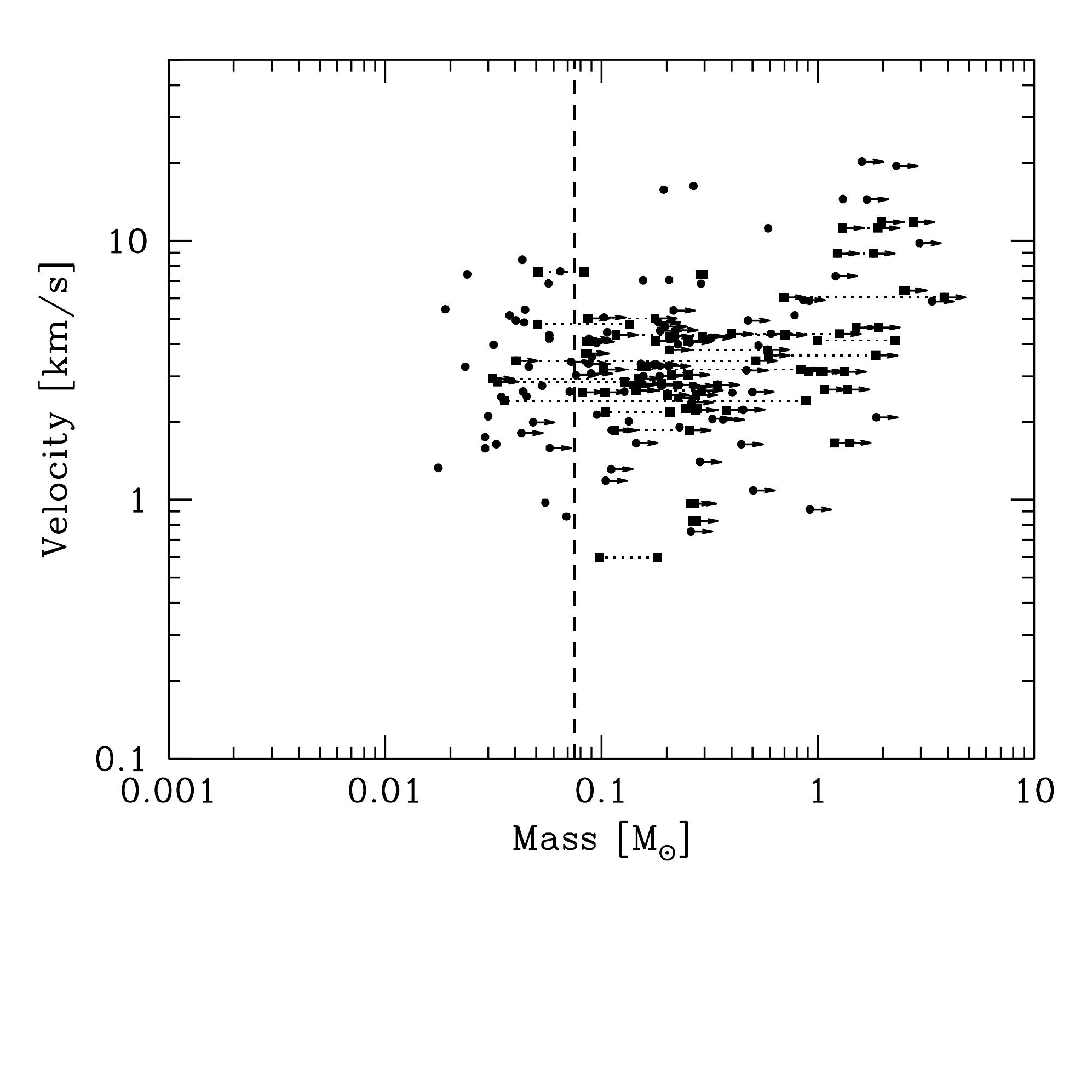}\vspace{-1.5cm}
\caption{The magnitudes of the velocities of each star and brown dwarf relative to the centre-of-mass velocity of the stellar system at the end of the radiation hydrodynamical calculation.  For binaries, the centre-of-mass velocity of the binary is given, and the two stars are connected by dotted lines and plotted as squares rather than circles.  Objects still accreting at the end of the calculation are denoted by horizontal arrows.  The root mean square velocity dispersion for the association (counting each binary once) is 5.5 km~s$^{-1}$ (3-D) or 3.2 km~s$^{-1}$ (1-D).  There is a dependence of the velocity dispersion on mass with VLM objects having a lower velocity dispersion than stars (see the main text).  Binaries are found to have a slightly lower velocity dispersion than single objects of only 4.6 km~s$^{-1}$ (3-D).  The vertical dashed line marks the star/brown dwarf boundary.
}
\label{veldisp}
\end{figure}

\subsection{Stellar kinematics}

At the end of the main barotropic calculation, most of the sub-clusters had merged together and the stellar distribution essentially consisted of a single cluster surrounded by a halo of ejected stars and brown dwarfs.  \cite{Bate2009a} analysed the radial properties of this cluster, looking for evidence of mass segregation and for radial variations of the stellar velocity dispersion and binarity.  Unfortunately, we are not able to follow the radiation hydrodynamical calculation until all the sub-clusters and filaments have collapsed into a single cluster and we have a lot fewer stars and brown dwarfs, so we restrict ourselves only to discussing the kinematics of the population as a whole without attempting to look for spatial variations.

In Fig.~\ref{veldisp}, we plot the magnitude of the velocity of every star or brown dwarf relative to the centre of mass of the stellar system at the end of the radiation hydrodynamical calculation.  For binaries (including those that are sub-components of triples and quadruples), we plot the two components with the centre of mass velocity of the binary using filled squares connected by a dotted line.  The overall root mean square (rms) velocity dispersion (counting each binary only once) is 5.5 km~s$^{-1}$ (3-D) or 3.2 km~s$^{-1}$ (1-D).  This is almost identical to the velocity dispersion found by \cite{Bate2009a} at the end of the main barotropic calculation, and larger than the velocity dispersions found from calculations of star formation in small 50-M$_\odot$ clouds \citep{BatBonBro2003,BatBon2005,Bate2005,Bate2009c}.  The velocity dispersion is 40\% higher than the 1D velocity dispersion of $2.3$ km s$^{-1}$ seen in the Orion Nebula Cluster \citep{JonWal1988, Tianetal1996}, consistent with the fact that the system formed here is lower in mass, but denser.

\cite{ReiCla2001} suggested that a greater velocity dispersion for brown dwarfs than stars may be a possible signature that brown dwarfs form as ejected stellar embryos. Past $N$-body simulations of the breakup of small-$N$ clusters with $N>3$ \citep{SteDur1998} and SPH calculations of $N=5$ clusters embedded in gas \citep{DelClaBat2003} found that there was no strong dependence of the velocity of an object on its mass, but both found that binaries should have a smaller velocity dispersion than single objects due to the recoil velocities of binaries being lower.  On the other hand, \cite{Delgadoetal2004} performed simulations of star formation in small turbulent clouds and found that the velocity dispersions of singles and binaries were indistinguishable, but that higher-order multiples had significantly lower velocity dispersions. 

From the large barotropic calculations of \cite{Bate2009a}, it was found that stars tend to have a slightly higher dispersion than VLM objects and that binaries have a lower velocity dispersion than single objects.  These same relations are also found from the radiation hydrodynamical calculation.  The rms velocity dispersion of VLM systems is 4.1 km~s$^{-1}$ (3-D) while for the stars (masses $\geq 0.1$ M$_\odot$) the rms velocity dispersion is 6.5 km~s$^{-1}$ (3-D).  Binaries (most of which have stellar primaries) have a velocity dispersion of 4.6 km~s$^{-1}$ (3-D), lower than both the velocity dispersion of all stars and the overall velocity dispersion.  

Observationally, while there is no strong evidence for VLM objects having a different velocity dispersion to stars, or binaries having a different velocity dispersion to single objects, studies of the radial velocities of stars and brown dwarfs in the Chamaeleon I dark cloud do find that brown dwarfs have a marginally lower velocity dispersion than the T Tauri stars \citep{JoeGue2001,Joergens2006}, in qualitative agreement with the results from both the barotropic calculations of \cite{Bate2009a} and radiation hydrodynamical calculation discussed here.

\begin{figure}
\centering
    \includegraphics[width=8.4cm]{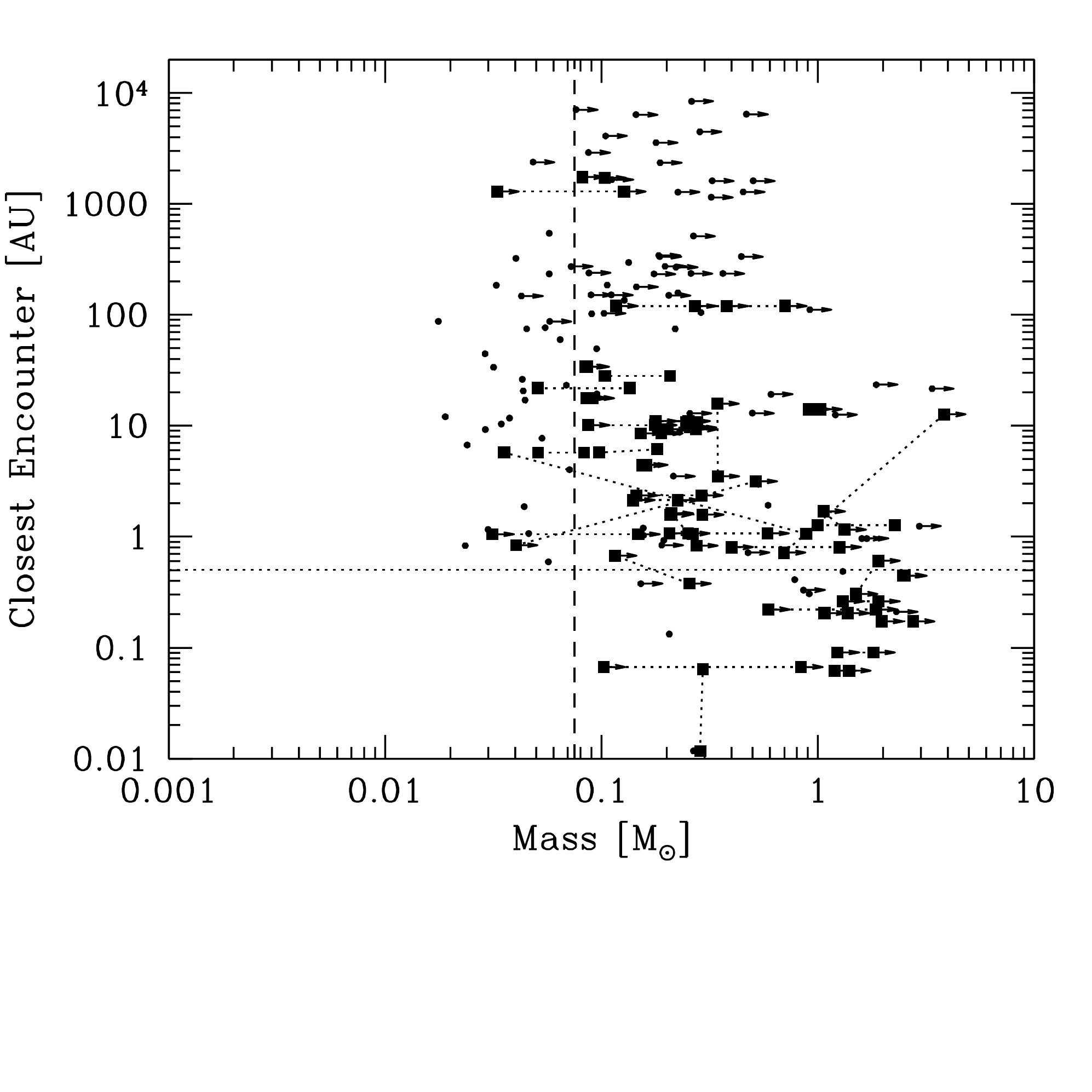}\vspace{-1.5cm}
\caption{The closest encounter distances of each star or brown dwarf during the radiation hydrodynamical calculation versus the final mass of each object.  Objects that are still accreting significantly at the end of the calculation are denoted with arrows indicating that they are still evolving and that their masses are lower limits.  Binaries are plotted with the two components connected by dotted lines and squares are used as opposed to circles.  The horizontal dotted line marks the size of the sink particle accretion radii, within which there is no gas.  The vertical dashed line marks the star/brown dwarf boundary.  Note that two objects almost merged with one another during the calculation (the merger radius was set to be 0.01~AU).
}
\label{truncate}
\end{figure}

\subsection{Stellar encounters and disc sizes}

\citet{ReiCla2001} also speculated that if brown dwarfs formed via ejection, they might have smaller, lower-mass discs than stars.  As mentioned in Section \ref{origin_imf}, the IMF in the radiation hydrodynamical calculation presented here and in the many past barotropic star cluster formation calculations originates through competition between accretion and ejection, but this applies both to stars and brown dwarfs.  The only difference is that brown dwarfs are ejected soon after they form, before they have accreted much mass, while stellar ejections occur after a longer period of accretion.  Discs around both stars and brown dwarfs may be truncated by dynamical encounters and ejections.

In Figure \ref{truncate}, we plot the distance of the closest encounter that every star and brown dwarf has had by the end of the radiation hydrodynamical calculation.  As in past barotropic calculations, there is a wide range of closest encounter distances (including two objects that almost merged), but the closest encounters tend to have occurred for stars rather than brown dwarfs.  Dynamical encounters between objects will truncate any circumstellar discs.  However, this plot cannot be taken to mean that many stars have small discs because of several reasons.  First, as will be seen in Section \ref{multiplicity_sec}, multiplicity is a strong function of primary mass.  In Figure \ref{truncate} it clear that binaries are responsible for many of the `closest encounters'.  Second, objects that are still accreting at the end of the calculation are still evolving and, since the mass of an object depends on its `age' more massive accreting objects are more likely to have had close encounters.  Finally, as noted by \cite{BatBonBro2003}, many stars that have close encounters have new discs formed from accretion {\em subsequent} to their closest dynamical encounter.

\begin{figure}
\centering
    \includegraphics[width=8.4cm]{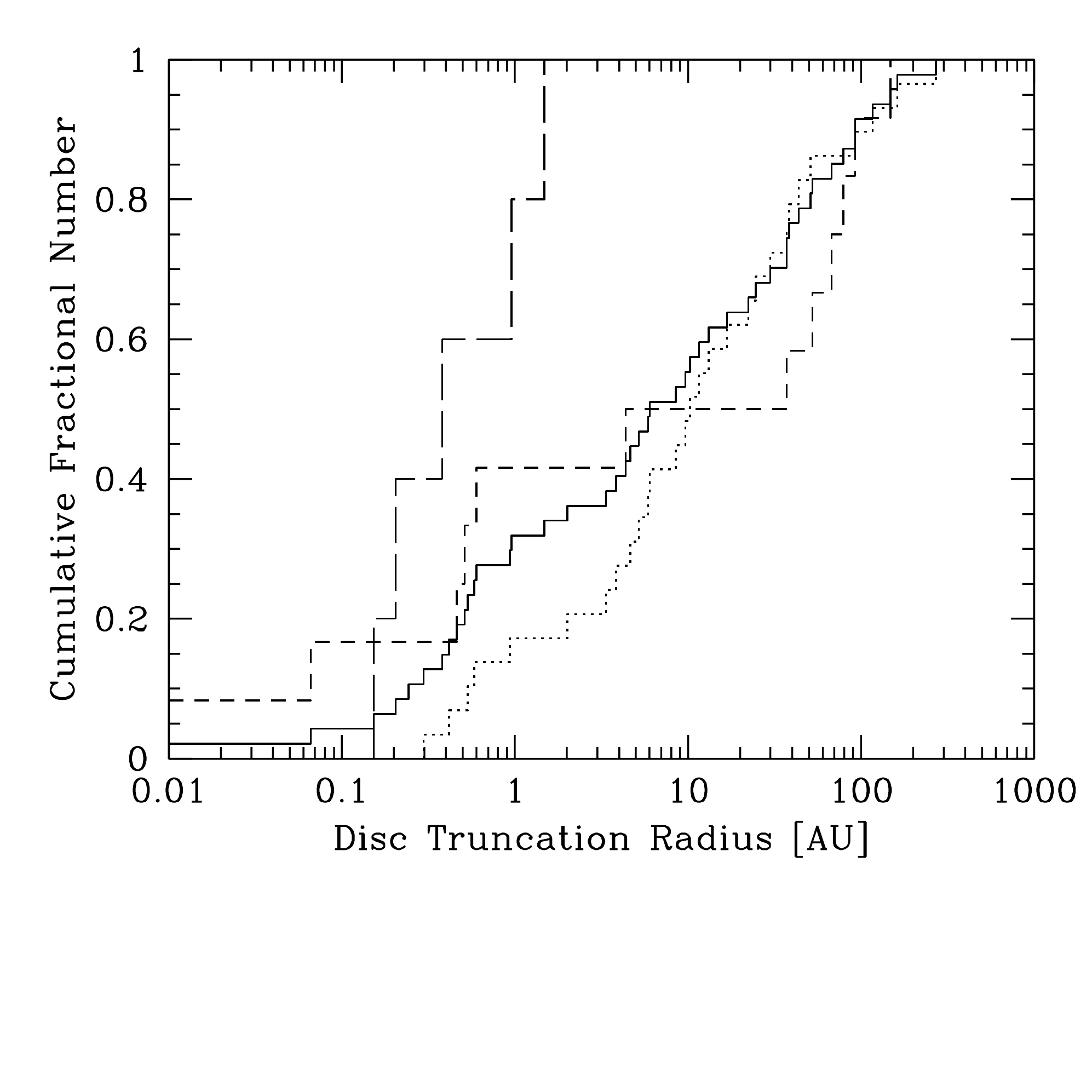}\vspace{-1.5cm}
\caption{Due to dynamical interactions, stars and brown dwarfs {\it potentially} have their discs truncated to approximately 1/2 of the periastron separation during the encounter (see also Figure \ref{truncate}).  At the end of the radiation hydrodynamical calculation, we plot the cumulative fraction objects as a function of the potential truncation radius. We exclude binaries and any objects that are still accreting at the end of the calculation.  The solid line gives the result for all stars and brown dwarfs, while the dotted, short-dashed, and long-dashed lines give the cumulative distributions for the mass ranges $M<0.1$, $0.1\leq M < 0.3$,  and $0.3\leq M < 1.0$, respectively.  There are no stars with masses $M \geq 1.0$ M$_\odot$ that are not either accreting or in multiple systems at the end of the calculation.
}
\label{cumtruncate}
\end{figure}

Despite these difficulties, if an object suffers a dynamical encounter that terminates its accretion this encounter will truncate any disc that is larger than approximately 1/2 of the periastron distance during the encounter \citep*{HalClaPri1996}.  Therefore, excluding binaries and objects that are still accreting, determining the distribution of 1/2 of the closest encounter distance should give us an indication of the disc size distribution around single objects that have reached their final masses.  Note that formally we have still included the wide components of triple and quadruple systems, but these constitute only 13 objects out of the 94 `single' non-accreting objects and all but one are in orbits with semi-major axes greater than 10~AU so their inclusion should not adversely affect our conclusions.

In Figure \ref{cumtruncate}, we plot the cumulative distributions of disc truncation radii (taken to be 1/2 of the closest encounter distance) for these objects.  The solid line gives the cumulative distribution for all 94 objects, while in the other distributions we break the sample into mass bins of $M_*<0.1$~M$_\odot$, $0.1~{\rm M}_\odot \leq M_* < 0.3$~M$_\odot$,  and $0.3~{\rm M}_\odot \leq M_* < 1.0$~M$_\odot$.  \cite{Bate2009a} found that in the main barotropic calculation there was a clear relation such that more massive stars tended to have had closer encounters.  Here, the statistics are not as good, but the stars in the highest mass bin have clearly typically had closer encounters.  

For VLM objects, 20\% have truncation radii greater than 40 AU, while 1/2 have truncation radii greater than 10 AU.  It has been known for a decade from infrared excess that young brown dwarfs have discs \citep{Muenchetal2001, NatTes2001, Apaietal2002, Nattaetal2002, Jayawardhanaetal2003, Luhmanetal2005a, Moninetal2010}, some of which also display evidence for accretion \citep{JayMohBas2002}.  At least some of these discs are inferred to have radii of 20-40~AU \citep[e.g.][]{Luhmanetal2007a}, while  \cite{SchJayWoo2006} estimate that at least 25\% of the brown dwarfs they survey in Taurus have discs larger than 10~AU in radius.  The cumulative distribution of truncation radii in Fig.~\ref{cumtruncate} is consistent with these observations, but {\em it should be used with caution}.  First, the simulation presented here produces a dense stellar cluster.  Disc truncation may be less important for setting disc sizes in a lower-density star-forming region like Taurus.  Second, Figure \ref{cumtruncate} does {\it not} give a disc size distribution.  At best, it is a distribution of {\it lower limits} to disc sizes because of the fact that stars can suffer a close dynamical encounter, but then accrete more material from the molecular cloud and form a new disc.  This happens frequently in the simulation, especially for the higher mass stars.  However, the distribution may be more useful for VLM objects because they tend to have their accretion terminated soon after they form by dynamical encounters and generally will not subsequently accrete significantly from the molecular cloud.

\begin{figure}
\centering
    \includegraphics[width=8.4cm]{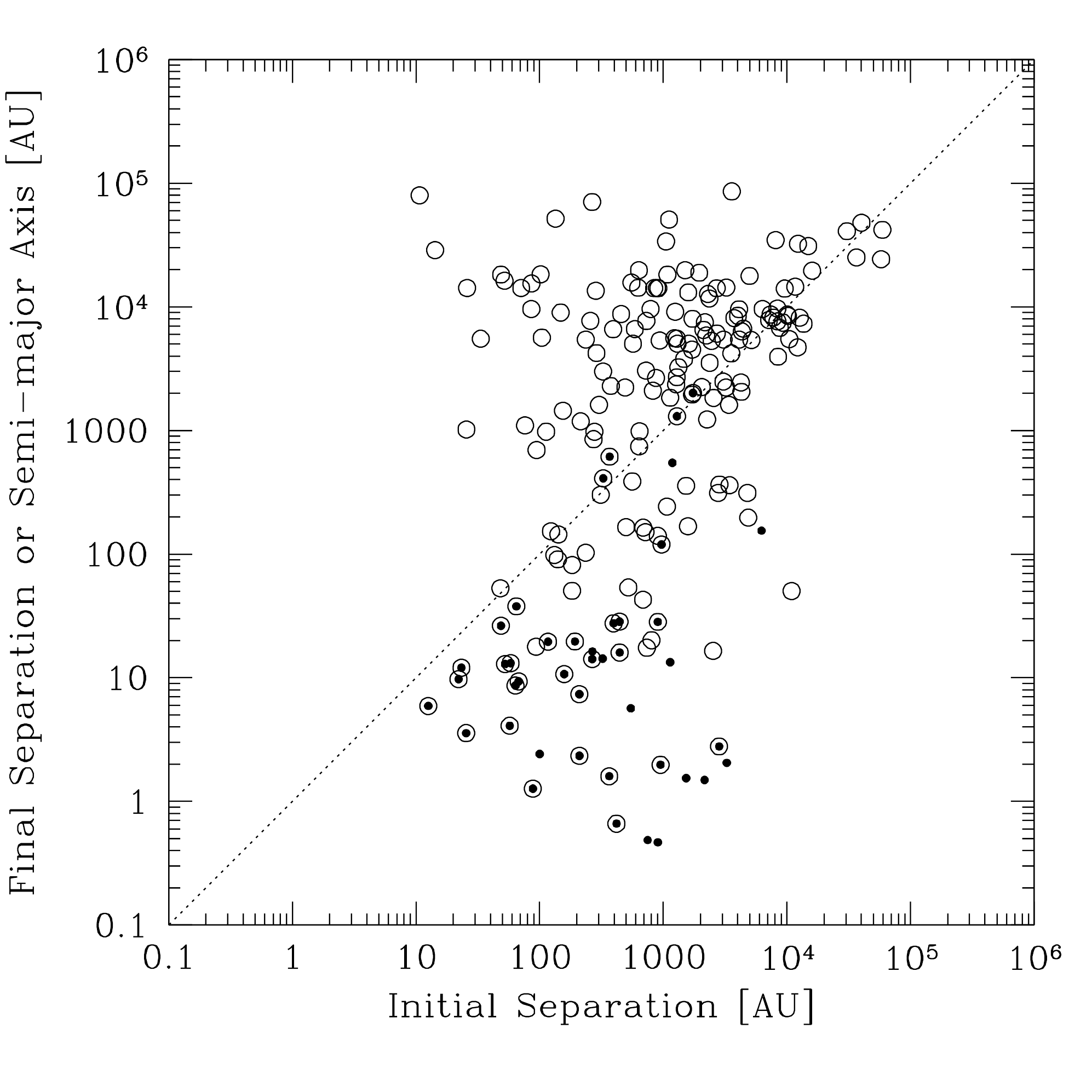}  
\caption{We find the closest object to each star or brown dwarf when it forms and plot their final versus initial separation (open circles).  We also plot the final semi-major axes versus the initial separations of all binaries at the end of the calculation (small filled circles).  Note that the closest object when a star or brown dwarf forms often does not remain close.  Also, many of the close multiple systems at the end of the calculation are composed of objects that formed at large distances from each other.  These results indicate the importance of dynamical interactions and orbital decay during the calculation.}
\label{separations} 
\end{figure}

\subsection{The formation of multiple systems}

The opacity limit for fragmentation
sets a minimum initial binary separation of $\approx 10$ AU since 
the size of a slowly-rotating first hydrostatic core is $\approx 5$~AU
\citep{Larson1969,MasInu2000,Tomidaetal2010a,Bate2011}.
In Fig.~\ref{separations}, we plot the distance to the closest other star or 
brown dwarf when each star or brown dwarf forms
 and the distance to this object at the end of the calculation (open
circles).  Also plotted is the initial separation and final semi-major axis of
all binaries at the end of the calculation (filled circles).  
No object forms closer than 10 AU
from an existing object, consistent with the expectations from
opacity limited for fragmentation.  

Examining Fig.~\ref{separations}, we find that some
objects that begin with close separations end up well separated.  Such situations 
occur when one of the objects is involved in a dynamical interaction (e.g. ejection from
a group or multiple system).
Alternately, objects that are initially widely separated ($100-10^4$~AU) can end
up in close bound systems (separations $<100$~AU).  In fact, most of the close
binaries at the end of the calculation are composed of mutual nearest
neighbours at the time of formation (i.e. in the figure the filled circles are
inside the open circles), but the separations of these objects may have been
reduced by up to 3 orders of magnitude during the evolution. 
In particular, despite the fact that no objects form closer than $\approx 10$~AU
from each other, at the end of the calculation there exist 21
binary systems and one triple system with separations $< 10$~AU.

The mechanisms by which close binaries form have been
discussed in detail by \cite{BatBonBro2002b}.  Briefly,
\citeauthor{BatBonBro2002b} found that,
rather than forming directly via fragmentation,
the close binary systems form from the orbital decay of wider systems
through a combination of dynamical interactions, accretion, 
and the interaction of binaries
and triples with circumbinary and circumtriple discs.  
Dynamical interactions can harden existing wide binaries by 
removing angular momentum and energy from their orbits.  
They also produce exchange interactions in which
a temporary unstable multiple system decays by ejecting one of the
components of the original binary (usually the lowest-mass object).  
However, dynamical interactions alone cannot produce the observed 
frequency of close binary systems, either beginning with stellar
clusters \citep{KroBur2001} or during the dissolution of small-N clusters \citep{SteDur1998}.  
The key is to have {\it dissipative} dynamical interactions,
where the presence of gas allows dynamical encounters to dissipate energy and 
transport angular momentum \citep{BatBonBro2002b}.  A good example is 
a star-disc encounter \citep*{Larson1990, ClaPri1991a, ClaPri1991b, McDCla1995, 
Heller1995, HalClaPri1996}. 
In addition to dynamical interactions, accretion onto a binary from an envelope decreases 
the binary's separation unless the specific angular momentum
of the accreted material is significantly greater than that of the binary
\citep{Artymowicz1983, Bate1997, BatBon1997, Bate2000}.
It can also change the relative separations of a triple system, 
destabilising it and forcing dynamical interactions \citep*{SmiBonBat1997}.
Circumbinary discs can remove large amounts of orbital angular 
momentum from an embedded binary system via gravitational torques,
thus tightening its orbit \citep{Pringle1991, Artymowiczetal1991}.

These mechanisms work in the calculation here to produce the close systems.
They also reduce the separations of wide $\sim 1000$~AU systems
to systems with intermediate separations $\sim 10-100$~AU.

\begin{figure*}
\centering
    \includegraphics[width=8.4cm]{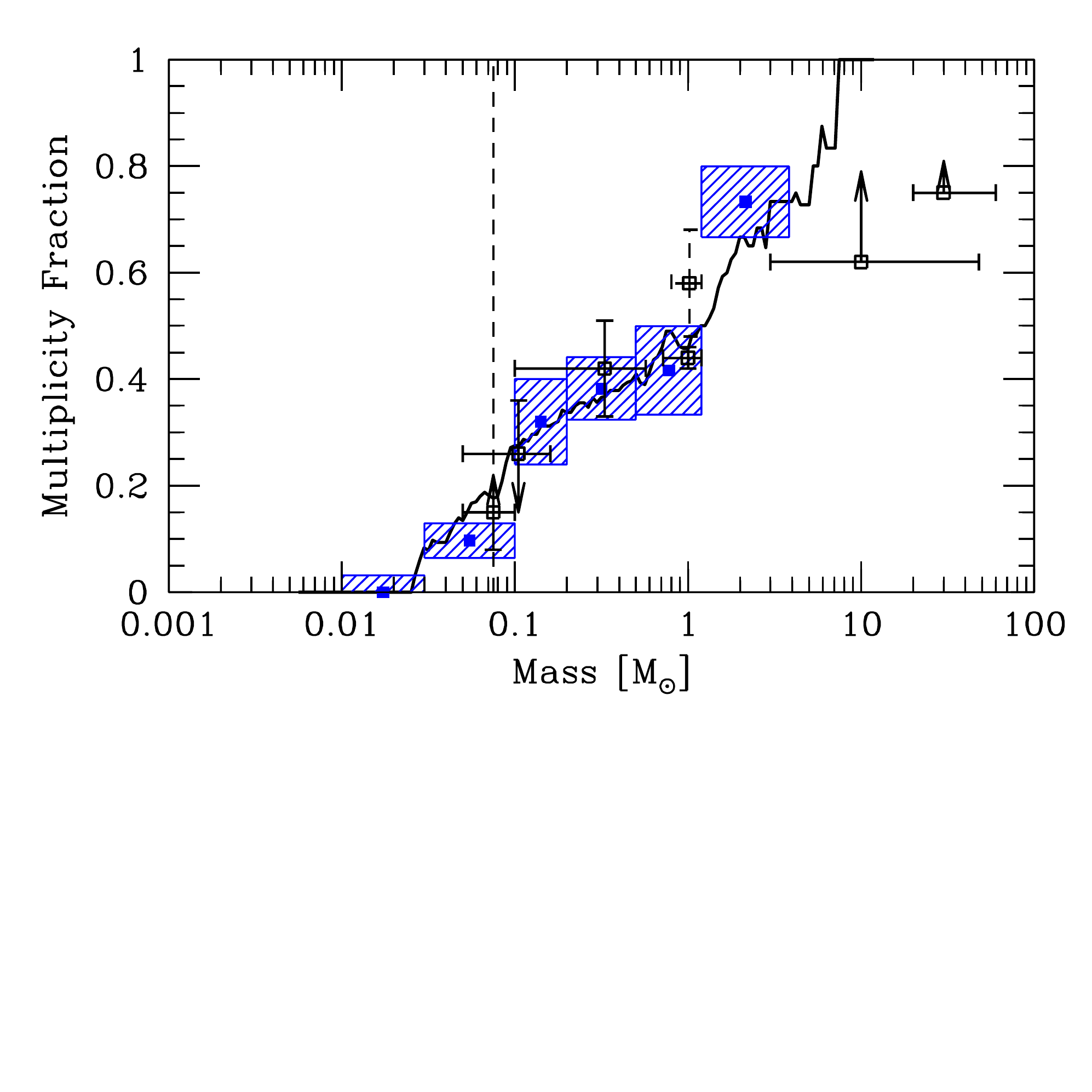}
    \includegraphics[width=8.4cm]{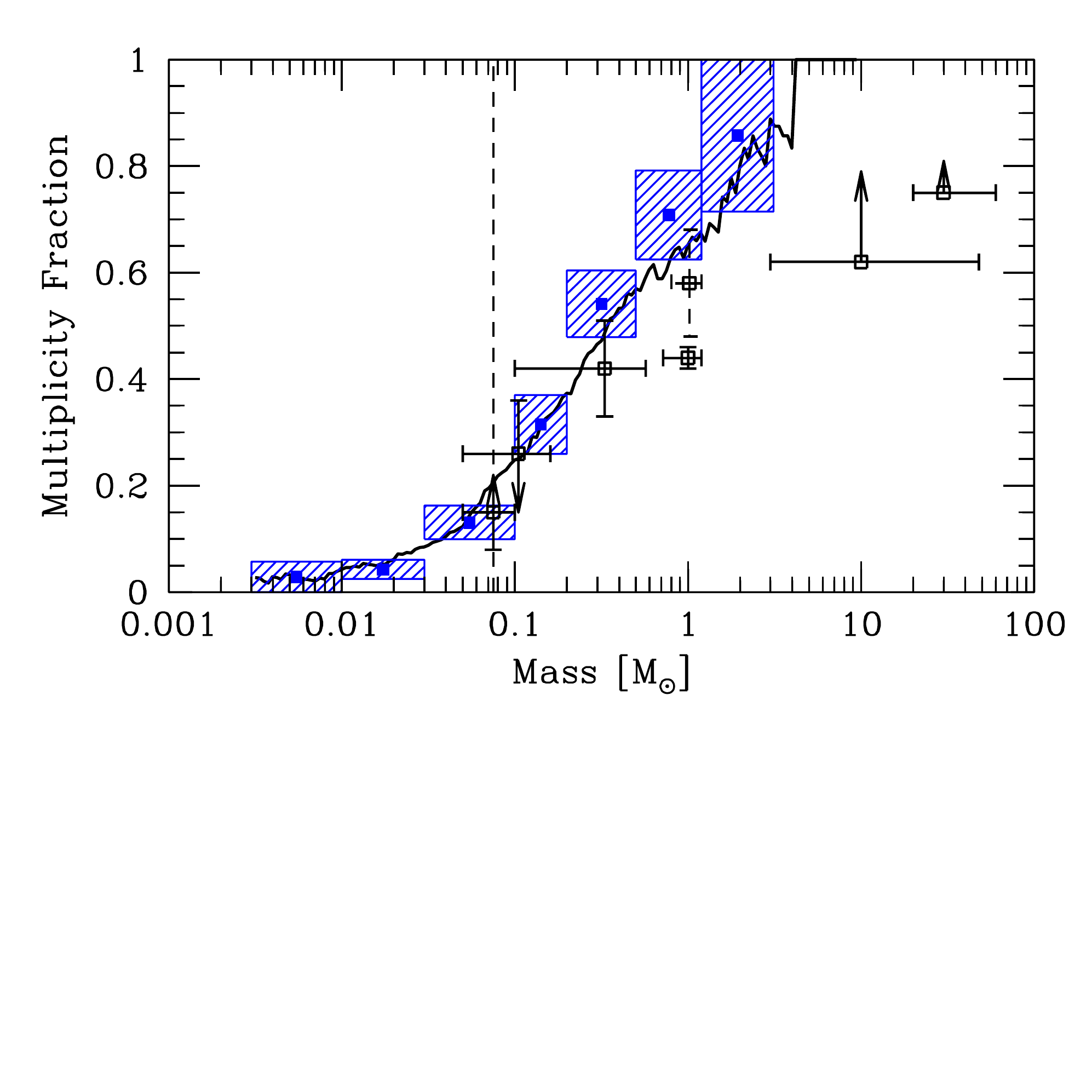}\vspace{-3cm}
\caption{Multiplicity fraction as a function of primary mass.  The left panel gives the result at the end of the radiation hydrodynamical calculation.  On the right, we give the result from the main barotropic calculation of Bate (2009a) at the same time.  The blue filled squares surrounded by shaded regions give the results from the calculations with their statistical uncertainties.  The thick solid lines give the continuous multiplicity fractions from the calculations computed using a boxcar average. The open black squares with error bars and/or upper/lower limits give the observed multiplicity fractions from the surveys of Close et al. (2003), Basri \& Reiners (2006), Fisher \& Marcy (1992), Raghavan et al. (2010), Duquennoy \& Mayor (1991), Preibisch et al. (1999) and Mason et al. (1998), from left to right.  Note that the error bars of the Duquennoy \& Mayor (1991) results have been plotted using dashed lines since this survey has been superseded by Raghavan et al. (2010).  The observed trend of increasing multiplicity with primary mass is well reproduced by both calculations.  Note that because the multiplicity is a steep function of primary mass it is important to ensure that similar mass ranges are used when comparing the simulation with observations. }
\label{multiplicity}
\end{figure*}

\subsection{Multiplicity as a function of primary mass}
\label{multiplicity_sec}

We turn now to the properties of the binary and higher-order multiple stars and brown dwarfs produced by the simulation.  Observationally, it is clear that the fraction of stars or brown dwarfs that are in multiple systems increases with stellar mass (massive stars: \citealt{Masonetal1998, Preibischetal1999, ShaTok2002, KobFry2007, Kouwenhovenetal2007, Masonetal2009}; intermediate-mass stars: \citealt{Patienceetal2002}; solar-type stars: \citealt{DuqMay1991,Raghavanetal2010}; M-dwarfs: \citealt{FisMar1992}; and very-low-mass stars and brown dwarfs: \citealt{Closeetal2003, Siegleretal2005,BasRei2006}).  It also seems that the multiplicity of young stars in low-density star-forming regions is somewhat higher than that of field stars \citep{Leinertetal1993, GheNeuMat1993,Simonetal1995,Ducheneetal2007}.  However, IC348 has a similar binary frequency to the field \citep{DucBouSim1999}.
In the Orion Nebula Cluster, \citet{Kohleretal2006} find that the binary frequency of low-mass stars is similar to that of field M dwarfs and lower than that of field solar-type stars, but that stars with masses $M>2$ M$_\odot$ have a higher binarity than stars with $0.1<M<2$ M$_\odot$ by a factor of 2.4 to 4.

\begin{table}
\begin{tabular}{lccccc}\hline
Mass Range ~ [M$_\odot$]& Single & Binary  & Triple & Quadruple  \\ \hline
\hspace{0.83cm}$M<0.03$       &      7     &     0     &      0      &     0    \\
$0.03\leq M<0.07$      &    20     &   0    &       0     &      0   \\
$0.07\leq M<0.10$      &      8      &    3      &     0     &      0   \\
$0.10\leq M<0.20$      &      17     &     7    &       1     &      0   \\
$0.20\leq M<0.50$      &      21     &     9    &       2     &     2   \\
$0.50\leq M<0.80$      &     5      &     2      &     0     &      1   \\
$0.80\leq M<1.2$        &       2      &     1      &     1     &      0   \\
\hspace{0.83cm}$M>1.2$        &       4       &    6      &     1     &      4   \\ \hline
All masses                    &   84   &     28     &      5   &      7           \\ \hline
\end{tabular}
\caption{\label{tablemult} The numbers of single and multiple systems for different primary mass ranges at the end of the radiation hydrodynamical calculation. }
\end{table}

\begin{table*}
\begin{tabular}{lcccccccccl}\hline
Object Numbers & $M_1$ & $M_2$  & $q$ & $a$  & $e$  & Initial & Relative Spin  & Spin$_1$-Orbit & Spin$_2$-Orbit & Comments \\
& & & & & & Separation & or Orbit Angle & Angle & Angle\\
        & [M$_\odot$] & [M$_\odot$] &  & [AU] & & [AU] & [deg] & [deg] & [deg] \\ \hline
 138,            117            & 0.29 & 0.29 & 0.97 &     0.39 & 0.61 &  907 &   91 &  129 &   64 \\
  32,             50            & 1.86 & 0.59 & 0.32 &        0.88 & 0.73 &  748 &  105 &   68 &  -94 \\
  20,             23            & 2.76 & 1.97 & 0.72 &        0.88 & 0.76 & 1533  &   13 &   49 &   62 \\
  25,             26            & 1.81 & 1.23 & 0.68 &        0.91 & 0.61 &  419 &    7 &   31 &   27 \\
  38,             45            & 1.40 & 1.20 & 0.86 &        1.08 & 0.60 &  88.3 &    4 &   69 &   73 \\
   6,             13            & 1.90 & 1.30 & 0.68 &        1.20 & 0.71 & 950  &   28 &   55 &   37 \\
   1,              7            & 1.37 & 1.07 & 0.78 &        1.25 & 0.63 & 366  &   87 &   96 & -169 \\
  64,             79            & 0.84 & 0.10 & 0.12 &        1.95 & 0.24 &  100 &   48 &   28 &   57 \\
  35,             33            & 1.91 & 1.50 & 0.79 &        1.96 & 0.67 & 2833  &   88 &   97 &   77 \\
  87,            121            & 1.26 & 0.40 & 0.32 &        2.02 & 0.26 & 2159  &   73 &   42 &  -36 \\
 118,            163            & 0.59 & 0.21 & 0.35 &        2.24 & 0.16 & 210  &   25 &   43 &   66 \\
   3,              8            & 2.27 & 1.00 & 0.44 &        2.56 & 0.45 &  3268 &   21 &   13 &   17 \\
  98,            109            & 0.29 & 0.21 & 0.71 &        3.01 & 0.44 & 25.4  &    4 &   68 &   66 \\
   5,             16            & 2.53 & 2.49 & 0.98 &        3.27 & 0.26 & 57.3  &    3 &   23 &   23 \\
 116,            132            & 0.25 & 0.12 & 0.45 &        4.82 & 0.52 & 547  &   19 &   59 &   56 \\
  63,            154            & 0.29 & 0.14 & 0.50 &        6.44 & 0.62 &  22.1 &    6 &    1 &   -5 \\
 115,            183            & 0.52 & 0.04 & 0.08 &        8.40 & 0.45 & 209  &   32 &   12 &   42 & Star/VLM triple \\
  90,            103            & 0.18 & 0.10 & 0.54 &        8.75 & 0.10 & 63.8  &   10 &   12 &   10 & Star/VLM binary  \\
 122,            145            & 0.23 & 0.14 & 0.62 &        8.81 & 0.23 & 158  &   13 &   21 &   13 \\
 102,            110            & 0.27 & 0.26 & 0.96 &        9.36 & 0.12 & 67.8  &    7 &   37 &   29 \\
 159,            150            & 0.16 & 0.15 & 0.96 &        9.91 & 0.35 & 1138  &   13 &   38 &   51 \\
  59,             68            & 0.08 & 0.05 & 0.61 &       10.6 & 0.46 & 12.6  &   13 &   17 &    4 & VLM binary \\
  19,             27            & 0.28 & 0.25 & 0.89 &       12.8 & 0.08 & 58.4  &    3 &   22 &   22 \\
  76,             83            & 0.25 & 0.21 & 0.84 &       12.8 & 0.42 & 267  &    9 &   30 &   22 \\
 164,            179            & 0.15 & 0.03 & 0.21 &       14.3 & 0.80 &  23.3 &   22 &    4 &   22  & Star/VLM triple  \\
  92,            133            & 0.27 & 0.20 & 0.74 &       14.3 & 0.16 & 52.6  &   20 &   19 &   -2 \\
  44,             82            & 1.03 & 0.91 & 0.88 &       14.3 & 0.01 & 267  &    8 &   35 &   31 \\
  41,             89            & 0.25 & 0.18 & 0.71 &       14.7 & 0.09 & 324  &   11 &   31 &   20 \\
  94,            129            & 0.18 & 0.09 & 0.49 &       14.8 & 0.09 &  445  &   56 &   57 &    3  & Star/VLM binary  \\
   4,             84            & 1.33 & 1.06 & 0.80 &       19.3 & 0.02 & 117  &   12 &   40 &   41 \\
 160,            168            & 0.19 & 0.15 & 0.81 &       19.5 & 0.45 &  902 &   15 &   29 &   34 \\
 104,             93            & 0.35 & 0.34 & 1.00 &       22.5 & 0.13 &  193 &    6 &   36 &   39 \\
 140,            147            & 0.09 & 0.09 & 0.94 &       26.1 & 0.01 & 48.7  &    5 &   23 &   19  & VLM binary \\
  65,             77            & 0.27 & 0.26 & 0.95 &       27.5 & 0.01 & 396  &    7 &   21 &   14 \\
  88,            127            & 0.21 & 0.10 & 0.50 &       29.7 & 0.05 &  443 &   41 &   24 &  -20 \\
 175,            181            & 0.09 & 0.08 & 0.98 &       36.4 & 0.06 & 64.8  &  106 &   53 &  -54 & VLM binary \\
  10,             17            & 3.84 & 0.70 & 0.18 &      165 & 0.16 & 6244  &   67 &  104 &   92 \\
  67,             74            & 0.14 & 0.05 & 0.38 &      356 & 0.35 & 327  &   74 &  144 &   77  & Star/VLM binary  \\
  49,             96            & 0.38 & 0.27 & 0.71 &      406 & 0.74 & 969  &   44 &   70 &   46 \\
 106,            148            & 0.71 & 0.12 & 0.17 &      474 & 0.70 &  369 &   53 &   70 &   26 \\
  12,            105            & 0.88 & 0.04 & 0.04 &      620 & 0.42 &  1185 &   77 &   91 &   93 & Star/VLM binary  \\
 153,            182            & 0.13 & 0.03 & 0.26 &      721 & 0.82 & 1289  &   44 &  128 &   93  & Star/VLM binary \\
 171,            174            & 0.10 & 0.08 & 0.79 &     8366 & 0.90 & 1743  &   94 &   36 &   99  & Star/VLM binary \\
\\
  (25,  26),        37            & (3.04) & 1.68 & 0.55 &        5.53 & 0.19 &    -- &   34 &    -- &    -- \\
  (64,  79),        55            & (0.94) & 0.86 & 0.91 &       18.1 & 0.10 &    -- &    4 &    -- &    -- \\
 (115, 183),       149            & (0.56) & 0.19 & 0.34 &       19.9 & 0.10 &    -- &   14 &    -- &    --   & Star/VLM triple\\
   (5,  16),        15            & (5.02) & 2.95 & 0.59 &       23.5 & 0.16 &    -- &   36 &    -- &    -- \\
 ( 87, 121),       100            & (1.66) & 1.21 & 0.73 &       36.8 & 0.24 &    -- &   32 &    -- &    -- \\
 (122, 145),       123            & (0.37) & 0.26 & 0.70 &       45.2 & 0.08 &    -- &    5 &    -- &    -- \\
 (116, 132),       131            & (0.37) & 0.15 & 0.41 &       57.8 & 0.19 &    -- &    6 &    -- &    -- \\
 (104,  93),       134            & (0.69) & 0.22 & 0.31 &      108 & 0.17 &    -- &   10 &    -- &    -- \\
 (164, 179),       173            & (0.18) & 0.07 & 0.40 &      194 & 0.49 &    -- &  126 &    -- &    --   & Star/VLM triple\\
\\
  ((87, 121), 100),  139            & (2.87) & 0.06 & 0.02 &      138 & 0.27 &    -- &  -- &    -- &    -- \\
   (4,  84),        (44,  82)       & (2.39) & (1.94) & 0.81 &      139 & 0.03 &    -- &  -- &    -- &    -- \\
  (38,  45),        (32,  50)       & (2.59) & (2.45) & 0.94 &      142 & 0.39 &    -- &  -- &    -- &    -- \\
 ((115, 183), 149),  126            & (0.75) & 0.50 & 0.67 &      145 & 0.09 &    -- &  -- &    -- &    -- \\
  (76,  83),        (41,  89)       & (0.46) & (0.43) & 0.93 &      161 & 0.45 &    -- &   -- &    -- &    -- \\
  ((25,  26),  37),   40            & (4.72) & 3.38 & 0.72 &      177 & 0.31 &    -- &  -- &    -- &    -- \\
 ((116, 132), 131),  119            & (0.52) & 0.11 & 0.21 &    10575 & 0.92 &    -- &  -- &    -- &    -- \\
\hline
\end{tabular}
\caption{\label{tablemultprop} The properties of the 40 multiple systems at the end of the calculation.  The structure of each system is described using a binary hierarchy.  For each `binary' we give the masses of the primary $M_1$ and secondary $M_2$, the mass ratio $q=M_2/M_1$, the semi-major axis $a$, the eccentricity $e$, the separation of the components when the second one first formed, the relative spin angle, and the angles between orbit and each of the primary's and secondary's spins.  For triples, we give the relative angle between the inner and outer orbital planes.  The combined masses of multiple systems are given in parentheses. }
\end{table*}

To quantify the fraction of stars and brown dwarfs that are in multiple systems we use the multiplicity fraction, $mf$, defined as a function of stellar mass.  We define this as
\begin{equation}
mf = \frac{B+T+Q}{S+B+T+Q},
\end{equation}
where $S$ is the number of single stars within a given mass range and, $B$, $T$, and $Q$ are the numbers of binary, triple, and quadruple systems, respectively, for which the primary has a mass in the same mass range.  Note that this differs from the companion star fraction, $csf$, that is also often used and where the numerator has the form $B+2T+3Q$.  We choose the multiplicity fraction following \citet{HubWhi2005} who point out that this measure is more robust observationally in the sense that if a new member of a multiple system is found (e.g. a binary is found to be a triple) the quantity remains unchanged.  We also note that it is more robust for simulations too in the sense that if a high-order system decays because it is unstable the numerator only changes if a quadruple decays into two binaries (which is quite rare).  Furthermore, if the denominator is much larger than the numerator (e.g. for brown dwarfs where the multiplicity fraction is low) the production of a few single objects does not result in a large change to the value of $mf$.  This is useful because many of the systems in existence at the end of the calculations presented here may undergo further dynamical evolution.  By using the multiplicity fraction our statistics are less sensitive to this later evolution.

The method we use for identifying multiple systems is the same as that used by \cite{Bate2009a}, and a full description of the algorithm is given in the method section of that paper.
When analysing the simulations, some subtleties arise.  For example, many `binaries' are in fact members of triple or quadruple systems and some `triple' systems are components of quadruple or higher-order systems.  From this point on, unless otherwise stated, we define the numbers of multiple systems as follows.  The number of binaries excludes those that are components of triples or quadruples.  The number of triples excludes those that are members of quadruples.  However, higher order systems are ignored (e.g. a quintuple system may consist of a triple and a binary in orbit around each other, but this would be counted as one binary and one triple).  We choose quadruple systems as a convenient point to stop as it is likely that most higher order systems will not be stable in the long-term and would decay if the cluster was evolved for many millions of years.  The numbers of single and multiple stars produced by the radiation hydrodynamical calculation are given in Table \ref{tablemult} following these definitions.  In Table \ref{tablemultprop}, we give the properties of the 40 multiple systems.

In the left panel of Fig.~\ref{multiplicity}, we compare the multiplicity fraction of the stars and brown dwarfs as a function of stellar mass obtained from the radiation hydrodynamical calculation with observations.  The results from a variety of observational surveys (see the figure caption) are plotted using black open boxes with associated error bars and/or upper/lower limits.  The data point from the survey of \cite{DuqMay1991} is plotted using dashed lines for the error bars since this survey has been recently superseded by that of \cite{Raghavanetal2010}.  The results from the radiation hydrodynamical simulation have been plotted in two ways.  First, using the numbers given in Table \ref{tablemult} we compute the multiplicity in six mass ranges: low-mass brown dwarfs (masses $<0.03$~M$_\odot$), VLM objects excluding the low-mass brown dwarfs (masses $0.03-0.10$ M$_\odot$), low-mass M-dwarfs (masses $0.10-0.20$ M$_\odot$), high-mass M-dwarfs (masses $0.20-0.50$ M$_\odot$), K-dwarfs and solar-type stars (masses $0.50-1.20$ M$_\odot$), and intermediate mass stars (masses $>1.2$ M$_\odot$). The filled blue squares give the multiplicity fractions in these mass ranges, while the surrounding blue hatched regions give the range in stellar masses over which the fraction is calculated and the $1\sigma$ (68\%) uncertainty on the multiplicity fraction.  In addition, a thick solid line gives the continuous multiplicity fraction computed using a boxcar average of the results from the radiation hydrodynamical simulation.  The width of the boxcar average is one order of magnitude in stellar mass.

The radiation hydrodynamical calculation clearly produces a multiplicity fraction that strongly increases with increasing primary mass.  Furthermore, the values in each mass range are in agreement with observation.  In the right panel of Fig.~\ref{multiplicity}, we provide the equivalent quantities obtained from the main barotropic calculation of \cite{Bate2009a} at the same time as the end of the radiation hydrodynamical calculation.  Those readers who wish to examine the multiplicity at the end of the barotropic calculation can find this in \cite{Bate2009a}.  The barotropic calculations also produce a multiplicity that is a strongly increasing function of mass.  In fact, the results using radiation hydrodynamics and a barotopic equation of state are very similar.  The main barotropic calculation gives multiplicities that are somewhat higher for primary masses $>0.2$~M$_\odot$ than those given by the radiation hydrodynamical calculation, but the results are consistent within the statistical uncertainties.

It is important to note that the surveys with which we are comparing the multiplicities are primarily of field stars rather than young stars.  This is necessary because surveys of young stars either do not sample a large range of separations and mass ratios, or the statistics are too poor.  However, there may be considerable evolution of the multiplicities between the age of the stars when the calculations were stopped ($\sim 10^5$ yrs) and a field population.  This question of the subsequent evolution of the clusters produced by hydrodynamical simulations was recently tackled by \cite{MoeBat2010} who took the end point of the main barotropic calculation of \cite{Bate2009a} and evolved it to an age of $10^7$~yrs using an N-body code under a variety of assumptions regarding the dispersal of the molecular cloud.  \citeauthor{MoeBat2010} found that the multiplicity distribution evolved very little during dispersal of the molecular cloud and was surprisingly robust to different assumptions regarding gas dispersal.  Even under the assumption of no gas removal at $10^7$~yrs, although the multiplicities were found to have decreased slightly compared with those at the end of the hydrodynamical calculation, they were still formally consistent.   They concluded that when star formation occurs in a clustered environment, the multiple systems that are produced are quite robust against dynamical disruption during continued evolution.  Therefore, we do not expect the multiplicities presented in Fig.~\ref{multiplicity} to evolve significantly as the stars evolve into a field population.

In detail, we find:
\begin{description}
\item[{\bf Solar-type stars:}] \citet{DuqMay1991} found an observed multiplicity fraction of $mf = 0.58 \pm 0.1$.  However, the recent larger survey carried out by \cite{Raghavanetal2010} revised this downwards to $0.44 \pm 0.02$ and they concluded that the higher value obtained by \citeauthor{DuqMay1991} was due to them overestimating their incompleteness correction.  The radiation hydrodynamical calculation gives a multiplicity fraction of $0.42\pm 0.08$ over the mass range $0.5-1.2$~M$_\odot$ which is in good agreement with the result of \cite{Raghavanetal2010}.

\item[{\bf M-dwarfs:}] \citet{FisMar1992} found an observed multiplicity fraction of $0.42 \pm 0.09$.  In the mass range $0.1-0.5$ M$_\odot$ we obtain $0.36 \pm 0.05$.  Fischer \& Marcy's sample contains stars with masses between 0.1 and 0.57 solar masses, but the vast majority have masses in the range $0.2-0.5$ M$_\odot$ whereas in the simulation almost half of the low-mass stars have masses less than 0.2 M$_\odot$.  In the $0.2-0.5$ M$_\odot$ mass range we obtain $0.38 \pm 0.06$.  All these values are consistent with the statistical uncertainties.

\item[{\bf VLM objects:}]  There has been much interest in the multiplicity of VLM objects in recent years (\citealt{Martinetal2000, Martinetal2003, Burgasseretal2003, Burgasseretal2006, Closeetal2003, Closeetal2007, Gizisetal2003, Pinfieldetal2003, Bouyetal2003, Bouyetal2006, Siegleretal2003, Siegleretal2005, Luhman2004a, MaxJef2005}; \citealt*{KraWhiHil2005, KraWhiHil2006}; \citealt{BasRei2006, Reidetal2006, Allenetal2007, Konopackyetal2007, Ahmicetal2007, Artigauetal2007, Reidetal2008}; \citealt*{LawHodMac2008}; \citealt{Maxtedetal2008, Luhmanetal2009b, Radiganetal2009, BurDhiWes2009, Fahertyetal2011}).  For a recent review, see \citet{Burgasseretal2007}.  Over the entire mass range of $0.018 - 0.10$ M$_\odot$, we find a very low multiplicity of just $0.08\pm 0.05$, although this is twice the value found from the main barotropic calculation of \cite{Bate2009a}.  However, the multiplicity drops rapidly with decreasing primary mass and the {\it observed} VLM objects tend to have high masses.  The calculation gives multiplicities of $0.32\pm 0.08$ for the mass range $0.1-0.2$ M$_\odot$, $0.27 \pm 0.15$ for the mass range $0.07-0.10$ M$_\odot$, and $0.0 \pm 0.05$ for the mass range $0.03-0.07$ M$_\odot$.  Therefore, to compare with observations it is very important to compare like with like.  The observed frequency of VLM binaries is typically found to be $\approx 15$\% \citep{Closeetal2003, Closeetal2007, Martinetal2003, Bouyetal2003, Gizisetal2003, Siegleretal2005, Reidetal2008}.  The surveys are most complete for binary separations greater than a couple of AU. \cite{BasRei2006} and \cite{Allen2007} estimated the total frequency (including spectroscopic systems) to be $\approx 20-25$\%.  These surveys typically targeted primaries with masses in the range $0.03-0.1$ M$_\odot$, but most of these objects in fact have masses greater than 0.07 M$_\odot$.  Thus, the closest comparison with our calculation is our frequency of $0.27 \pm 0.15$ for the mass range $0.07-0.10$ M$_\odot$.  This is in good agreement with observations, but the uncertainty is large because of the small number of objects.  Taking the average over the larger range of $0.03-0.20$ M$_\odot$ gives $0.20 \pm 0.05$ which is also in good agreement. Because the statistics from the radiation hydrodynamical calculation are not as good as those of the barotropic simulations it is difficult to determine whether or not including radiative feedback has an effect on the VLM multiplicity.  However, over the range $0.03-0.20$~M$_\odot$, the multiplicity from the rerun barotropic calculation of \cite{Bate2009a}, which has the same sink particle accretion radius size as the radiation hydrodynamical calculation, is $0.17 \pm 0.03$ which is in good agreement with the value obtained from the radiation hydrodynamical calculation.  Therefore, the use of radiation hydrodynamics rather than a barotropic equation of state does not seem to alter the VLM multiplicity significantly, and both are in good agreement with observations \citep[if small sink particle accretion radii are used;][]{Bate2009a}.

\item[{\bf Low-mass brown dwarfs:}]  The frequency of low-mass binary brown dwarfs (primary masses less than 30 Jupiter masses) is observationally unconstrained.  \cite{Bate2009a} predicted that the multiplicity would continue to fall as the primary mass is decreased and that the binary frequency in the mass range $0.01-0.03$ M$_\odot$ should be $\lsim 7$\%.  In the radiation hydrodynamical calculation, out of 27 systems with primary masses $<0.07$~M$_\odot$ there are no multiple systems.  Thus, although the statistics are not as good, the radiation hydrodynamical calculations also predict a very low multiplicity for low-mass brown dwarfs.
\end{description}

\subsubsection{Star-VLM binaries}

We turn now to the issue of VLM/brown dwarf companions to stars.  As in the previous section, we do not consider brown dwarf companions as such, rather we consider VLM companions ($< 0.1$ M$_\odot$) to stars ($\geq 0.1$ M$_\odot$).   The radiation hydrodynamical calculation produced 8 stellar-VLM systems out of 86 stellar systems, a frequency of $9 \pm 2$\%.  This is indistinguishable from the frequency given by the main barotropic calculation ($9.0\pm 1.6$\%).  One of the 8 systems is a 0.15-M$_\odot$ star with two VLM companions with with semi-major axes of 14 and 194 AU (Table \ref{tablemultprop}).  Another is a 0.52-M$_\odot$ star with a VLM companion at 8.4~AU and a 0.19-M$_\odot$ star at 20~AU.  The other six are binaries.  Five of the binaries have primary masses in the range $0.1-0.2$ M$_\odot$ (semi-major axes of 9, 15, 356, 721, and 8366 AU), and the sixth is a 0.88~M$_\odot$ star with a 36 M$_{\rm J}$ companion at 620 AU.  

Although the statistics are not as good as the main barotropic calculation (which produced 26 stellar-VLM systems), the properties are in good agreement.  In the main barotropic calculation, 14 of the primaries had masses between $0.1-0.2$ M$_\odot$, 7 had primary masses in the range $0.2-0.5$ M$_\odot$, and 3 had primary masses between 0.5 and 0.8 M$_\odot$ \citep{Bate2009a}.   However, within the statistical uncertainties, the frequency of stellar-VLM systems was not found to vary with primary mass.  \cite{Bate2009a} found a dependence of the separations of stellar-VLM binaries on primary mass.  He found a wide range of separations for primary masses $<0.2$~M$_\odot$, from close ($<20$~AU) to wide ($>1000$~AU) systems, but that the median separation increasing strongly with primary mass ($<30$~AU for stellar masses $<0.2$~M$_\odot$ to $\gsim 1000$~AU for solar-type primaries).  Although we cannot confirm this trend of median separation with primary mass from the radiation hydrodynamical simulation due to the smaller numbers of systems, the results are consistent with the barotropic results.

There has been much discussion over the past decade of the observed ``brown dwarf desert" for close brown dwarf companions solar-type stars \citep[frequency $\approx 1$\%;][]{MarBut2000, GreLin2006} and how this changes for wider separations and different primary masses.  \citet{McCZuc2004} found that the frequency of wide brown dwarfs to G, K, and M stars between 75-300 AU was 1\%$\pm$1\%.  The frequencies of wide brown dwarf companions to A and B stars \citep{KouBroKap2007}, M dwarfs \citep{Gizisetal2003}, and other brown dwarfs appears to be similarly low, although the frequency of wide binary brown dwarfs may be higher when they are very young \citep{Closeetal2007}.  Our results are consistent with these observations in the sense that we do not find brown dwarf companions to solar-type stars in close orbits and we find a low frequency of VLM companions at larger separations.  However, with our small numbers of objects we are unable to place strong limits.

\subsubsection{The frequencies of triple and quadruple systems}
\label{freq_high_order}

Consulting Table \ref{tablemult}, we find that the radiation hydrodynamical calculation produced 84 single stars/brown dwarfs, 28 binaries, 5 triples and 7 quadruples.  This gives an overall frequency of triple and quadruple systems of only $4\pm 2$\% and $6\pm 2$\%, respectively.  These are upper limits because some of these systems may be disrupted if the calculation were followed longer.  \cite{Bate2009a} found slightly lower values from the barotropic calculations, although the radiation hydrodynamic results agree within the uncertainties.

For the barotropic calculations, the frequencies of high-order multiples were found to increase strongly with primary mass.  Although the statistics are much poorer for the radiation hydrodynamical calculation, this also appears to be the case.  For VLM primaries, there are no triples or quadruples out of 38 systems.  For M-dwarf primaries ($0.10-0.50$ M$_\odot$) the frequency of triples/quadruples is $8\pm 4$\%, while for solar-type and intermediate mass stars the frequency is $\approx 26 \pm 10$\%.

Comparing these results with observations,  \cite{FisMar1992} found 7 triples and 1 quadruple amongst 99 M-star primaries giving a frequency of $8\pm 3$\%, in good agreement.  \citet{Raghavanetal2010} found the frequency of triple and higher-order multiple systems with solar-type primaries to be $11\pm 1$\%.  For primaries in the mass range $0.5-1.2$~M$_\odot$, the radiation hydrodynamical simulation gives a frequency of $17\pm 12$\%.  In summary, the frequencies of triples/quadruples obtained from the radiation hydrodynamical calculation are consistent with current observational surveys, but the statistical uncertainties are large.

\subsection{Separation distributions of multiples}
\label{sec:separations}

The main barotropic calculation of \cite{Bate2009a} produced the first reasonably large sample of multiple systems from a single hydrodynamical calculation: 58 stellar and 32 VLM binaries, in addition to 19 stellar and 4 VLM triple systems and 23 stellar and 2 VLM quadruple systems.  The radiation hydrodynamical calculation produces fewer multiple systems due to the effects of radiative feedback and the because we are not able to follow the evolution as far (Table \ref{table1}).  However, because the characteristic stellar mass increases with radiative feedback, it still provides nearly half as many multiple stellar systems (25 binaries, 5 triples and 7 quadruples).  The main difficulty is because the number of VLM objects is more than an order of magnitude lower, so the number of VLM multiples is very small.  The calculation only produces 3 VLM binaries.  Despite this, it is of interest to examine the distributions of semi-major axes.

\begin{figure}
\centering
    \includegraphics[width=8.4cm]{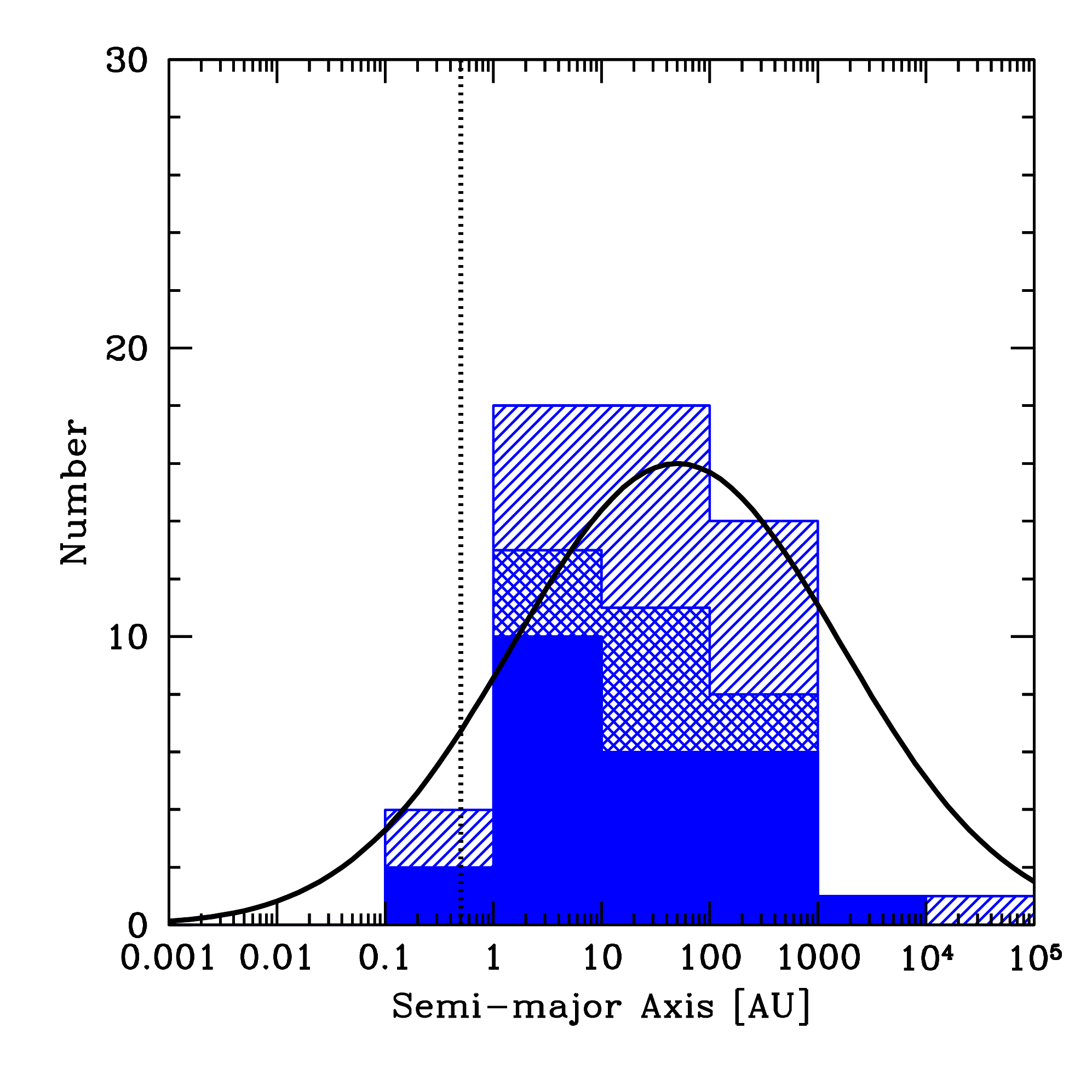}  
\caption{The distributions of separations (semi-major axes) of multiple systems with stellar primaries ($M_*>0.1$~M$_\odot$) produced by the radiation hydrodynamical calculation.  The solid, double hatched, and single hatched histograms give the orbital separations of binaries, triples, and quadruples, respectively (each triple contributes two separations, each quadruple contributes three separations).  The curve gives the solar-type separation distribution (scaled to match the area) from Raghavan et al.\ (2010). 
The vertical dotted line gives the resolution limit of the calculations as determined by the accretion radii of the sink particles.}
\label{separation_dist}
\end{figure}

Observationally, the median separation of binaries is found to depend on primary mass.  \citet{DuqMay1991} found that the median separation of solar-type binaries was $\approx 30$ AU.  In the recent larger survey of solar-type stars, \cite{Raghavanetal2010} found $\approx 40$~AU.   \citet{FisMar1992} found indications of a smaller median separation of $\approx 10 $ AU for M-dwarf binaries.  Finally, VLM binaries are found to have a median separation $\lsim 4$ AU \citep{Closeetal2003, Closeetal2007, Siegleretal2005}, with few VLM binaries found to have separations greater than 20 AU, particularly in the field \citep{Allenetal2007}.  A list of VLM multiple systems can be found at http://vlmbinaries.org/.  \citet{Closeetal2007} estimated that young VLM objects have a wide ($>100$ AU) binary frequency of $\sim 6$\%$\pm$3\% for ages less than 10 Myr, but only 0.3\%$\pm$0.1\% for field VLM objects.

Although we are able to follow binaries as close as 0.01~R$_\odot$ before they are assumed to merge in the radiation hydrodynamical calculation, the sink particle accretion radii are 0.5~AU.  Thus, dissipative interactions between stars and gas are omitted on these scales which likely affects the formation of close systems \citep{BatBonBro2002a}.  

In Figure \ref{separation_dist}, we present the separation (semi-major axis) distributions of the stellar (primary masses greater than 0.10 M$_\odot$) multiples.  We do not plot the distribution of VLM binaries because there are only three systems.  The distribution is compared with the log-normal distribution from the survey of solar-type stars of \cite{Raghavanetal2010} (which is very similar to that of \citealt{DuqMay1991}).  The filled histogram gives the separations of binary systems, while the double hatched region adds the separations from triple systems (two separations for each triple, determined by decomposing a triple into a binary with a wider companion), and the single hatched region includes the separations of quadruple systems (three separations for each quadruple which may be comprised of two binary components or a triple with a wider companion).  The vertical dotted line denotes the sink particle accretion radius.

The median separation (including separations from binary, triple, and quadruple systems) of the stellar systems is 15 AU with a standard deviation of 0.97 dex (i.e. 1 order of magnitude).  Given the smaller number of systems, this is in reasonable agreement with the value of 26 AU with a standard deviation of 1.15 dex obtained by \cite{Bate2009a} for the main barotropic calculation.  Both the median separation (40~AU) and the dispersion (1.52 dex) obtained by \cite{Raghavanetal2010} for solar-type stars are larger.  However, for the median, it is important to note that most of the primaries from the calculation are M-dwarfs, not solar-type stars and \cite{FisMar1992} found that M-dwarf binaries have smaller separations (median $\approx 10$~AU).  For the width of the distribution, the number of close systems is likely underestimated because of the lack of dissipation on small scales \citep[see][]{Bate2009a}.  The number of wide systems is low because the stellar cluster is dense and wide binaries can not exist within the cluster.  There appears to be a similar deficit of wide binaries in the Orion Nebula Cluster (\citealt*{BatClaMcC1998, ScaClaMcC1999}; \citealt{Reipurthetal2007}).  Furthermore, \cite{MoeBat2010} and \cite{Kouwenhovenetal2010} have shown that wide systems can be formed as a star cluster disperses \citep[see also][]{MoeCla2011}.

\begin{figure*}
\centering
    \includegraphics[width=5.8cm]{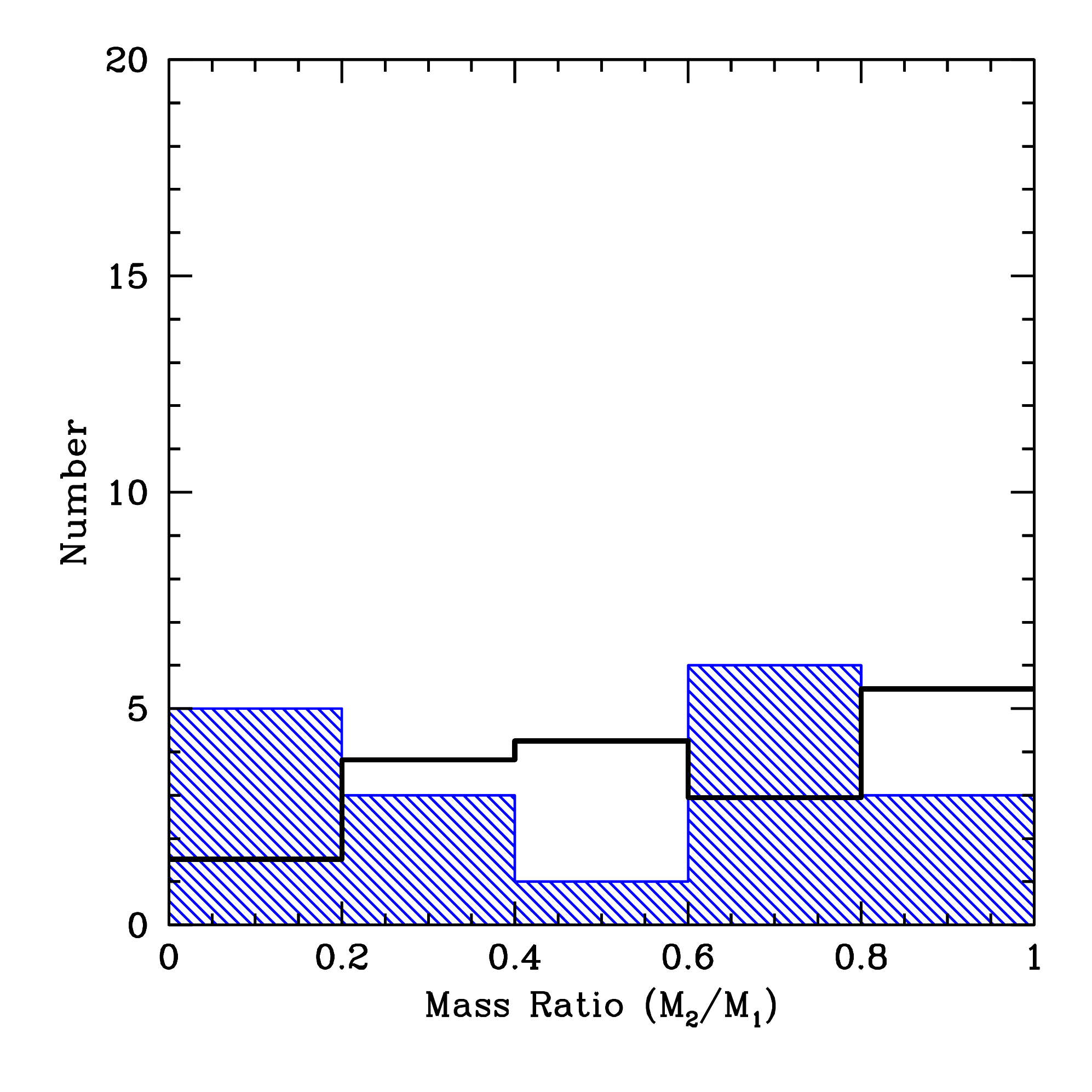} 
    \includegraphics[width=5.8cm]{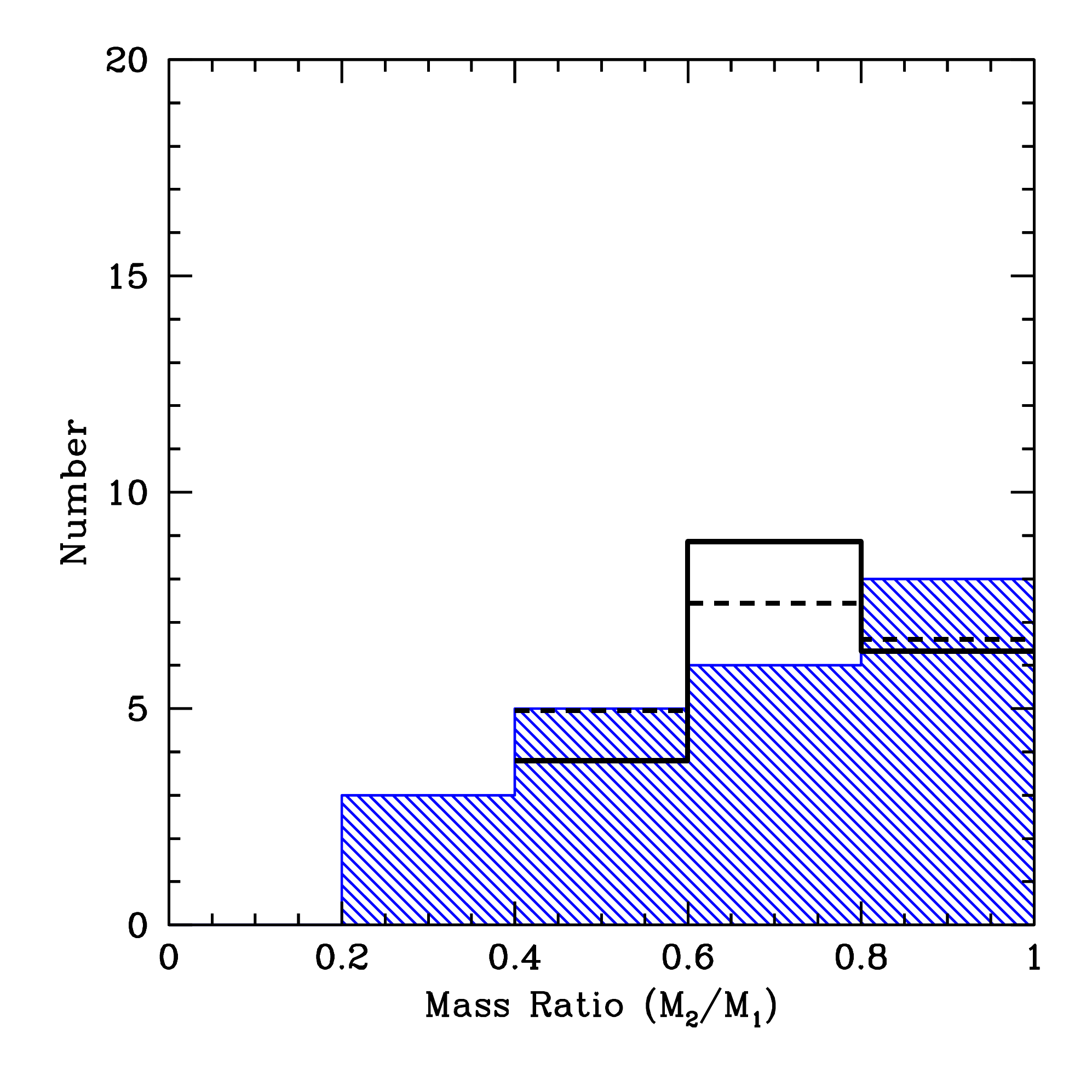} 
    \includegraphics[width=5.8cm]{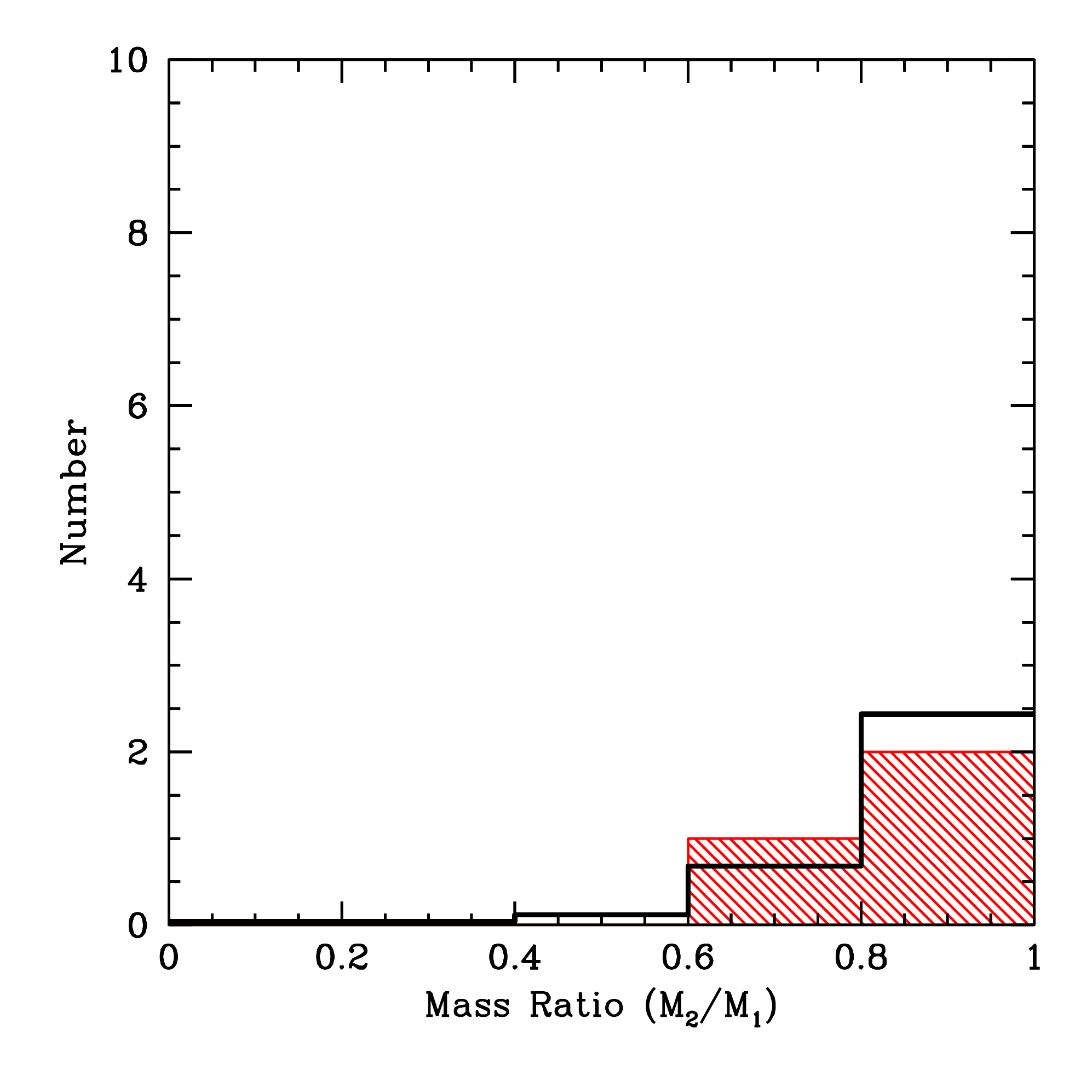}
\caption{The mass ratio distributions of binary systems with stellar primaries in the mass ranges $M_1>0.5$ M$_\odot$ (left) and $M_1=0.1-0.5$ M$_\odot$ (centre) and VLM primaries (right; $M_1<0.1$ M$_\odot$) produced by the radiation hydrodynamical calculation.  The solid black lines give the observed mass ratio distributions of \citet{Raghavanetal2010} for pairs with solar-type primaries (left), \citet{FisMar1992} for $M_1=0.3-0.57$ M$_\odot$ (centre, solid line) and $M_1=0.2-0.57$ M$_\odot$ (centre, dashed line), and of the known very-low-mass binary systems from the list at http://vlmbinaries.org/ (right).  The observed mass ratio distributions have been scaled so that the areas under the distributions ($M_2/M_1=0.4-1.0$ only for the centre panel) match those from the simulation results.  The VLM binaries produced by the simulation are biased towards equal masses when compared with M-dwarf and solar-type binaries.  All three of the VLM binaries have $M_2/M_1>0.6$ while for the M-dwarf binaries the fraction is only 63\% and for the more massive primaries the fraction is only 50\%. }
\label{massratios}
\end{figure*}

The three VLM binaries have semi-major axes of 10.6, 26.1, and 36.4 AU.  With only three systems, no firm conclusions can be drawn.  However, these semi-major axes are consistent with the fact that most observed VLM binaries have projected separations $\lsim 20$~AU and that wide systems ($>100$~AU) are rare.  Furthermore, \cite{Bate2009a}, who obtained 32 VLM multiples from the main barotropic calculation, found that the median separation of VLM multiples decreased as the calculation was evolved from $\approx 30$~AU at $1.04~t_{\rm ff}$ to $\approx 10$~AU at $1.50~t_{\rm ff}$ because many were still accreting gas and interacting with other systems early on.  He concluded that VLM binaries may form with reasonably wide separations and evolve to smaller separations \citep[c.f.][]{BatBonBro2002b}.   It is interesting to note that of the three VLM binaries found here, only the closest (10.6~AU) has stopped accreting.  Soon after the binary was formed at 1.034~$t_{\rm ff}$, a third object formed nearby making it a VLM triple.  The binary's initial separation was 12.6~AU which grew to 16.5~AU while the binary was accreting, and then was reduced to 10.8~AU in a dynamical encounter at $t=1.10~t_{\rm ff}$ that terminated its accretion.  The triple survived until another dynamical encounter at $t=1.17~t_{\rm ff}$ which striped off the wider VLM object and reduced the separation of the binary slightly to 10.6~AU.  For the other two VLM binaries which were still accreting at the end of the simulation, the 36.4~AU VLM binary was formed at 1.194~$t_{\rm ff}$ with a separation of 65~AU which decreased continually until the simulation was stopped.  The 26.1~AU binary formed at 1.13 with an initial separation of 48.6~AU which quickly decreased to 19.7~AU and then grew under the action of accretion to its final value.  In addition to the theoretical evidence from  \cite{Bate2009a} that the separation distribution of VLM binaries may evolve with time, the observational studies of \cite{Closeetal2007} and \cite{Burgasseretal2007} suggest that young wide VLM binaries are disrupted, leading to the observed paucity of old wide VLM systems.

In summary, the radiation hydrodynamical simulation produces a stellar separation distribution that is broad with a median separation that is in reasonable agreement with field systems.  It lacks very close systems (presumably due to the lack of dissipation on small scales).  It also lacks very wide systems, which may be formed as the cluster disperses.  The VLM binaries are consistent with the observation that most VLM binaries are close, but with only three systems, two of which are still evolving, no stronger conclusions can be drawn.

\subsection{Mass ratios distributions of multiples}
\label{sec:massratios}

Along with the separation distributions of the multiple systems we can investigate the mass ratio distributions.  We begin by considering only binaries, but we include binaries that are components of triple and quadruple systems.  A triple system composed of a binary with a wider companion contributes the mass ratio from the binary, as does a quadruple composed of a triple with a wider companion.  A quadruple composed of two binaries orbiting each other contributes two mass ratios --- one from each of the binaries.

Observationally, the mass ratio distribution of binaries also is found to depend on primary mass.  \citet{DuqMay1991} found that the mass ratio distribution of solar-type binaries peaked at $M_2/M_1 \approx 0.2$.  However, the recent survey of \cite{Raghavanetal2010} overturns this result.  \cite{Raghavanetal2010} found a flat mass ratio distribution for solar-type primaries in the range $M_2/M_1=0.2-0.95$, with a drop-off at lower mass ratios and a strong peak at nearly equal masses (so-called twins; \citealt{Tokovinin2000b}).  They find the mass ratios of pairs in higher-order systems follow the same distribution.  These results are consistent with the earlier study of \citet{Halbwachsetal2003} who found a bi-modal distribution for spectroscopic binaries with primary masses in the mass range $0.6-1.9$ M$_\odot$ and periods $\lsim 10$ years with a broad peak in the range $M_2/M_1=0.2-0.7$ and a peak for equal-mass systems.  \citet{Mazehetal2003} found a flat mass ratio distribution for spectroscopic binaries with primaries in the mass range $0.6-0.85$ M$_\odot$.   \citet{FisMar1992} also found a flat mass ratio distribution in the range $M_2/M_1 = 0.4-1.0$ for M-dwarf binaries with all periods.  In the Taurus-Auriga star-forming region, \cite{Krausetal2011} report a flat mass ratio distribution for primaries in the range $0.7-2.5$~M$_\odot$, but for primaries in the mass range $0.25-0.7$~M$_\odot$ they find a bias toward equal-mass systems.  This change becomes even more extreme for VLM binaries, which are found to have a strong preference for equal-mass systems \citep{Closeetal2003, Siegleretal2005,Reidetal2006}.

\begin{figure}
\centering
    \includegraphics[width=8.4cm]{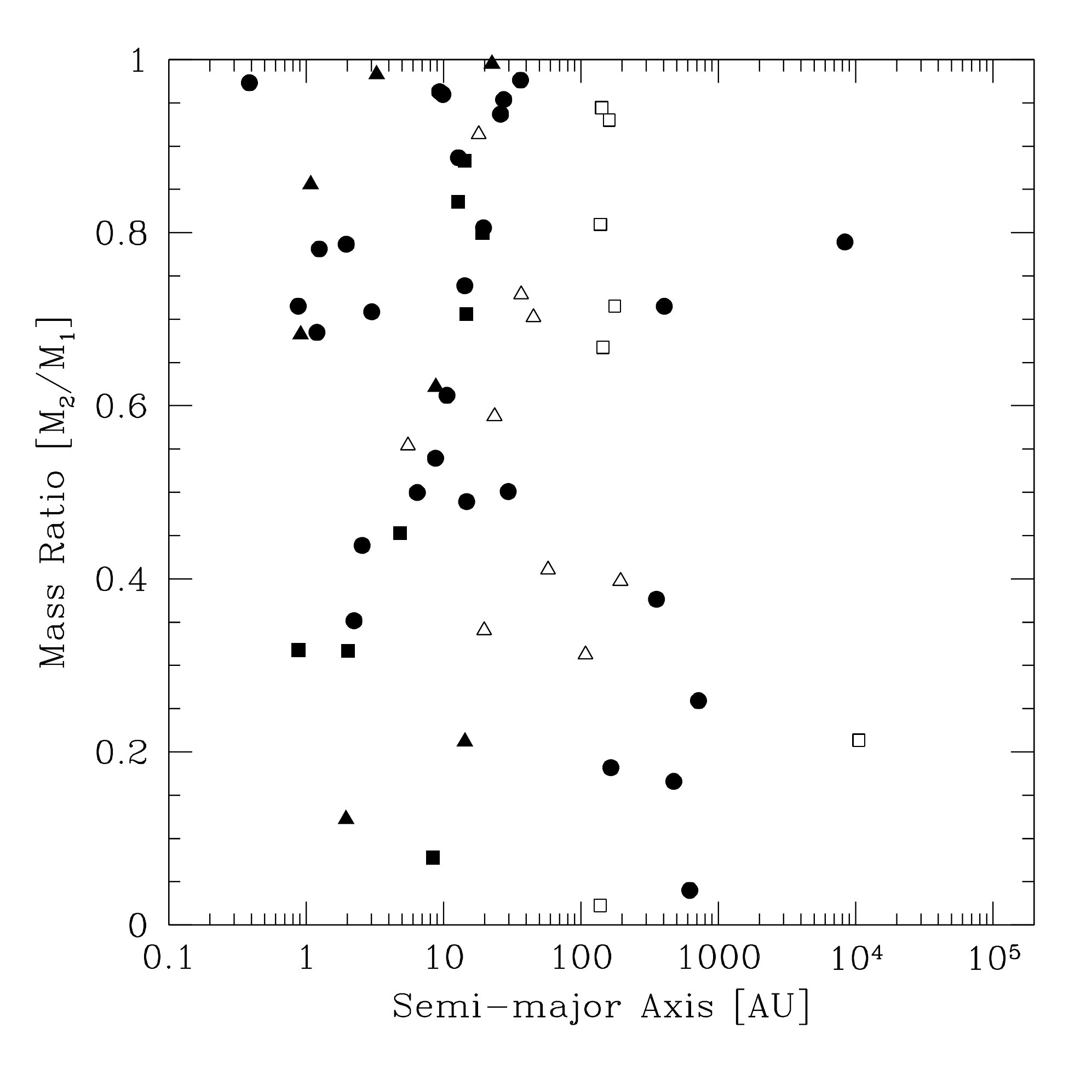}
\caption{The mass ratios of binaries (filled circles), pairs in triples (filled triangles), pairs in quadruples (filled squared), the wide components of triples (open triangles), and the widest components of quadruples (open squares) as a function of semi-major axis at the end of the radiation hydrodynamical calculation.  For the wide components of triples, the mass ratio compares the mass of the widest component to the sum of the masses of the two closest components (the pair).  For quadruples involving a two binary components (pairs), the mass ratio is between the two pairs, and for quadruples involving a triple, the mass ratio is between the mass of the fourth component and the triple.  There is a clear relationship between mass ratio and separation with closer binaries having a greater fraction of near equal-mass systems.}
\label{a_q}
\end{figure}

In Figure \ref{massratios}, we present the mass ratio distributions of the stars with masses $\geq 0.5$ M$_\odot$ (left panel), M-dwarfs with masses $0.1~{\rm M}_\odot \leq M_1<0.5$ M$_\odot$ (centre panel), and VLM objects (right panel).  We compare the M-dwarf mass ratio distribution to that of \citet{FisMar1992}, and the higher mass stars to the mass ratio distribution of pairs with solar-type primaries obtained by \citet{Raghavanetal2010}.  The VLM mass ratio distribution is compared with the listing of VLM multiples at http://vlmbinaries.org/.  

We find that the ratio of near-equal mass systems to systems with dissimilar masses decreases going from VLM objects to solar-type stars in a similar way to the observed mass ratio distributions, although the statistical significance is not strong.  Specifically, all three of the VLM binaries have $M_2/M_1>0.6$ while for primary masses $0.1-0.5$ M$_\odot$ the fraction is only 63\%, and for solar-type stars ($>0.5$~M$_\odot$) the fraction is 50\%.  The M-dwarf mass ratio distribution is consistent with Fischer \& Marcy's distribution.  There are only three VLM binaries, but they are consistent with the observation most VLM systems have mass ratios greater than 0.6.  For solar-type systems, \cite{Raghavanetal2010} obtained a generally flat mass ratio distribution, again consistent with the results obtained here.

The barotropic calculations of \cite{Bate2009a} also gave higher proportions of near equal-mass binaries for VLM binaries than M-dwarf binaries (with greater statistical significance), but gave similar fractions for solar-type and M-dwarf binaries.  At the time, \cite{Bate2009a} concluded that the barotropic calculations did not produce enough unequal-mass solar-type binaries because \cite{DuqMay1991} found that the mass ratio distribution peaked at $M_2/M_1 \approx 0.2$.  But the new mass ration distribution obtained by \cite{Raghavanetal2010} has reduced the discrepancy.

As with the VLM separation distribution, \cite{Bate2009a} found that the VLM binary mass ratio distribution evolved with time, becoming more biased towards equal-mass systems.  Both the apparent evolution of VLM binary separations and mass ratios are consistent with the evolution of close binaries discussed by \citet{BatBonBro2002b}.  Dynamical exchange interactions between binaries and single objects tend to produce more equal-mass components, as does accretion of gas from circumbinary discs or the accretion of infalling gas with high specific angular momentum.  The evolution seen in the radiation hydrodynamical calculation is consistent with this: the 10.6~AU VLM binary had a mass ratio of 0.49 at $1.05~t_{\rm ff}$ which grew to its final value of 0.61 before the binary stopped accreting at $1.10~t_{\rm ff}$, while the mass ratios of the two VLM binaries which were still accreting at the end of the calculation ($M_2/M_1=0.94, 0.98$) were being equalised by this accretion.

\subsubsection{Mass ratio versus separation}

In Figure \ref{a_q}, we plot mass ratios against separation (semi-major axis) for the binaries, triples, and quadruples at the end of the main calculation.  Note that for this figure we include systems that are sub-components of higher-order systems, using filled symbols to denote pairs that are binaries (circles), or are components of triples (triangles) or quadruples (squares).  We also include the mass ratios of the wide components of triples and quadruples.

\cite{Bate2009a} found a clear relation between mass ratio and separation from the barotropic calculations, with closer binaries having a preference for equal masses.  He obtained median mass ratios for {\it binary} separations in the ranges $1-10$, $10-100$, $100-1000$ and $1000-10^4$ AU of $M_2/M_1=0.74, 0.57, 0.68, 0.17$, respectively.  Including the mass ratios of triples and quadruples (as defined in the caption of Figure \ref{a_q}), these median values became 0.74, 0.41, 0.15, and 0.07, respectively.  

In the radiation hydrodynamical calculation, we find a similar trend (but again, with poorer statistical significance).  The median binary mass ratios in the separation ranges $1-10$, $10-100$ and $100-1000$ AU are  $M_2/M_1=0.62, 0.81, 0.22$, respectively.  Including the mass ratios of triples and quadruples, the median values become $M_2/M_1=0.59, 0.74, 0.39$, respectively.  There is only 1 binary with a separation $>1000$~AU.  Thus, there seems to be a trend for systems with separations $\lsim 100$~AU to have more equal masses than for wider systems.  A trend of more equal-mass binaries with decreasing separation is expected from the evolution of protobinary systems accreting gas from an envelope \citep{Bate2000}.  Furthermore, dynamical interactions between binaries and single stars tend to tighten binaries at the same time as increasing the binary mass ratio through exchange interactions.

In terms of the so-called twins, the radiation hydrodynamical calculation produced 43 binaries (including those in triple and quadruple systems), of which there are 7 twins (pairs with mass ratios $M_2/M_1>0.95$) and all have semi-major axes less than 40~AU.  Furthermore, two of these twins are components of triple systems.  This is in good agreement with observations that consistently find that closer binaries have a higher fraction of twins \citep{Soderhjelm1997, Tokovinin2000b, Halbwachsetal2003}.  \citet{Tokovinin2000b} found evidence for the frequency of twins falling off for orbital periods greater than 40 days, while \citet{Halbwachsetal2003} found that the fraction of near equal-mass systems ($M_2/M_1>0.8$) is always larger for shorter period binaries than longer period binaries regardless of the dividing value of the period (from just a few days up to 10 years).  The most recent study of \cite{Raghavanetal2010} found the mass ratio distribution depends on period, with less than 1/4 of twins having periods longer than 200 years (separations $\approx 40$~AU) and no twins having separations greater than periods of 1000 years (separations of $\approx 115$~AU). 

Finally, we note \cite{Raghavanetal2010} find that more than half of the pairs with periods less than 100 days (separation $\approx 0.5$~AU) are components of triples, suggesting that dynamical interactions may be important for their formation \citep[see][]{BatBonBro2002b}.  The closest pair formed in the radiation hydrodynamical calculation is a 0.4-AU binary.  However, of the 15 pairs with semi-major axes less than 5~AU, 8 are binaries while 4 are components of triples and 3 are components of quadruples.  Thus, the radiation hydrodynamical calculation also results in approximately half of the closest pairs being members of higher-order systems.

\begin{figure*}
\centering
    \includegraphics[width=8.4cm]{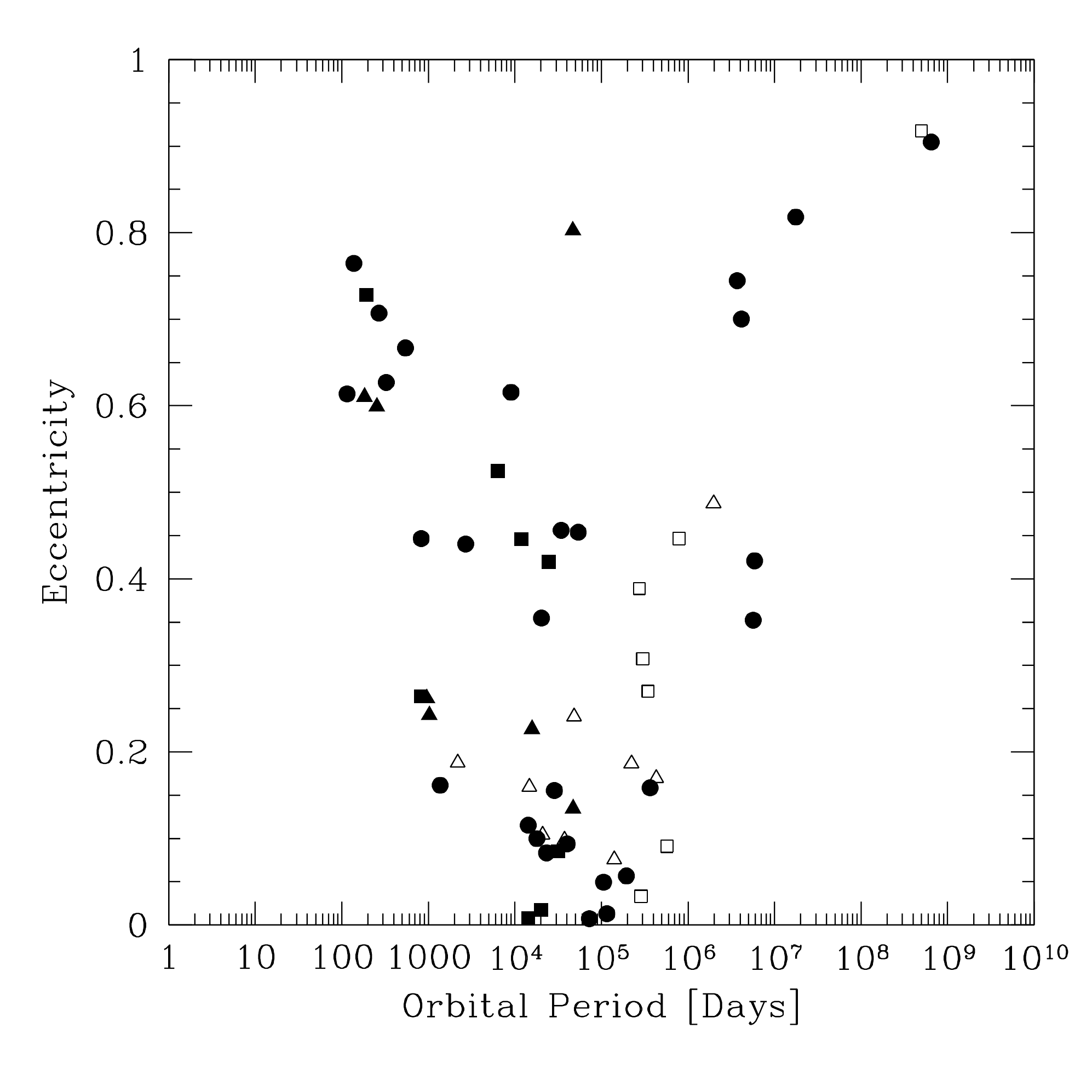}
    \includegraphics[width=8.4cm]{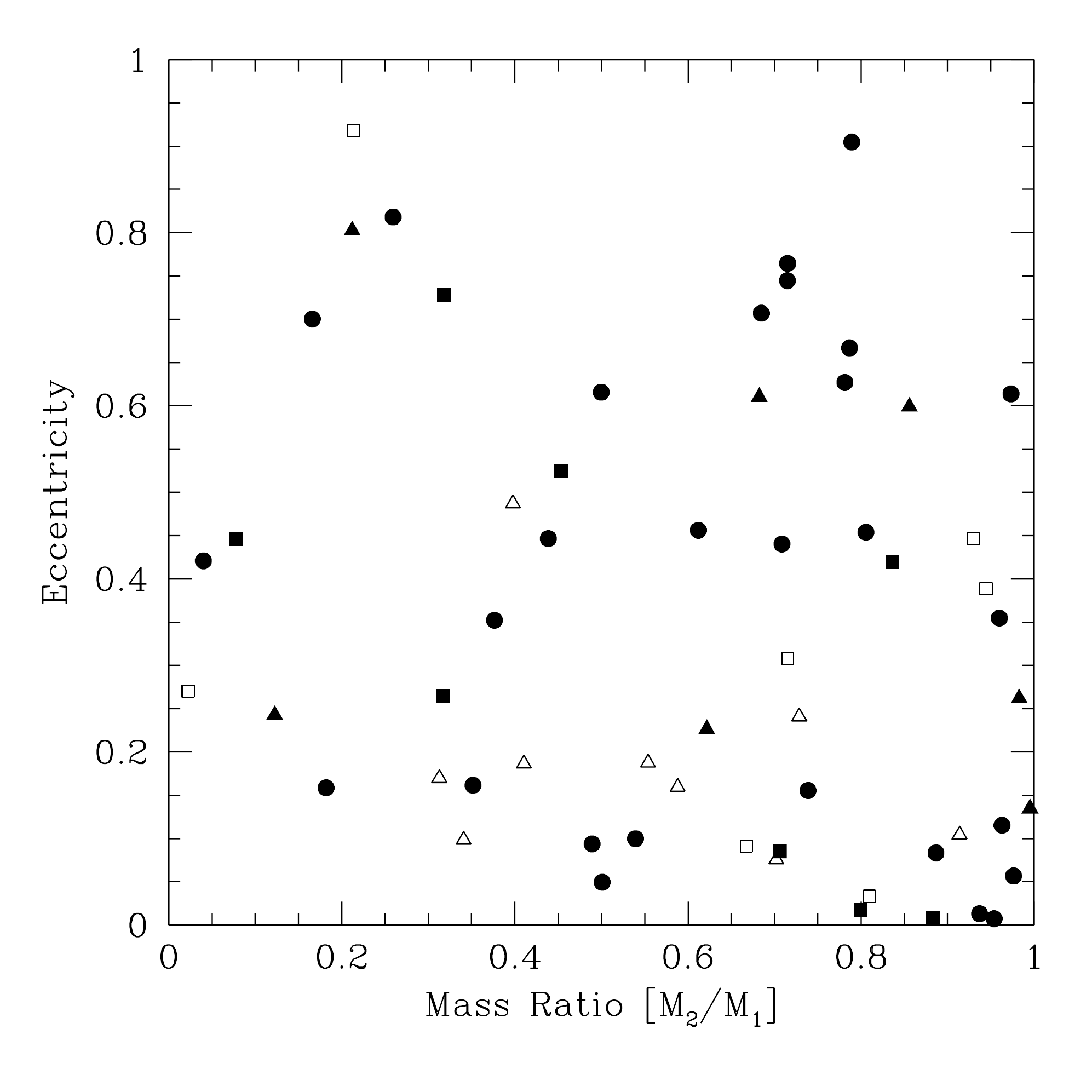}
\caption{The eccentricities of binaries (filled circles), pairs in triples (filled triangles), pairs in quadruples (filled squared), the wide components of triples (open triangles), and the widest components of quadruples (open squares) as a function of orbital period (left) and mass ratio (right) at the end of the radiation hydrodynamical calculation. The distributions look reasonable, but the eight binaries/pairs with the shortest periods all have large eccentricities which may be due to the absence of gas dynamics inside 0.5 AU of each sink particle.}
\label{eccentricity}
\end{figure*}

\subsubsection{Mass ratios of triples and quadruples}

For stellar triple and quadruple systems, \citet{Tokovinin2008} reports that triples are observed to have a median outer mass ratio of 0.39 independent of the outer orbital period while quadruples involving two binary sub-components have a similar median outer mass ratio of $\approx 0.45$, but there appears to be a dependence on the outer orbital period with systems with shorter outer periods having higher mass ratios.  Of 9 triple systems, we obtain a median mass ratio of 0.55 (0.59 excluding triples which are members of quadruple systems).  There are only 3 quadruple systems consisting of two pairs, all with outer mass ratios $>0.8$ and outer periods $5.4 < \log_{10}(P_{\rm d})<5.9$ (measured in days).  \cite{Tokovinin2008} finds no outer mass ratios $<0.6$ for orbital periods $\log_{10}(P_{\rm d})<5.4$ in this orbital period range, but a wide range of mass ratios for longer periods.  Since we only have three systems and they all fall near the apparent observed step change it is not possible to draw any firm conclusions. 

\cite{Bate2009a} found that quadruples composed of a triple and a wide fourth component out number quadruples composed of two binaries by 2:1 in the main barotopic calculation.  Observationally, \cite{Tokovinin2000a} finds roughly equal numbers of such quadruples, and the radiation hydrodynamical calculation produces a ratio of 4:3, consistent with the observations.

In summary, there is no detectable change in the mass ratio distributions of binary and higher-order multiple systems when going from barotropic to radiation hydrodynamical calculations.  In both cases, the calculations are consistent with observed trends such as a preference for equal-mass binaries when moving to lower primary masses and a preference for twins to have close separations.

\subsection{Orbital eccentricities}

Observationally, there is an upper envelope to binary eccentricities at periods less than about one year, and binaries with periods less than 12 days are almost exclusively found to have circular orbits due to tidal circularisation \citep{DuqMay1991, Halbwachsetal2003, Raghavanetal2010}.  However, the radiation hydrodynamical calculation does not allow us to probe such small separations due to the absence of dissipation on scales $<0.5$~AU.  Observations also indicate that eccentricities $e<0.1$ are rare for periods greater than $\approx 100$ days (separations $\gsim 1$ AU).  \cite{Raghavanetal2010} finds no binaries with $e<0.1$ and orbital periods greater than 100 days, though they do find that the outer orbits of two triples and one quadruple have $e<0.1$.  \cite{DuqMay1991} and \cite{Raghavanetal2010} also find that the upper-eccentricity envelope is dominated by components of triple systems, possibly due to the action of the Kozai mechanism \citep{Kozai1962}.  Finally, \citet{Halbwachsetal2003} find that the eccentricities of binaries with mass ratios $M_2/M_1>0.8$ with periods greater than $\approx 10$ days (the tidal circularisation radius) are lower than for more unequal mass ratio systems.

In the left panel of Fig.~\ref{eccentricity} we plot the eccentricities versus orbital period for the binaries, triples and quadruples from the radiation hydrodynamical calculation.  The symbols have the same meaning as in Fig.~\ref{a_q}. \cite{Bate2009a} found that when using sink particle radii of 5~AU in barotropic calculations, that there was an excess of high eccentricity ($e>0.7$) binaries with separations $<10$~AU.  This excess disappeared when the simulation was rerun with small accretion radii of 0.5~AU.  Following the gas to smaller scales allowed dissipative interactions between closer multiple systems.  Indeed, this was part of the reason that accretion radii of only 0.5~AU were used for this paper.  In the radiation hydrodynamical calculation, the 8 shortest period binaries all have eccentricities between 0.6 and 0.8.  These are also the 8 closest systems, with semi-major axes ranging from 0.4-2.0~AU.  Therefore, despite the small accretion radii, it is likely that their high eccentricities are due, at least in part, to the lack of dissipative interactions with the gas.  We do note, however, that 3 of the 8 systems are also binary components of higher-order systems and, therefore, their high eccentricities may also be related to the observed upper eccentricity envelope of binary components of higher-order systems \citep{Raghavanetal2010}.

The mean eccentricity of all 59 orbits is $e=0.35 \pm 0.04$ with a standard deviation of 0.27.  The median is $e=0.27$. The mean eccentricity of pairs (including components of triples and quadruples) is $e=0.38 \pm 0.04$ with a standard deviation of 0.27.  The mean eccentricity of the triples and quadruples is $e=0.26 \pm 0.06$ with a standard deviation of 0.22.   The mean eccentricity obtained by \cite{Bate2009a} for the rerun barotropic calculation (with accretion radii of 0.5~AU) was $e=0.45$.  The median eccentricity from \cite{Raghavanetal2010} is about $e=0.4$, so there is reasonable agreement.  

\begin{figure*}
\centering
    \includegraphics[width=8.4cm]{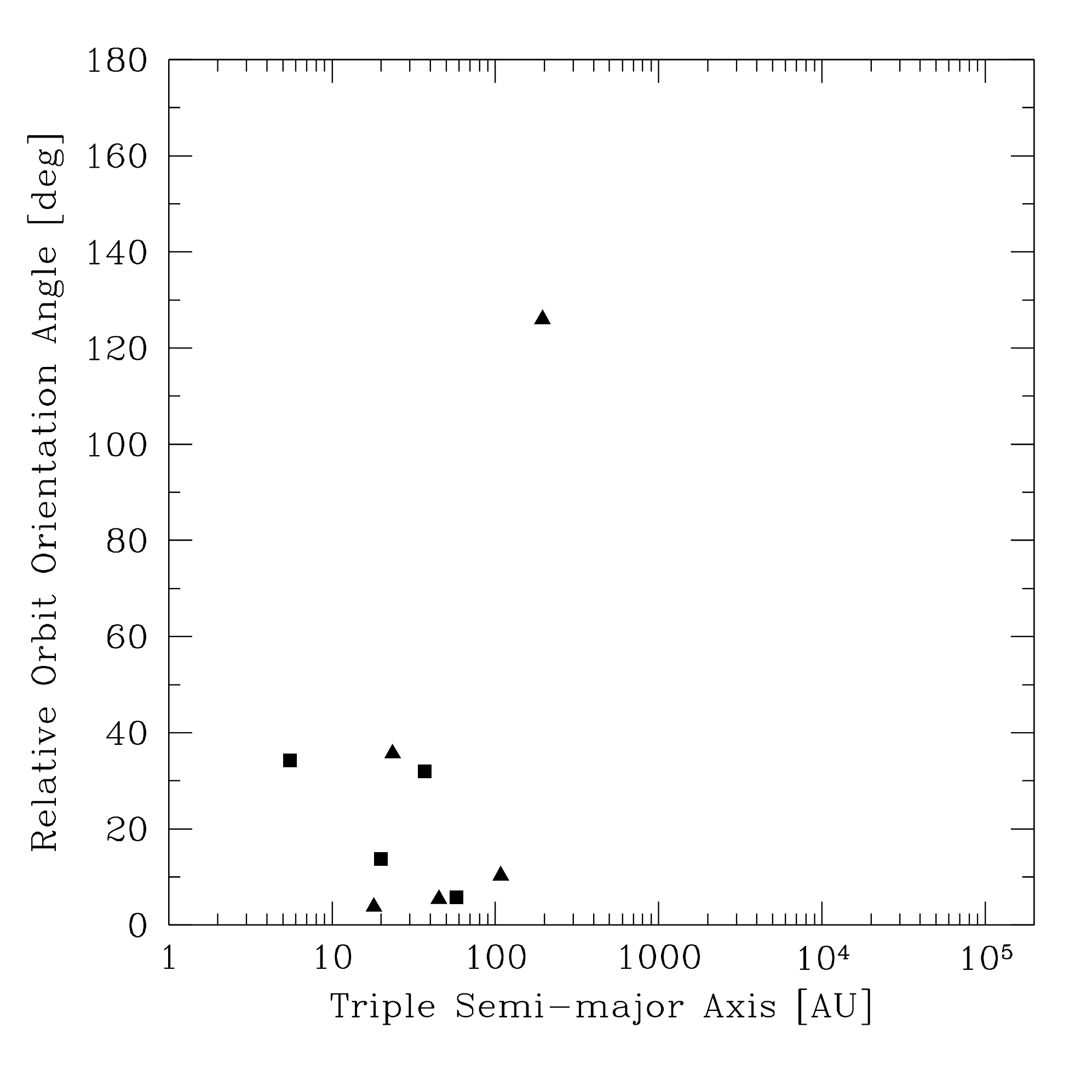}
    \includegraphics[width=8.4cm]{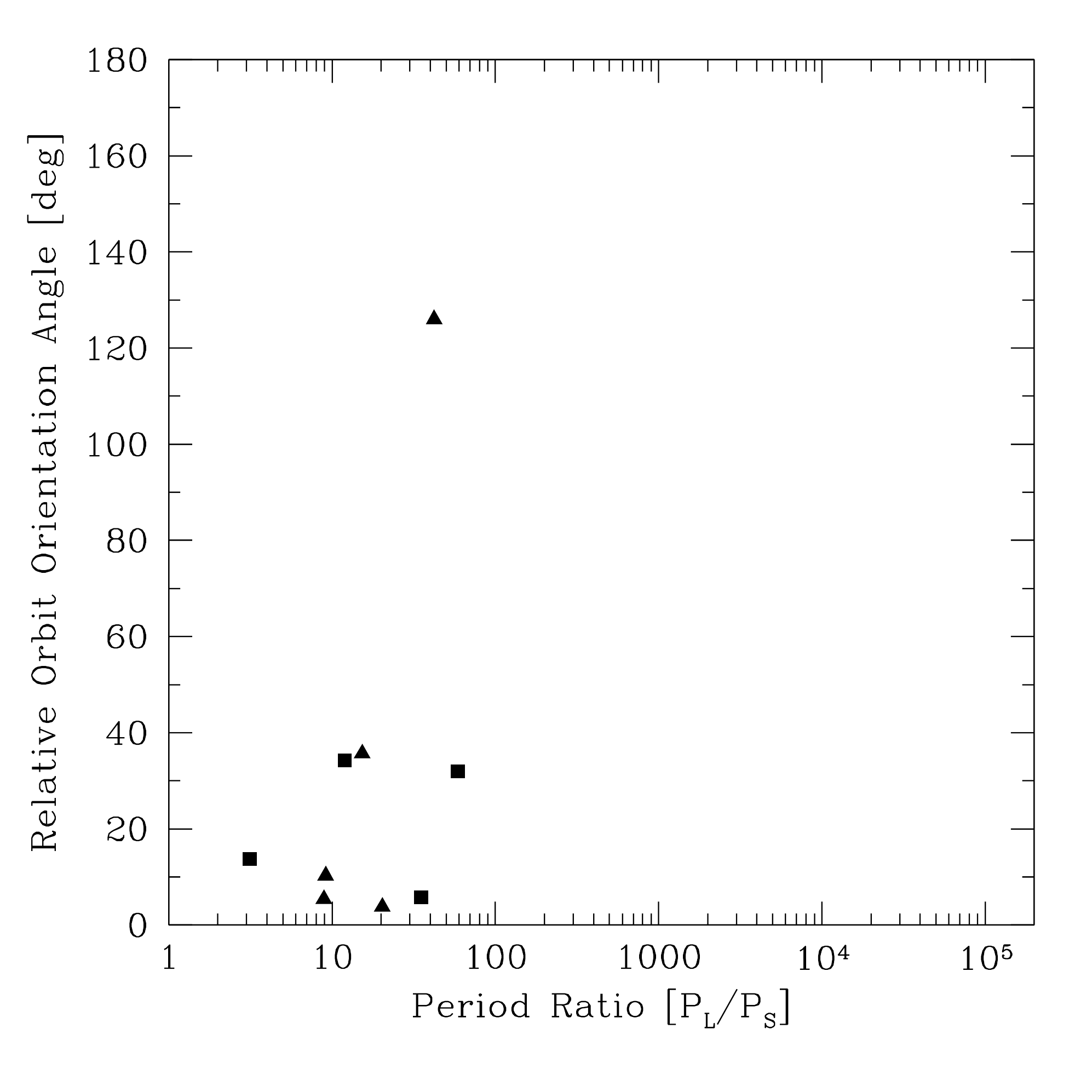}
\caption{The relative inclinations of the two orbital planes for the 9 triple systems produced by the radiation hydrodynamical calculation.  Triples that are sub-components of quadruples are plotted as squares. We give plots of the relative orbital orientation angle versus the semi-major axis of the third component (left) and versus the period ratio of the long and short period orbits (right).  Only the widest triple has a large relative orbital angle.  Note that the one system with period ratio $P_{\rm L}/P_{\rm S}\approx 3$ is likely to be dynamically unstable and to undergo further evolution.}
\label{triples:a_periodratio}
\end{figure*}

However, \cite{Raghavanetal2010} report a flat distribution of eccentricities for periods longer than 12 days out to $e=0.6$, whereas the radiation hydrodynamical calculation produces more than twice as many orbits with $e<0.2$ compared to the the intervals $0.2 \leq e < 0.4$ and $0.4 \leq e < 0.6$.  In particular, there are six binaries with $e<0.08$ (along with the outer orbits of one triple and one quadruple), where as observed systems with $e<0.1$ are rare. Examining Fig.~\ref{eccentricity} it can be seen that two of the 6 binaries are the components of triple systems, which may be related to the finding of \cite{Raghavanetal2010} that components of higher-order multiple systems can have low eccentricities.  Furthermore, all but one of these six binaries has a mass ratio $M_2/M_1>0.8$ (right panel of Fig.~\ref{eccentricity}) which is in qualitative agreement with the finding of \citet{Halbwachsetal2003} that near-equal mass binaries have smaller eccentricities than more unequal mass ratio systems.  \cite{Bate2009a} also found evidence that near-equal mass binaries had smaller eccentricities in the barotropic calculations.  In the radiation hydrodynamical calculation, the median eccentricity of binaries with mass ratios $M_2/M_1<0.8$ is $e=0.45$ (29 orbits) while for $M_2/M_1>0.8$ the median is $e=0.13$ (14 orbits).  Excluding the 8 shortest period systems (since they likely have high eccentricities due to the absence of dissipation on small scales)
 the median eccentricity of binaries with mass ratios $M_2/M_1<0.8$ is $e=0.42$ (23 orbits) while for $M_2/M_1>0.8$ the median is $e=0.10$ (12 orbits). Thus, we also find evidence for a link between mass ratio and eccentricity such that near-equal mass systems have lower eccentricities, as is observed.  As possible explanation for this is that accretion, which drives binaries towards equal masses \citep{Artymowicz1983,Bate1997,BatBon1997,Bate2000}, may also provide dissipation which damps eccentricity.

Finally, we note that VLM binaries are observed to have a preference for low eccentricities with a median value of 0.34 \citep{DupLiu2011}.  The barotropic calculation of \cite{Bate2009a} with small accretion radii also produced low-eccentricity VLM binaries \citep{Bate2010b}, with those VLM binaries with separations less than 30 AU having a mean eccentricity of 0.23.  Unfortunately, the radiation hydrodynamical calculation only produces three VLM binaries.  Two of the three do have small eccentricities (the 10.6-AU binary has an eccentricity of 0.46, the 26-AU binary has an eccentricity of 0.013, and the 36-AU binary has an eccentricity of 0.06), but they are also still accreting so no firm conclusions can be drawn.

\subsection{Relative alignment of orbital planes for triples}

For a hierarchical triple system there are two orbital planes, one corresponding to the short-period orbit and one to the long-period orbit.  Observationally, it is difficult to determine the relative orientation angle, $\Phi$, of the two orbits of a triple system due to the number of quantities that must be measured to fully characterise the orbits. However, the mean value of $\Phi$ can be measured simply from knowledge of the number of co-rotating and counter-rotating systems \citep{Worley1967, Tokovinin1993, SteTok2002}.

The first studies \citep{Worley1967,vanAlbada1968b} of the relative orbital orientations of triple systems found a small tendency towards alignment of the angular momentum vectors of the orbits.  Of 54 systems with known directions of the relative motions, 39 showed co-revolution and 15 counter-revolution resulting in a mean relative inclination angle of $\langle\Phi\rangle\approx 50^\circ$.  For 10 visual systems with known orbits, 5 systems were found to have $\Phi<90^\circ$, 2 had $\Phi>90^\circ$ and 3 were ambiguous.   \citet{Fekel1981} examined 20 systems with known orbits and periods of less than 100 years (for the wide orbit).  He found that 1/3 had non-coplanar orbits.  Finally, \citet{SteTok2002} performed the most detailed study to date.  From 135 visual triple systems for which the relative directions of the orbital motions are known they found $\langle\phi\rangle=67^\circ\pm 9^\circ$ and this result was also consistent with 22 systems for which the orbits were known.  They also found a tendency for the mean relative orbital angular momentum angle to increase with increasing orbital period ratio (i.e.\ systems with more similar orbital periods tend to be more closely aligned).

The main barotropic calculation of \cite{Bate2009a} produced 40 triple systems (17 of which were sub-components of quadruple systems), with a mean relative orbital orientation angle of $\langle\Phi\rangle=65^\circ\pm 6^\circ$, in good agreement with the observed value.  

The radiation hydrodynamical calculation only produced nine triple systems, four of which are components of quadruple systems.  The mean relative orbital orientation angle of the all these triple systems is $\langle\Phi\rangle=30^\circ\pm 13^\circ$, which is about 1.7$\sigma$ lower than the observed value.  For the five pure triples, $\langle\Phi\rangle=36^\circ\pm 26^\circ$.  The relative angles are plotted in Fig.~\ref{triples:a_periodratio} as functions of semi-major axis and period ratio.  It can be seen that all have small relative angles, except the widest system.  We conclude that both the observed and simulated triple systems have a tendency towards orbital coplanarity, but the small number of systems produced by the radiation hydrodynamical calculation prohibits us from making a stronger statement.

\begin{figure*}
\centering
    \includegraphics[width=8.4cm]{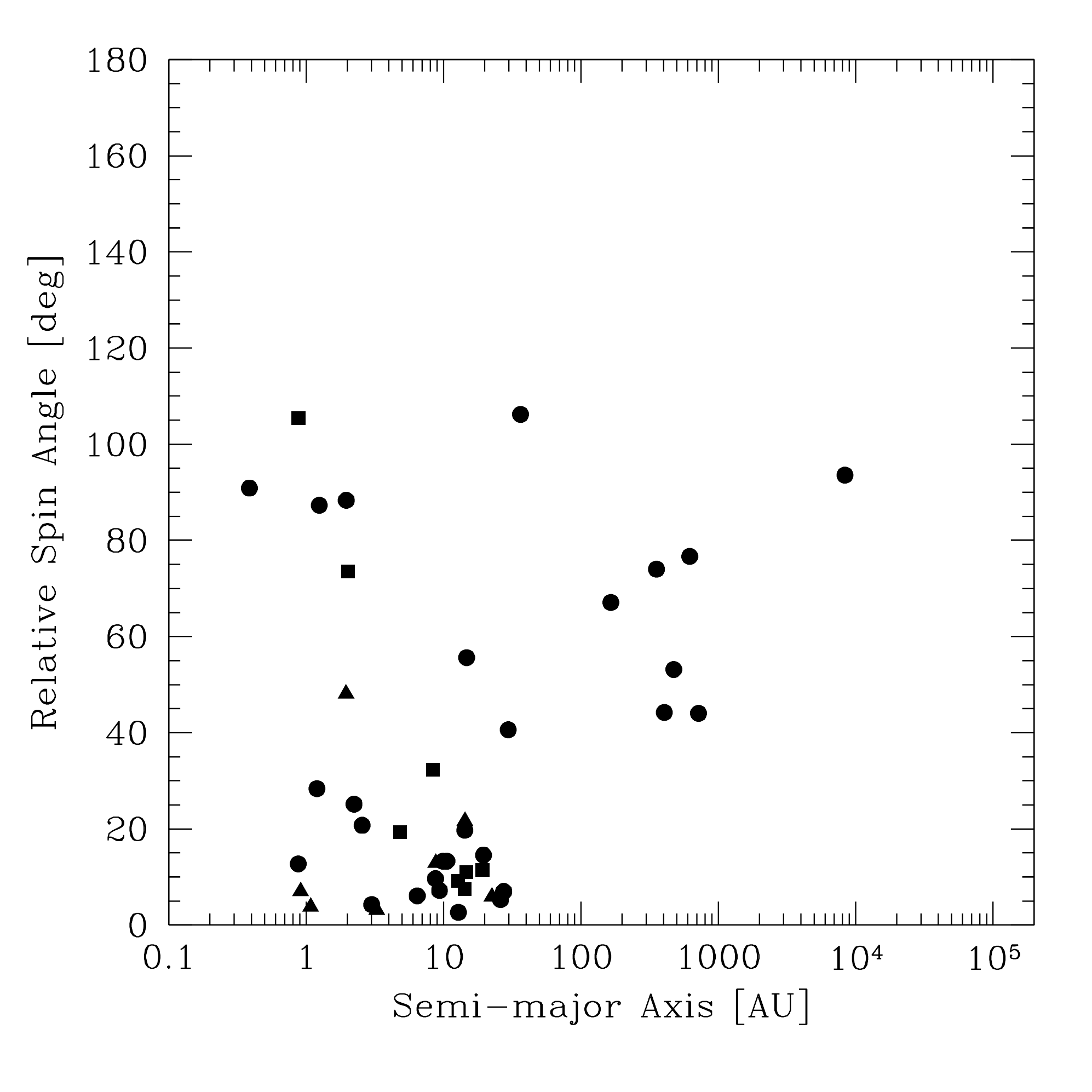}
    \includegraphics[width=8.4cm]{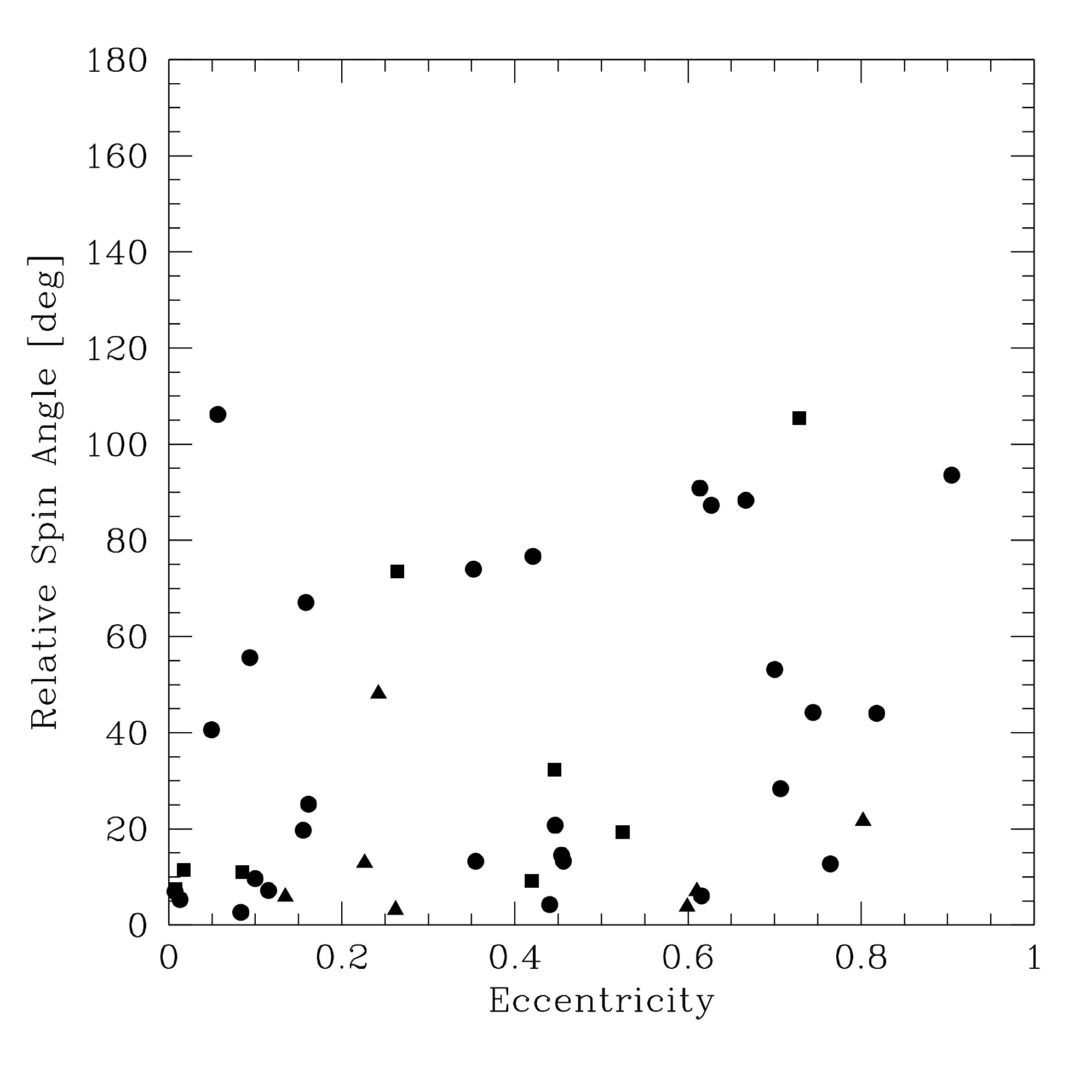}
\caption{The relative inclinations of the rotation axes of the sink particles (modelling stars and their inner discs) of the binary systems produced by the radiation hydrodynamical calculation as functions of the binary's separation (left) and eccentricity (right).  We include binaries that are sub-components of triples (triangles) and quadruples (squares).  All binaries for which the spins are closely aligned have semi-major axes $\lsim 30$~AU. 
}
\label{spinspin}
\end{figure*}

\subsection{Relative alignment of discs and orbits}

Finally, we consider the relative alignment of the spins of the sink particles in binary systems.  Unfortunately there is not a direct analogy with real binary systems in this case because the sink particles are larger than stars and yet smaller than a typical disc.  The orientation of the sink particle spin thus represents the orientation of the total angular momentum of the star and the inner part of its surrounding disc.  This distinction is important because during the formation of an object the angular momentum usually varies with time as gas falls on to it from the turbulent cloud.  Thus, the orientation of the sink particle frequently differs substantially from the orientation of its resolved disc (if one exists) and, furthermore, the orientations of both the sink particles and their discs change with time while the object continues to accrete gas. The orientations may evolve with time due to gravitational torques \citep{Bateetal2000}.

Observationally, \citet{Weis1974} found a tendency for alignment between the stellar equatorial planes and orbital planes among primaries in F star binaries, but not A star binaries.  The orbital separations were mainly in the $10-100$ AU range.  Similarly, \citet{Guthrie1985} found no correlation for 23 A star binaries with separations 10-70 AU.  More recently, \citet{Hale1994} considered 73 binary and multiple systems containing solar-type stars and found evidence for approximate coplanarity between the orbital plane and the stellar equatorial planes for binary systems with separations less than $\approx 30$ AU and apparently uncorrelated stellar rotation and orbital axes for wider systems.  For higher-order multiple systems, however, non-coplanar systems were found to exist for both wide and close orbits.  Hale found no evidence to support a difference dependent on spectral type, eccentricity or age.  In terms of circumstellar discs, there is evidence for misaligned discs from observations of misaligned jets from protostellar objects \citep*{DavMunEis1994}, inferred jet precession \citep{Eisloffeletal1996,Davisetal1997}, and direct observations \citep{Koresko1998, Stapelfeldtetal1998}.  However, these are not statistically useful samples.  \citet*{MonMenDuc1998}, \citet*{DonJenMat1999}, \citet{Jensenetal2004}, \citet*{WolSteHen2001}, and \citet*{MonMenPer2006} used polarimetry to study the relative disc alignment in T Tauri wide binary and multiple systems and all found a preference for disc alignment in binaries. However, \citet{Jensenetal2004} also found that the wide components of triples and quadruples appear to have random orientations.  For more massive Herbig Ae/Be binaries, \cite{ Bainesetal2006} found that the circumprimary disc was preferentially aligned with the orbit and the larger study of \cite{Wheelwrightetal2011} also finds that the discs are preferentially aligned with the orbit.  

The barotropic calculations of \cite{Bate2009a} produced ambiguous results.  The main barotropic calculation with large accretion radii produced a strong tendency for alignment between sink particle spins, but the rerun calculation with smaller accretion radii did not show any tendency for alignment \citep{Bate2011}.

At the end of the radiation hydrodynamical calculation we plot the relative spin angles for the 43 binaries (including those that are components of triple and quadruple systems) in Fig.~\ref{spinspin} as functions of semi-major axis and orbital eccentricity.  There are no relative spin angles greater than $110^\circ$, and only 4 of the 43 systems have angles greater than $90^\circ$, indicating a strong tendency for alignment.  The mean relative spin angle is $35^\circ\pm 5^\circ$, while the median angle is $20^\circ$. For the 28 pure binaries, the mean is $40^\circ\pm 6^\circ$ and the median is $27^\circ$, while for the binaries that are components of higher-order systems the mean is $25^\circ\pm 8^\circ$ and the median is $12^\circ$, so there is an indication that the spins of binaries that are components of higher-order multiples may be more aligned.

Examining the left panel of Fig.~\ref{spinspin} it is clear that the tendency for alignment depends on separation:  all the binaries that have relative spin angles less than 30$^\circ$ have separations less than $\approx 30$~AU or orbital periods less than $\approx 400$ years.  Taking binaries with semi-major axes less than 30~AU, the mean relative spin angle is $26^\circ\pm 5^\circ$, while those with longer periods have a mean of $70^\circ\pm 8^\circ$.  The right panel of Fig.~\ref{spinspin} indicates that there may be a weak relation between the relative spin angle and the eccentricity, with more circular systems having a stronger tendency for alignment.  Such a relation may come about through accretion onto a binary system or gravitational torques between the stars and discs \cite[e.g.][]{Bateetal2000}, either of which would tend to align the components of the binary and may damp eccentricity.  However, we also note that the distribution of relative spin angles seems to be independent of the total mass of the binary.

If the spins of the components of close binaries tend to be aligned with one another, one might also expect the spins to be aligned with the orbital plane of the binary.  Indeed, this is the case, though the alignment is not as strong as for the individual spins.  Taking the 26 binaries with relative spin angles less than 30$^\circ$, the mean spin-orbit angle is $31^\circ\pm 3^\circ$ with a standard deviation of $19^\circ$.  All of these systems have separations less than 30~AU.  For the remaining 17 systems for which the spins are only weakly aligned, the mean spin-orbit angle is $48^\circ\pm 11^\circ$ and the standard deviation is much larger ($64^\circ$). 

In summary, for binaries with separations $\lsim 30$~AU, the radiation hydrodynamical calculation gives strong tendencies for alignment between the spins of the components of binaries and for coplanarity of the orbital plane and the equatorial planes of the components for binaries.  These results are in good agreement with the observed coplanarity of observed binaries \citep{Hale1994} and in qualitative agreement with the many observational studies examining disc alignment mentioned above.

\section{Conclusions}
\label{conclusions}

We have presented results from the largest radiation hydrodynamical simulation of star cluster formation to date that resolves the opacity limit for fragmentation.  It also resolves protoplanetary discs (radii $\geq 1$ AU), binaries, and multiple systems.  The calculation uses sink particles to model the stars and brown dwarfs.  We discuss in some detail (Section \ref{limitations}) the problems associated with trying to include the luminosity coming from inside a sink particle's accretion radius, concluding that attempts made in the literature to date most likely overestimate the luminosity.  Although we omit the luminosity originating from within each sink particle, we use small accretion radii of only 0.5~AU and argue that because protostars are {\it observed} to be under-luminous (the so-called `luminosity problem') the level of radiative feedback included in the simulation presented here may be more realistic than if the extra luminosity was included.

The calculation produced 183 stars and brown dwarfs.  This number of objects is not as large as that produced from the same initial conditions using a barotropic equation of state \citep{Bate2009a} because of the impact of radiative feedback.  However, it is still sufficient to allow comparison of the statistical properties of the stars, brown dwarfs and multiple systems with the results of observational surveys.  \cite{Bate2009a} obtained good agreement between observations and barotropic simulations for the properties of multiple stellar systems, but obtained a brown-dwarf dominated IMF. Overall, the radiation hydrodynamical calculation displays good agreement with a wide range of observed stellar properties with no obvious deficiencies.  Together, the barotropic and radiation hydrodynamical calculations imply that the main physical processes involved in determining the properties multiple stellar systems are gravity and gas dynamics (i.e.~dissipative $N$-body dynamics), while obtaining a realistic IMF also requires radiative feedback.  We note, however, that the star formation rate in the calculations is much higher than observed.  To solve this problem may require globally unbound molecular clouds and/or the inclusion of magnetic fields and kinetic feedback.
Our detailed conclusions are as follows.

\begin{enumerate}
\item The calculation produces an IMF that is statistically indistinguishable from the parameterisation of the observed IMF by \cite{Chabrier2005}, and  a ratio of brown dwarfs to stars which is also in good agreement with observations.  The use of a realistic equation of state and radiation hydrodynamics rather than a barotropic equation of state decreases the number of brown dwarfs formed by an order of magnitude, while having less of an impact on the number of stars formed.  This corrects the over-production of brown dwarfs that is obtained when using a barotropic equation of state.  We find that the median mass and form of the IMF do not evolve significantly during the simulation.
\item As in previous, smaller calculations, the IMF originates from competition between accretion and dynamical interactions which terminate the accretion and sets an object's final mass.  Stars and brown dwarfs form the same way, with similar accretion rates from the molecular cloud, but stars accrete for longer than brown dwarfs before undergoing the dynamical interactions that terminate their accretion.  The higher characteristic stellar mass that is obtained when radiative feedback is included comes about because the typical distance between objects is larger so that the timescale between dynamical interactions is longer and, thus, the objects typically accretion to higher masses before their accretion is terminated.
\item We find that stars have a slightly higher velocity dispersion than VLM objects, and binaries have a lower velocity dispersion than single objects.  
\item We examine the potential effect of dynamical interactions on protoplanetary disc sizes.  We find that more massive stars have had closer encounters.  It is difficult to directly associate the closest encounter with the radii of protostellar discs because many stars accrete new discs after suffering a close encounter.  This is particularly true for the more massive stars.  However, for VLM objects, dynamical encounters usually occur soon after their formation and terminate their accretion so their truncation radii may more closely reflect their disc radii.  Under this assumption we find that at least 20\% of VLM objects should have disc radii $>40$ AU. In lower density star-forming environments this fraction may be expected to be larger.
\item We find that multiplicity strongly increases with primary mass.  The results are in good agreement with the observed multiplicities of G, K, and M dwarfs and VLM objects.  For objects with primary masses in the range $0.03-0.20$ M$_\odot$ the multiplicity fraction is $0.20\pm 0.05$.  We predict that the multiplicity continues to drop for lower-mass brown dwarfs. We find very low frequencies of VLM companions to stars, in agreement with observations.
\item We examine the separation distributions of binaries, triples and quadruples.  We find a broad separation distribution for stars with a median separation of $\approx 15$ AU and a standard deviation of 1 dex.  Unfortunately the calculation only produces three VLM binaries, two of which are still evolving.  However, all of them have separations less than 70~AU and the VLM binary that has reached its final state has a close separation of 11~AU in qualitative agreement with observations.
\item The mass ratio distributions of solar-type and M-dwarf binaries are roughly flat, consistent with observations.  However, the VLM binaries have near-equal masses as appears to be the case for observed systems.  We find that closer binaries tend to have a higher proportion of equal mass components in broad agreement with observed trends.
\item The eccentricity distribution is broad with no obvious dependence on period.  There may be an excess of short-period highly eccentric binaries because of the absence of dissipation on small scales due to the use of sink particles.  There may also be a weak link between mass ratio and eccentricity such that `twins' have lower eccentricities, as is observed.
\item We investigate the relative orientation of the orbital planes of triple systems.  We find a tendency for orbital alignment, in qualitative agreement with observations.
\item Finally, we study the relative orientations of sink particle spins (angular momentum vectors) in binaries (analogous to the rotation axes of stars and their inner discs).  We find that binaries with separations $\lsim 40$~AU have a strong tendency for spin alignment, in good agreement with observations.  We also find that binaries in which the spins are closely aligned also have a tendency for alignment of the stellar spins with the orbit.
\end{enumerate}

\section*{Acknowledgments}

MRB thanks the referee, C.F. McKee, and A.\ Tokovinin for thoughtfully reading the manuscript and making suggestions which helped improve the final paper.
MRB is grateful for the hospitality and financial support he received while on study leave at the Monash Centre for Astrophysics where this paper was written up.
The calculations for this paper were performed on the University of Exeter Supercomputer, 
a DiRAC Facility jointly funded by STFC, the Large Facilities Capital Fund of BIS, and the University of Exeter.
This publication has made use of the Very-Low-Mass Binaries Archive housed at http://www.vlmbinaries.org and maintained by Nick Siegler, Chris Gelino, and Adam Burgasser.
This work, conducted as part of the award ``The formation of stars and planets: Radiation hydrodynamical and magnetohydrodynamical simulations" made under the European Heads of Research Councils and European Science Foundation EURYI (European Young Investigator) Awards scheme, was supported by funds from the Participating Organisations of EURYI and the EC Sixth Framework Programme. 

\bibliography{mbate}

\begin{thebibliography}{}

\bibitem[\protect\citeauthoryear{{Ahmic}, {Jayawardhana}, {Brandeker},
  {Scholz}, {van Kerkwijk}, {Delgado-Donate} \& {Froebrich}}{{Ahmic}
  et~al.}{2007}]{Ahmicetal2007}
{Ahmic} M.,  {Jayawardhana} R.,  {Brandeker} A.,  {Scholz} A.,  {van Kerkwijk}
  M.~H.,  {Delgado-Donate} E.,    {Froebrich} D.,  2007, \apj, 671, 2074

\bibitem[\protect\citeauthoryear{{Alexander}}{{Alexander}}{1975}]{Alexander197%
5}
{Alexander} D.~R.,  1975, \apjs, 29, 363

\bibitem[\protect\citeauthoryear{{Allen}}{{Allen}}{2007}]{Allen2007}
{Allen} P.~R.,  2007, \apj, 668, 492

\bibitem[\protect\citeauthoryear{{Allen}, {Koerner}, {McElwain}, {Cruz} \&
  {Reid}}{{Allen} et~al.}{2007}]{Allenetal2007}
{Allen} P.~R.,  {Koerner} D.~W.,  {McElwain} M.~W.,  {Cruz} K.~L.,    {Reid}
  I.~N.,  2007, \aj, 133, 971

\bibitem[\protect\citeauthoryear{{Andersen}, {Meyer}, {Greissl} \&
  {Aversa}}{{Andersen} et~al.}{2008}]{Andersenetal2008}
{Andersen} M.,  {Meyer} M.~R.,  {Greissl} J.,    {Aversa} A.,  2008, \apjl,
  683, L183

\bibitem[\protect\citeauthoryear{{Apai}, {Pascucci}, {Henning}, {Sterzik},
  {Klein}, {Semenov}, {G{\"u}nther} \& {Stecklum}}{{Apai}
  et~al.}{2002}]{Apaietal2002}
{Apai} D.,  {Pascucci} I.,  {Henning} T.,  {Sterzik} M.~F.,  {Klein} R.,
  {Semenov} D.,  {G{\"u}nther} E.,    {Stecklum} B.,  2002, \apjl, 573, L115

\bibitem[\protect\citeauthoryear{{Artigau}, {Lafreni{\`e}re}, {Doyon},
  {Albert}, {Nadeau} \& {Robert}}{{Artigau} et~al.}{2007}]{Artigauetal2007}
{Artigau} {\'E}.,  {Lafreni{\`e}re} D.,  {Doyon} R.,  {Albert} L.,  {Nadeau}
  D.,    {Robert} J.,  2007, \apjl, 659, L49

\bibitem[\protect\citeauthoryear{{Artymowicz}}{{Artymowicz}}{1983}]{Artymowicz%
1983}
{Artymowicz} P.,  1983, \actaa, 33, 223

\bibitem[\protect\citeauthoryear{{Artymowicz}, {Clarke}, {Lubow} \&
  {Pringle}}{{Artymowicz} et~al.}{1991}]{Artymowiczetal1991}
{Artymowicz} P.,  {Clarke} C.~J.,  {Lubow} S.~H.,    {Pringle} J.~E.,  1991,
  \apjl, 370, L35

\bibitem[\protect\citeauthoryear{{Baines}, {Oudmaijer}, {Porter} \&
  {Pozzo}}{{Baines} et~al.}{2006}]{Bainesetal2006}
{Baines} D.,  {Oudmaijer} R.~D.,  {Porter} J.~M.,    {Pozzo} M.,  2006, \mnras,
  367, 737

\bibitem[\protect\citeauthoryear{{Baraffe}, {Chabrier} \& {Gallardo}}{{Baraffe}
  et~al.}{2009}]{BarChaGal2009}
{Baraffe} I.,  {Chabrier} G.,    {Gallardo} J.,  2009, \apjl, 702, L27

\bibitem[\protect\citeauthoryear{{Basri} \& {Reiners}}{{Basri} \&
  {Reiners}}{2006}]{BasRei2006}
{Basri} G.,  {Reiners} A.,  2006, \aj, 132, 663

\bibitem[\protect\citeauthoryear{{Bastian}, {Covey} \& {Meyer}}{{Bastian}
  et~al.}{2010}]{BasCovMey2010}
{Bastian} N.,  {Covey} K.~R.,    {Meyer} M.~R.,  2010, \araa, 48, 339

\bibitem[\protect\citeauthoryear{{Bate}}{{Bate}}{1997}]{Bate1997}
{Bate} M.~R.,  1997, \mnras, 285, 16

\bibitem[\protect\citeauthoryear{{Bate}}{{Bate}}{1998}]{Bate1998}
{Bate} M.~R.,  1998, \apjl, 508, L95

\bibitem[\protect\citeauthoryear{{Bate}}{{Bate}}{2000}]{Bate2000}
{Bate} M.~R.,  2000, \mnras, 314, 33

\bibitem[\protect\citeauthoryear{{Bate}}{{Bate}}{2005}]{Bate2005}
{Bate} M.~R.,  2005, \mnras, 363, 363

\bibitem[\protect\citeauthoryear{{Bate}}{{Bate}}{2009a}]{Bate2009a}
{Bate} M.~R.,  2009a, \mnras, 392, 590

\bibitem[\protect\citeauthoryear{{Bate}}{{Bate}}{2009b}]{Bate2009c}
{Bate} M.~R.,  2009b, \mnras, 397, 232

\bibitem[\protect\citeauthoryear{{Bate}}{{Bate}}{2009c}]{Bate2009b}
{Bate} M.~R.,  2009c, \mnras, 392, 1363

\bibitem[\protect\citeauthoryear{{Bate}}{{Bate}}{2010a}]{Bate2010}
{Bate} M.~R.,  2010a, \mnras, 404, L79

\bibitem[\protect\citeauthoryear{{Bate}}{{Bate}}{2010b}]{Bate2010b}
{Bate} M.~R.,  2010b, Highlights of Astronomy, 15, 769

\bibitem[\protect\citeauthoryear{{Bate}}{{Bate}}{2011}]{Bate2011}
{Bate} M.~R.,  2011, \mnras, 0, submitted

\bibitem[\protect\citeauthoryear{{Bate} \& {Bonnell}}{{Bate} \&
  {Bonnell}}{1997}]{BatBon1997}
{Bate} M.~R.,  {Bonnell} I.~A.,  1997, \mnras, 285, 33

\bibitem[\protect\citeauthoryear{{Bate} \& {Bonnell}}{{Bate} \&
  {Bonnell}}{2005}]{BatBon2005}
{Bate} M.~R.,  {Bonnell} I.~A.,  2005, MNRAS, 356, 1201

\bibitem[\protect\citeauthoryear{{Bate}, {Bonnell} \& {Bromm}}{{Bate}
  et~al.}{2002a}]{BatBonBro2002a}
{Bate} M.~R.,  {Bonnell} I.~A.,    {Bromm} V.,  2002a, MNRAS, 332, L65

\bibitem[\protect\citeauthoryear{{Bate}, {Bonnell} \& {Bromm}}{{Bate}
  et~al.}{2002b}]{BatBonBro2002b}
{Bate} M.~R.,  {Bonnell} I.~A.,    {Bromm} V.,  2002b, MNRAS, 336, 705

\bibitem[\protect\citeauthoryear{{Bate}, {Bonnell} \& {Bromm}}{{Bate}
  et~al.}{2003}]{BatBonBro2003}
{Bate} M.~R.,  {Bonnell} I.~A.,    {Bromm} V.,  2003, MNRAS, 339, 577

\bibitem[\protect\citeauthoryear{{Bate}, {Bonnell}, {Clarke}, {Lubow},
  {Ogilvie}, {Pringle} \& {Tout}}{{Bate} et~al.}{2000}]{Bateetal2000}
{Bate} M.~R.,  {Bonnell} I.~A.,  {Clarke} C.~J.,  {Lubow} S.~H.,  {Ogilvie}
  G.~I.,  {Pringle} J.~E.,    {Tout} C.~A.,  2000, \mnras, 317, 773

\bibitem[\protect\citeauthoryear{{Bate}, {Bonnell} \& {Price}}{{Bate}
  et~al.}{1995}]{BatBonPri1995}
{Bate} M.~R.,  {Bonnell} I.~A.,    {Price} N.~M.,  1995, MNRAS, 277, 362

\bibitem[\protect\citeauthoryear{{Bate} \& {Burkert}}{{Bate} \&
  {Burkert}}{1997}]{BatBur1997}
{Bate} M.~R.,  {Burkert} A.,  1997, \mnras, 288, 1060

\bibitem[\protect\citeauthoryear{{Bate}, {Clarke} \& {McCaughrean}}{{Bate}
  et~al.}{1998}]{BatClaMcC1998}
{Bate} M.~R.,  {Clarke} C.~J.,    {McCaughrean} M.~J.,  1998, \mnras, 297, 1163

\bibitem[\protect\citeauthoryear{{Benz}}{{Benz}}{1990}]{Benz1990}
{Benz} W.,  1990, in {Buchler} J.~R.,  ed., Numerical Modelling of Nonlinear
  Stellar Pulsations Problems and Prospects. Kluwer, Dordrecht, p.~269

\bibitem[\protect\citeauthoryear{{Benz}, {Cameron}, {Press} \& {Bowers}}{{Benz}
  et~al.}{1990}]{Benzetal1990}
{Benz} W.,  {Cameron} A.~G.~W.,  {Press} W.~H.,    {Bowers} R.~L.,  1990, \apj,
  348, 647

\bibitem[\protect\citeauthoryear{{Bihain} \& {et al.}}{{Bihain} \& {et
  al.}}{2009}]{Bihainetal2009}
{Bihain} G.,  {et al.} 2009, \aap, 506, 1169

\bibitem[\protect\citeauthoryear{{Boley}, {Hartquist}, {Durisen} \&
  {Michael}}{{Boley} et~al.}{2007}]{Boleyetal2007}
{Boley} A.~C.,  {Hartquist} T.~W.,  {Durisen} R.~H.,    {Michael} S.,  2007,
  \apjl, 656, L89

\bibitem[\protect\citeauthoryear{{Bonnell} \& {Bate}}{{Bonnell} \&
  {Bate}}{2002}]{BonBat2002}
{Bonnell} I.~A.,  {Bate} M.~R.,  2002, \mnras, 336, 659

\bibitem[\protect\citeauthoryear{{Bonnell}, {Bate}, {Clarke} \&
  {Pringle}}{{Bonnell} et~al.}{2001}]{Bonnelletal2001a}
{Bonnell} I.~A.,  {Bate} M.~R.,  {Clarke} C.~J.,    {Pringle} J.~E.,  2001,
  MNRAS, 323, 785

\bibitem[\protect\citeauthoryear{{Bonnell}, {Bate} \& {Vine}}{{Bonnell}
  et~al.}{2003}]{BonBatVin2003}
{Bonnell} I.~A.,  {Bate} M.~R.,    {Vine} S.~G.,  2003, MNRAS, 343, 413

\bibitem[\protect\citeauthoryear{{Bonnell}, {Larson} \& {Zinnecker}}{{Bonnell}
  et~al.}{2007}]{BonLarZin2007}
{Bonnell} I.~A.,  {Larson} R.~B.,    {Zinnecker} H.,  2007, in {Reipurth} B.,
  {Jewitt} D.,   {Keil} K.,  eds, Protostars and Planets V {The Origin of the
  Initial Mass Function}.
pp 149--164

\bibitem[\protect\citeauthoryear{{Boss}, {Fisher}, {Klein} \& {McKee}}{{Boss}
  et~al.}{2000}]{Bossetal2000}
{Boss} A.~P.,  {Fisher} R.~T.,  {Klein} R.~I.,    {McKee} C.~F.,  2000, \apj,
  528, 325

\bibitem[\protect\citeauthoryear{{Bouy}, {Brandner}, {Mart{\'{\i}}n},
  {Delfosse}, {Allard} \& {Basri}}{{Bouy} et~al.}{2003}]{Bouyetal2003}
{Bouy} H.,  {Brandner} W.,  {Mart{\'{\i}}n} E.~L.,  {Delfosse} X.,  {Allard}
  F.,    {Basri} G.,  2003, \aj, 126, 1526

\bibitem[\protect\citeauthoryear{{Bouy}, {Mart{\'{\i}}n}, {Brandner},
  {Zapatero-Osorio}, {B{\'e}jar}, {Schirmer}, {Hu{\'e}lamo} \& {Ghez}}{{Bouy}
  et~al.}{2006}]{Bouyetal2006}
{Bouy} H.,  {Mart{\'{\i}}n} E.~L.,  {Brandner} W.,  {Zapatero-Osorio} M.~R.,
  {B{\'e}jar} V.~J.~S.,  {Schirmer} M.,  {Hu{\'e}lamo} N.,    {Ghez} A.~M.,
  2006, \aap, 451, 177

\bibitem[\protect\citeauthoryear{{Boyd} \& {Whitworth}}{{Boyd} \&
  {Whitworth}}{2005}]{BoyWhi2005}
{Boyd} D.~F.~A.,  {Whitworth} A.~P.,  2005, \aap, 430, 1059

\bibitem[\protect\citeauthoryear{{Burgasser}, {Dhital} \& {West}}{{Burgasser}
  et~al.}{2009}]{BurDhiWes2009}
{Burgasser} A.~J.,  {Dhital} S.,    {West} A.~A.,  2009, \aj, 138, 1563

\bibitem[\protect\citeauthoryear{{Burgasser}, {Kirkpatrick}, {Cruz}, {Reid},
  {Leggett}, {Liebert}, {Burrows} \& {Brown}}{{Burgasser}
  et~al.}{2006}]{Burgasseretal2006}
{Burgasser} A.~J.,  {Kirkpatrick} J.~D.,  {Cruz} K.~L.,  {Reid} I.~N.,
  {Leggett} S.~K.,  {Liebert} J.,  {Burrows} A.,    {Brown} M.~E.,  2006,
  \apjs, 166, 585

\bibitem[\protect\citeauthoryear{{Burgasser}, {Kirkpatrick}, {Reid}, {Brown},
  {Miskey} \& {Gizis}}{{Burgasser} et~al.}{2003}]{Burgasseretal2003}
{Burgasser} A.~J.,  {Kirkpatrick} J.~D.,  {Reid} I.~N.,  {Brown} M.~E.,
  {Miskey} C.~L.,    {Gizis} J.~E.,  2003, \apj, 586, 512

\bibitem[\protect\citeauthoryear{{Burgasser}, {Reid}, {Siegler}, {Close},
  {Allen}, {Lowrance} \& {Gizis}}{{Burgasser} et~al.}{2007}]{Burgasseretal2007}
{Burgasser} A.~J.,  {Reid} I.~N.,  {Siegler} N.,  {Close} L.,  {Allen} P.,
  {Lowrance} P.,    {Gizis} J.,  2007, in {Reipurth} B.,  {Jewitt} D.,   {Keil}
  K.,  eds, Protostars and Planets V pp 427--441

\bibitem[\protect\citeauthoryear{{Burgess}, {Moraux}, {Bouvier}, {Marmo},
  {Albert} \& {Bouy}}{{Burgess} et~al.}{2009}]{Burgessetal2009}
{Burgess} A.~S.~M.,  {Moraux} E.,  {Bouvier} J.,  {Marmo} C.,  {Albert} L.,
  {Bouy} H.,  2009, \aap, 508, 823

\bibitem[\protect\citeauthoryear{{Chabrier}}{{Chabrier}}{2003}]{Chabrier2003}
{Chabrier} G.,  2003, \pasp, 115, 763

\bibitem[\protect\citeauthoryear{{Chabrier}}{{Chabrier}}{2005}]{Chabrier2005}
{Chabrier} G.,  2005, in {E.~Corbelli, F.~Palla, \& H.~Zinnecker} ed., The
  Initial Mass Function 50 Years Later Vol.~327 of Astrophysics and Space
  Science Library, {The Initial Mass Function: from Salpeter 1955 to 2005}.
pp 41--50

\bibitem[\protect\citeauthoryear{{Clark} \& {Bonnell}}{{Clark} \&
  {Bonnell}}{2004}]{ClaBon2004}
{Clark} P.~C.,  {Bonnell} I.~A.,  2004, \mnras, 347, L36

\bibitem[\protect\citeauthoryear{{Clark}, {Bonnell}, {Zinnecker} \&
  {Bate}}{{Clark} et~al.}{2005}]{Clarketal2005}
{Clark} P.~C.,  {Bonnell} I.~A.,  {Zinnecker} H.,    {Bate} M.~R.,  2005,
  \mnras, 359, 809

\bibitem[\protect\citeauthoryear{{Clarke} \& {Pringle}}{{Clarke} \&
  {Pringle}}{1991a}]{ClaPri1991a}
{Clarke} C.~J.,  {Pringle} J.~E.,  1991a, \mnras, 249, 584

\bibitem[\protect\citeauthoryear{{Clarke} \& {Pringle}}{{Clarke} \&
  {Pringle}}{1991b}]{ClaPri1991b}
{Clarke} C.~J.,  {Pringle} J.~E.,  1991b, \mnras, 249, 588

\bibitem[\protect\citeauthoryear{{Close}, {Siegler}, {Freed} \&
  {Biller}}{{Close} et~al.}{2003}]{Closeetal2003}
{Close} L.~M.,  {Siegler} N.,  {Freed} M.,    {Biller} B.,  2003, \apj, 587,
  407

\bibitem[\protect\citeauthoryear{{Close}, {Zuckerman}, {Song}, {Barman},
  {Marois}, {Rice}, {Siegler}, {Macintosh}, {Becklin}, {Campbell}, {Lyke},
  {Conrad} \& {Le Mignant}}{{Close} et~al.}{2007}]{Closeetal2007}
{Close} L.~M.,  {Zuckerman} B.,  {Song} I.,  {Barman} T.,  {Marois} C.,  {Rice}
  E.~L.,  {Siegler} N.,  {Macintosh} B.,  {Becklin} E.~E.,  {Campbell} R.,
  {Lyke} J.~E.,  {Conrad} A.,    {Le Mignant} D.,  2007, \apj, 660, 1492

\bibitem[\protect\citeauthoryear{{Davis}, {Eisloeffel}, {Ray} \&
  {Jenness}}{{Davis} et~al.}{1997}]{Davisetal1997}
{Davis} C.~J.,  {Eisloeffel} J.,  {Ray} T.~P.,    {Jenness} T.,  1997, \aap,
  324, 1013

\bibitem[\protect\citeauthoryear{{Davis}, {Mundt} \& {Eisloeffel}}{{Davis}
  et~al.}{1994}]{DavMunEis1994}
{Davis} C.~J.,  {Mundt} R.,    {Eisloeffel} J.,  1994, \apjl, 437, L55

\bibitem[\protect\citeauthoryear{{Delgado-Donate}, {Clarke} \&
  {Bate}}{{Delgado-Donate} et~al.}{2003}]{DelClaBat2003}
{Delgado-Donate} E.~J.,  {Clarke} C.~J.,    {Bate} M.~R.,  2003, \mnras, 342,
  926

\bibitem[\protect\citeauthoryear{{Delgado-Donate}, {Clarke} \&
  {Bate}}{{Delgado-Donate} et~al.}{2004}]{DelClaBat2004}
{Delgado-Donate} E.~J.,  {Clarke} C.~J.,    {Bate} M.~R.,  2004, \mnras, 347,
  759

\bibitem[\protect\citeauthoryear{{Delgado-Donate}, {Clarke}, {Bate} \&
  {Hodgkin}}{{Delgado-Donate} et~al.}{2004}]{Delgadoetal2004}
{Delgado-Donate} E.~J.,  {Clarke} C.~J.,  {Bate} M.~R.,    {Hodgkin} S.~T.,
  2004, \mnras, 351, 617

\bibitem[\protect\citeauthoryear{{Donar}, {Jensen} \& {Mathieu}}{{Donar}
  et~al.}{1999}]{DonJenMat1999}
{Donar} A.,  {Jensen} E.~L.~N.,    {Mathieu} R.~D.,  1999, in American
  Astronomical Society Meeting Abstracts Vol.~195 of American Astronomical
  Society Meeting Abstracts, {Protoplanetary Disks in Young Binaries: Testing
  Coplanarity}.
p. 79.04

\bibitem[\protect\citeauthoryear{{Duch{\^e}ne}, {Bontemps}, {Bouvier},
  {Andr{\'e}}, {Djupvik} \& {Ghez}}{{Duch{\^e}ne}
  et~al.}{2007}]{Ducheneetal2007}
{Duch{\^e}ne} G.,  {Bontemps} S.,  {Bouvier} J.,  {Andr{\'e}} P.,  {Djupvik}
  A.~A.,    {Ghez} A.~M.,  2007, \aap, 476, 229

\bibitem[\protect\citeauthoryear{{Duch{\^e}ne}, {Bouvier} \&
  {Simon}}{{Duch{\^e}ne} et~al.}{1999}]{DucBouSim1999}
{Duch{\^e}ne} G.,  {Bouvier} J.,    {Simon} T.,  1999, \aap, 343, 831

\bibitem[\protect\citeauthoryear{{Dupuy} \& {Liu}}{{Dupuy} \&
  {Liu}}{2011}]{DupLiu2011}
{Dupuy} T.~J.,  {Liu} M.~C.,  2011, \apj, 733, 122

\bibitem[\protect\citeauthoryear{{Duquennoy} \& {Mayor}}{{Duquennoy} \&
  {Mayor}}{1991}]{DuqMay1991}
{Duquennoy} A.,  {Mayor} M.,  1991, \aap, 248, 485

\bibitem[\protect\citeauthoryear{{Eisloffel}, {Smith}, {Davis} \&
  {Ray}}{{Eisloffel} et~al.}{1996}]{Eisloffeletal1996}
{Eisloffel} J.,  {Smith} M.~D.,  {Davis} C.~J.,    {Ray} T.~P.,  1996, \aj,
  112, 2086

\bibitem[\protect\citeauthoryear{{Evans} II, {Dunham}, {J{\o}rgensen}, {Enoch},
  {Mer{\'{\i}}n}, {van Dishoeck}, {Alcal{\'a}}, {Myers}, {Stapelfeldt},
  {Huard}, {Allen}, {Harvey}, {van Kempen} \& {et al.}}{{Evans}
  et~al.}{2009}]{Evansetal2009}
{Evans} II N.~J.,  {Dunham} M.~M.,  {J{\o}rgensen} J.~K.,  {Enoch} M.~L.,
  {Mer{\'{\i}}n} B.,  {van Dishoeck} E.~F.,  {Alcal{\'a}} J.~M.,  {Myers}
  P.~C.,  {Stapelfeldt} K.~R.,  {Huard} T.~L.,  {Allen} L.~E.,  {Harvey} P.~M.,
   {van Kempen} T.,    {et al.} 2009, \apjs, 181, 321

\bibitem[\protect\citeauthoryear{{Faherty}, {Burgasser}, {Bochanski}, {Looper},
  {West} \& {van der Bliek}}{{Faherty} et~al.}{2011}]{Fahertyetal2011}
{Faherty} J.~K.,  {Burgasser} A.~J.,  {Bochanski} J.~J.,  {Looper} D.~L.,
  {West} A.~A.,    {van der Bliek} N.~S.,  2011, \aj, 141, 71

\bibitem[\protect\citeauthoryear{{Fekel} Jr.}{{Fekel}}{1981}]{Fekel1981}
{Fekel} Jr. F.~C.,  1981, \apj, 246, 879

\bibitem[\protect\citeauthoryear{{Fischer} \& {Marcy}}{{Fischer} \&
  {Marcy}}{1992}]{FisMar1992}
{Fischer} D.~A.,  {Marcy} G.~W.,  1992, \apj, 396, 178

\bibitem[\protect\citeauthoryear{{Fletcher} \& {Stahler}}{{Fletcher} \&
  {Stahler}}{1994a}]{FleSta1994a}
{Fletcher} A.~B.,  {Stahler} S.~W.,  1994a, \apj, 435, 313

\bibitem[\protect\citeauthoryear{{Fletcher} \& {Stahler}}{{Fletcher} \&
  {Stahler}}{1994b}]{FleSta1994b}
{Fletcher} A.~B.,  {Stahler} S.~W.,  1994b, \apj, 435, 329

\bibitem[\protect\citeauthoryear{{Fumagalli}, {da Silva} \&
  {Krumholz}}{{Fumagalli} et~al.}{2011}]{FumdaSKru2011}
{Fumagalli} M.,  {da Silva} R.~L.,    {Krumholz} M.~R.,  2011, ArXiv e-prints

\bibitem[\protect\citeauthoryear{{Ghez}, {Neugebauer} \& {Matthews}}{{Ghez}
  et~al.}{1993}]{GheNeuMat1993}
{Ghez} A.~M.,  {Neugebauer} G.,    {Matthews} K.,  1993, \aj, 106, 2005

\bibitem[\protect\citeauthoryear{{Girichidis}, {Federrath}, {Banerjee} \&
  {Klessen}}{{Girichidis} et~al.}{2011}]{Girichidisetal2011}
{Girichidis} P.,  {Federrath} C.,  {Banerjee} R.,    {Klessen} R.~S.,  2011,
  \mnras, 413, 2741

\bibitem[\protect\citeauthoryear{{Gizis}, {Reid}, {Knapp}, {Liebert},
  {Kirkpatrick}, {Koerner} \& {Burgasser}}{{Gizis}
  et~al.}{2003}]{Gizisetal2003}
{Gizis} J.~E.,  {Reid} I.~N.,  {Knapp} G.~R.,  {Liebert} J.,  {Kirkpatrick}
  J.~D.,  {Koerner} D.~W.,    {Burgasser} A.~J.,  2003, \aj, 125, 3302

\bibitem[\protect\citeauthoryear{{Goodwin}, {Whitworth} \&
  {Ward-Thompson}}{{Goodwin} et~al.}{2004a}]{GooWhiWar2004c}
{Goodwin} S.~P.,  {Whitworth} A.~P.,    {Ward-Thompson} D.,  2004a, \aap, 419,
  543

\bibitem[\protect\citeauthoryear{{Goodwin}, {Whitworth} \&
  {Ward-Thompson}}{{Goodwin} et~al.}{2004b}]{GooWhiWar2004a}
{Goodwin} S.~P.,  {Whitworth} A.~P.,    {Ward-Thompson} D.,  2004b, \aap, 414,
  633

\bibitem[\protect\citeauthoryear{{Goodwin}, {Whitworth} \&
  {Ward-Thompson}}{{Goodwin} et~al.}{2004c}]{GooWhiWar2004b}
{Goodwin} S.~P.,  {Whitworth} A.~P.,    {Ward-Thompson} D.,  2004c, \aap, 423,
  169

\bibitem[\protect\citeauthoryear{{Goodwin}, {Whitworth} \&
  {Ward-Thompson}}{{Goodwin} et~al.}{2006}]{GooWhiWar2006}
{Goodwin} S.~P.,  {Whitworth} A.~P.,    {Ward-Thompson} D.,  2006, \aap, 452,
  487

\bibitem[\protect\citeauthoryear{{Greissl}, {Meyer}, {Wilking}, {Fanetti},
  {Schneider}, {Greene} \& {Young}}{{Greissl} et~al.}{2007}]{Greissletal2007}
{Greissl} J.,  {Meyer} M.~R.,  {Wilking} B.~A.,  {Fanetti} T.,  {Schneider} G.,
   {Greene} T.~P.,    {Young} E.,  2007, \aj, 133, 1321

\bibitem[\protect\citeauthoryear{{Grether} \& {Lineweaver}}{{Grether} \&
  {Lineweaver}}{2006}]{GreLin2006}
{Grether} D.,  {Lineweaver} C.~H.,  2006, \apj, 640, 1051

\bibitem[\protect\citeauthoryear{{Guthrie}}{{Guthrie}}{1985}]{Guthrie1985}
{Guthrie} B.~N.~G.,  1985, \mnras, 215, 545

\bibitem[\protect\citeauthoryear{{Halbwachs}, {Mayor}, {Udry} \&
  {Arenou}}{{Halbwachs} et~al.}{2003}]{Halbwachsetal2003}
{Halbwachs} J.~L.,  {Mayor} M.,  {Udry} S.,    {Arenou} F.,  2003, \aap, 397,
  159

\bibitem[\protect\citeauthoryear{{Hale}}{{Hale}}{1994}]{Hale1994}
{Hale} A.,  1994, \aj, 107, 306

\bibitem[\protect\citeauthoryear{{Hall}, {Clarke} \& {Pringle}}{{Hall}
  et~al.}{1996}]{HalClaPri1996}
{Hall} S.~M.,  {Clarke} C.~J.,    {Pringle} J.~E.,  1996, \mnras, 278, 303

\bibitem[\protect\citeauthoryear{{Hartmann}, {Zhu} \& {Calvet}}{{Hartmann}
  et~al.}{2011}]{HarZhuCal2011}
{Hartmann} L.,  {Zhu} Z.,    {Calvet} N.,  2011, ArXiv e-prints

\bibitem[\protect\citeauthoryear{{Heller}}{{Heller}}{1995}]{Heller1995}
{Heller} C.~H.,  1995, \apj, 455, 252

\bibitem[\protect\citeauthoryear{{Hosokawa}, {Offner} \& {Krumholz}}{{Hosokawa}
  et~al.}{2011}]{HosOffKru2011}
{Hosokawa} T.,  {Offner} S.~S.~R.,    {Krumholz} M.~R.,  2011, \apj, 738, 140

\bibitem[\protect\citeauthoryear{{Hosokawa} \& {Omukai}}{{Hosokawa} \&
  {Omukai}}{2009}]{HosOmu2009}
{Hosokawa} T.,  {Omukai} K.,  2009, \apj, 691, 823

\bibitem[\protect\citeauthoryear{{Hubber}, {Goodwin} \& {Whitworth}}{{Hubber}
  et~al.}{2006}]{HubGooWhi2006}
{Hubber} D.~A.,  {Goodwin} S.~P.,    {Whitworth} A.~P.,  2006, \aap, 450, 881

\bibitem[\protect\citeauthoryear{{Hubber} \& {Whitworth}}{{Hubber} \&
  {Whitworth}}{2005}]{HubWhi2005}
{Hubber} D.~A.,  {Whitworth} A.~P.,  2005, \aap, 437, 113

\bibitem[\protect\citeauthoryear{{Jayawardhana}, {Ardila}, {Stelzer} \&
  {Haisch} Jr.}{{Jayawardhana} et~al.}{2003}]{Jayawardhanaetal2003}
{Jayawardhana} R.,  {Ardila} D.~R.,  {Stelzer} B.,    {Haisch} Jr. K.~E.,
  2003, \aj, 126, 1515

\bibitem[\protect\citeauthoryear{{Jayawardhana}, {Mohanty} \&
  {Basri}}{{Jayawardhana} et~al.}{2002}]{JayMohBas2002}
{Jayawardhana} R.,  {Mohanty} S.,    {Basri} G.,  2002, \apjl, 578, L141

\bibitem[\protect\citeauthoryear{{Jensen}, {Mathieu}, {Donar} \&
  {Dullighan}}{{Jensen} et~al.}{2004}]{Jensenetal2004}
{Jensen} E.~L.~N.,  {Mathieu} R.~D.,  {Donar} A.~X.,    {Dullighan} A.,  2004,
  \apj, 600, 789

\bibitem[\protect\citeauthoryear{{Joergens}}{{Joergens}}{2006}]{Joergens2006}
{Joergens} V.,  2006, \aap, 448, 655

\bibitem[\protect\citeauthoryear{{Joergens} \& {Guenther}}{{Joergens} \&
  {Guenther}}{2001}]{JoeGue2001}
{Joergens} V.,  {Guenther} E.,  2001, \aap, 379, L9

\bibitem[\protect\citeauthoryear{{Jones} \& {Walker}}{{Jones} \&
  {Walker}}{1988}]{JonWal1988}
{Jones} B.~F.,  {Walker} M.~F.,  1988, \aj, 95, 1755

\bibitem[\protect\citeauthoryear{{Kenyon}, {Hartmann}, {Strom} \&
  {Strom}}{{Kenyon} et~al.}{1990}]{Kenyonetal1990}
{Kenyon} S.~J.,  {Hartmann} L.~W.,  {Strom} K.~M.,    {Strom} S.~E.,  1990,
  \aj, 99, 869

\bibitem[\protect\citeauthoryear{{Kirkpatrick}, {Barman}, {Burgasser},
  {McGovern}, {McLean}, {Tinney} \& {Lowrance}}{{Kirkpatrick}
  et~al.}{2006}]{Kirkpatricketal2006}
{Kirkpatrick} J.~D.,  {Barman} T.~S.,  {Burgasser} A.~J.,  {McGovern} M.~R.,
  {McLean} I.~S.,  {Tinney} C.~G.,    {Lowrance} P.~J.,  2006, \apj, 639, 1120

\bibitem[\protect\citeauthoryear{{Kirkpatrick}, {Dahn}, {Monet}, {Reid},
  {Gizis}, {Liebert} \& {Burgasser}}{{Kirkpatrick}
  et~al.}{2001}]{Kirkpatricketal2001}
{Kirkpatrick} J.~D.,  {Dahn} C.~C.,  {Monet} D.~G.,  {Reid} I.~N.,  {Gizis}
  J.~E.,  {Liebert} J.,    {Burgasser} A.~J.,  2001, \aj, 121, 3235

\bibitem[\protect\citeauthoryear{{Klessen}}{{Klessen}}{2001}]{Klessen2001}
{Klessen} R.~S.,  2001, \apj, 556, 837

\bibitem[\protect\citeauthoryear{{Klessen} \& {Burkert}}{{Klessen} \&
  {Burkert}}{2000}]{KleBur2000}
{Klessen} R.~S.,  {Burkert} A.,  2000, \apjs, 128, 287

\bibitem[\protect\citeauthoryear{{Klessen} \& {Burkert}}{{Klessen} \&
  {Burkert}}{2001}]{KleBur2001}
{Klessen} R.~S.,  {Burkert} A.,  2001, \apj, 549, 386

\bibitem[\protect\citeauthoryear{{Klessen}, {Burkert} \& {Bate}}{{Klessen}
  et~al.}{1998}]{KleBurBat1998}
{Klessen} R.~S.,  {Burkert} A.,    {Bate} M.~R.,  1998, ApJ, 501, L205+

\bibitem[\protect\citeauthoryear{{Kobulnicky} \& {Fryer}}{{Kobulnicky} \&
  {Fryer}}{2007}]{KobFry2007}
{Kobulnicky} H.~A.,  {Fryer} C.~L.,  2007, \apj, 670, 747

\bibitem[\protect\citeauthoryear{{K{\"o}hler}, {Petr-Gotzens}, {McCaughrean},
  {Bouvier}, {Duch{\^e}ne}, {Quirrenbach} \& {Zinnecker}}{{K{\"o}hler}
  et~al.}{2006}]{Kohleretal2006}
{K{\"o}hler} R.,  {Petr-Gotzens} M.~G.,  {McCaughrean} M.~J.,  {Bouvier} J.,
  {Duch{\^e}ne} G.,  {Quirrenbach} A.,    {Zinnecker} H.,  2006, \aap, 458, 461

\bibitem[\protect\citeauthoryear{{Konopacky}, {Ghez}, {Rice} \&
  {Duch{\^e}ne}}{{Konopacky} et~al.}{2007}]{Konopackyetal2007}
{Konopacky} Q.~M.,  {Ghez} A.~M.,  {Rice} E.~L.,    {Duch{\^e}ne} G.,  2007,
  \apj, 663, 394

\bibitem[\protect\citeauthoryear{{Koresko}}{{Koresko}}{1998}]{Koresko1998}
{Koresko} C.~D.,  1998, \apjl, 507, L145

\bibitem[\protect\citeauthoryear{{Kouwenhoven}, {Brown} \&
  {Kaper}}{{Kouwenhoven} et~al.}{2007}]{KouBroKap2007}
{Kouwenhoven} M.~B.~N.,  {Brown} A.~G.~A.,    {Kaper} L.,  2007, \aap, 464, 581

\bibitem[\protect\citeauthoryear{{Kouwenhoven}, {Brown}, {Portegies Zwart} \&
  {Kaper}}{{Kouwenhoven} et~al.}{2007}]{Kouwenhovenetal2007}
{Kouwenhoven} M.~B.~N.,  {Brown} A.~G.~A.,  {Portegies Zwart} S.~F.,    {Kaper}
  L.,  2007, \aap, 474, 77

\bibitem[\protect\citeauthoryear{{Kouwenhoven}, {Goodwin}, {Parker}, {Davies},
  {Malmberg} \& {Kroupa}}{{Kouwenhoven} et~al.}{2010}]{Kouwenhovenetal2010}
{Kouwenhoven} M.~B.~N.,  {Goodwin} S.~P.,  {Parker} R.~J.,  {Davies} M.~B.,
  {Malmberg} D.,    {Kroupa} P.,  2010, \mnras, 404, 1835

\bibitem[\protect\citeauthoryear{{Kozai}}{{Kozai}}{1962}]{Kozai1962}
{Kozai} Y.,  1962, \aj, 67, 591

\bibitem[\protect\citeauthoryear{{Kratter} \& {Matzner}}{{Kratter} \&
  {Matzner}}{2006}]{KraMat2006}
{Kratter} K.~M.,  {Matzner} C.~D.,  2006, \mnras, 373, 1563

\bibitem[\protect\citeauthoryear{{Kratter}, {Matzner} \& {Krumholz}}{{Kratter}
  et~al.}{2008}]{KraMatKru2008}
{Kratter} K.~M.,  {Matzner} C.~D.,    {Krumholz} M.~R.,  2008, \apj, 681, 375

\bibitem[\protect\citeauthoryear{{Kratter}, {Matzner}, {Krumholz} \&
  {Klein}}{{Kratter} et~al.}{2010}]{Kratteretal2010}
{Kratter} K.~M.,  {Matzner} C.~D.,  {Krumholz} M.~R.,    {Klein} R.~I.,  2010,
  \apj, 708, 1585

\bibitem[\protect\citeauthoryear{{Kraus}, {Ireland}, {Martinache} \&
  {Hillenbrand}}{{Kraus} et~al.}{2011}]{Krausetal2011}
{Kraus} A.~L.,  {Ireland} M.~J.,  {Martinache} F.,    {Hillenbrand} L.~A.,
  2011, \apj, 731, 8

\bibitem[\protect\citeauthoryear{{Kraus}, {White} \& {Hillenbrand}}{{Kraus}
  et~al.}{2005}]{KraWhiHil2005}
{Kraus} A.~L.,  {White} R.~J.,    {Hillenbrand} L.~A.,  2005, \apj, 633, 452

\bibitem[\protect\citeauthoryear{{Kraus}, {White} \& {Hillenbrand}}{{Kraus}
  et~al.}{2006}]{KraWhiHil2006}
{Kraus} A.~L.,  {White} R.~J.,    {Hillenbrand} L.~A.,  2006, \apj, 649, 306

\bibitem[\protect\citeauthoryear{{Kroupa}}{{Kroupa}}{2001}]{Kroupa2001}
{Kroupa} P.,  2001, \mnras, 322, 231

\bibitem[\protect\citeauthoryear{{Kroupa} \& {Burkert}}{{Kroupa} \&
  {Burkert}}{2001}]{KroBur2001}
{Kroupa} P.,  {Burkert} A.,  2001, \apj, 555, 945

\bibitem[\protect\citeauthoryear{{Krumholz}, {Klein} \& {McKee}}{{Krumholz}
  et~al.}{2011}]{KruKleMcK2011}
{Krumholz} M.~R.,  {Klein} R.~I.,    {McKee} C.~F.,  2011, ArXiv e-prints

\bibitem[\protect\citeauthoryear{{Krumholz}, {Matzner} \& {McKee}}{{Krumholz}
  et~al.}{2006}]{KruMatMcK2006}
{Krumholz} M.~R.,  {Matzner} C.~D.,    {McKee} C.~F.,  2006, \apj, 653, 361

\bibitem[\protect\citeauthoryear{{Lamb}, {Oey}, {Werk} \& {Ingleby}}{{Lamb}
  et~al.}{2010}]{Lambetal2010}
{Lamb} J.~B.,  {Oey} M.~S.,  {Werk} J.~K.,    {Ingleby} L.~D.,  2010, \apj,
  725, 1886

\bibitem[\protect\citeauthoryear{{Larson}}{{Larson}}{1969}]{Larson1969}
{Larson} R.~B.,  1969, \mnras, 145, 271

\bibitem[\protect\citeauthoryear{{Larson}}{{Larson}}{1981}]{Larson1981}
{Larson} R.~B.,  1981, \mnras, 194, 809

\bibitem[\protect\citeauthoryear{{Larson}}{{Larson}}{1990}]{Larson1990}
{Larson} R.~B.,  1990, in {R.~Capuzzo-Dolcetta, C.~Chiosi, \& A.~di Fazio} ed.,
  Physical Processes in Fragmentation and Star Formation Vol.~162 of
  Astrophysics and Space Science Library, {Formation of star clusters}.
pp 389--399

\bibitem[\protect\citeauthoryear{{Law}, {Hodgkin} \& {Mackay}}{{Law}
  et~al.}{2008}]{LawHodMac2008}
{Law} N.~M.,  {Hodgkin} S.~T.,    {Mackay} C.~D.,  2008, \mnras, 384, 150

\bibitem[\protect\citeauthoryear{{Leinert}, {Zinnecker}, {Weitzel}, {Christou},
  {Ridgway}, {Jameson}, {Haas} \& {Lenzen}}{{Leinert}
  et~al.}{1993}]{Leinertetal1993}
{Leinert} C.,  {Zinnecker} H.,  {Weitzel} N.,  {Christou} J.,  {Ridgway} S.~T.,
   {Jameson} R.,  {Haas} M.,    {Lenzen} R.,  1993, \aap, 278, 129

\bibitem[\protect\citeauthoryear{{Li}, {Norman}, {Mac Low} \& {Heitsch}}{{Li}
  et~al.}{2004}]{Lietal2004}
{Li} P.~S.,  {Norman} M.~L.,  {Mac Low} M.-M.,    {Heitsch} F.,  2004, \apj,
  605, 800

\bibitem[\protect\citeauthoryear{{Lodieu}, {Hambly}, {Jameson} \&
  {Hodgkin}}{{Lodieu} et~al.}{2008}]{Lodieuetal2008}
{Lodieu} N.,  {Hambly} N.~C.,  {Jameson} R.~F.,    {Hodgkin} S.~T.,  2008,
  \mnras, 383, 1385

\bibitem[\protect\citeauthoryear{{Low} \& {Lynden-Bell}}{{Low} \&
  {Lynden-Bell}}{1976}]{LowLyn1976}
{Low} C.,  {Lynden-Bell} D.,  1976, \mnras, 176, 367

\bibitem[\protect\citeauthoryear{{Luhman}}{{Luhman}}{2004}]{Luhman2004a}
{Luhman} K.~L.,  2004, \apj, 614, 398

\bibitem[\protect\citeauthoryear{{Luhman}, {Adame}, {D'Alessio}, {Calvet},
  {McLeod}, {Bohac}, {Forrest}, {Hartmann}, {Sargent} \& {Watson}}{{Luhman}
  et~al.}{2007}]{Luhmanetal2007a}
{Luhman} K.~L.,  {Adame} L.,  {D'Alessio} P.,  {Calvet} N.,  {McLeod} K.~K.,
  {Bohac} C.~J.,  {Forrest} W.~J.,  {Hartmann} L.,  {Sargent} B.,    {Watson}
  D.~M.,  2007, \apj, 666, 1219

\bibitem[\protect\citeauthoryear{{Luhman}, {Allen}, {Allen}, {Gutermuth},
  {Hartmann}, {Mamajek}, {Megeath}, {Myers} \& {Fazio}}{{Luhman}
  et~al.}{2008}]{Luhmanetal2008}
{Luhman} K.~L.,  {Allen} L.~E.,  {Allen} P.~R.,  {Gutermuth} R.~A.,  {Hartmann}
  L.,  {Mamajek} E.~E.,  {Megeath} S.~T.,  {Myers} P.~C.,    {Fazio} G.~G.,
  2008, \apj, 675, 1375

\bibitem[\protect\citeauthoryear{{Luhman}, {Lada}, {Hartmann}, {Muench},
  {Megeath}, {Allen}, {Myers}, {Muzerolle}, {Young} \& {Fazio}}{{Luhman}
  et~al.}{2005}]{Luhmanetal2005a}
{Luhman} K.~L.,  {Lada} C.~J.,  {Hartmann} L.,  {Muench} A.~A.,  {Megeath}
  S.~T.,  {Allen} L.~E.,  {Myers} P.~C.,  {Muzerolle} J.,  {Young} E.,
  {Fazio} G.~G.,  2005, \apjl, 631, L69

\bibitem[\protect\citeauthoryear{{Luhman}, {Mamajek}, {Allen} \&
  {Cruz}}{{Luhman} et~al.}{2009}]{Luhmanetal2009a}
{Luhman} K.~L.,  {Mamajek} E.~E.,  {Allen} P.~R.,    {Cruz} K.~L.,  2009, \apj,
  703, 399

\bibitem[\protect\citeauthoryear{{Luhman}, {Mamajek}, {Allen}, {Muench} \&
  {Finkbeiner}}{{Luhman} et~al.}{2009}]{Luhmanetal2009b}
{Luhman} K.~L.,  {Mamajek} E.~E.,  {Allen} P.~R.,  {Muench} A.~A.,
  {Finkbeiner} D.~P.,  2009, \apj, 691, 1265

\bibitem[\protect\citeauthoryear{{Machida}, {Inutsuka} \&
  {Matsumoto}}{{Machida} et~al.}{2010}]{MacInuMat2010}
{Machida} M.~N.,  {Inutsuka} S.-i.,    {Matsumoto} T.,  2010, \apj, 724, 1006

\bibitem[\protect\citeauthoryear{{Machida} \& {Matsumoto}}{{Machida} \&
  {Matsumoto}}{2011}]{MacMat2011}
{Machida} M.~N.,  {Matsumoto} T.,  2011, \mnras, pp 284--+

\bibitem[\protect\citeauthoryear{{Marcy} \& {Butler}}{{Marcy} \&
  {Butler}}{2000}]{MarBut2000}
{Marcy} G.~W.,  {Butler} R.~P.,  2000, \pasp, 112, 137

\bibitem[\protect\citeauthoryear{{Mart{\'{\i}}n}, {Barrado y Navascu{\'e}s},
  {Baraffe}, {Bouy} \& {Dahm}}{{Mart{\'{\i}}n} et~al.}{2003}]{Martinetal2003}
{Mart{\'{\i}}n} E.~L.,  {Barrado y Navascu{\'e}s} D.,  {Baraffe} I.,  {Bouy}
  H.,    {Dahm} S.,  2003, \apj, 594, 525

\bibitem[\protect\citeauthoryear{{Mart{\'{\i}}n}, {Brandner}, {Bouvier},
  {Luhman}, {Stauffer}, {Basri}, {Zapatero Osorio} \& {Barrado y
  Navascu{\'e}s}}{{Mart{\'{\i}}n} et~al.}{2000}]{Martinetal2000}
{Mart{\'{\i}}n} E.~L.,  {Brandner} W.,  {Bouvier} J.,  {Luhman} K.~L.,
  {Stauffer} J.,  {Basri} G.,  {Zapatero Osorio} M.~R.,    {Barrado y
  Navascu{\'e}s} D.,  2000, \apj, 543, 299

\bibitem[\protect\citeauthoryear{{Maschberger}, {Clarke}, {Bonnell} \&
  {Kroupa}}{{Maschberger} et~al.}{2010}]{Maschbergeretal2010}
{Maschberger} T.,  {Clarke} C.~J.,  {Bonnell} I.~A.,    {Kroupa} P.,  2010,
  \mnras, 404, 1061

\bibitem[\protect\citeauthoryear{{Mason}, {Gies}, {Hartkopf}, {Bagnuolo} Jr.,
  {ten Brummelaar} \& {McAlister}}{{Mason} et~al.}{1998}]{Masonetal1998}
{Mason} B.~D.,  {Gies} D.~R.,  {Hartkopf} W.~I.,  {Bagnuolo} Jr. W.~G.,  {ten
  Brummelaar} T.,    {McAlister} H.~A.,  1998, \aj, 115, 821

\bibitem[\protect\citeauthoryear{{Mason}, {Hartkopf}, {Gies}, {Henry} \&
  {Helsel}}{{Mason} et~al.}{2009}]{Masonetal2009}
{Mason} B.~D.,  {Hartkopf} W.~I.,  {Gies} D.~R.,  {Henry} T.~J.,    {Helsel}
  J.~W.,  2009, \aj, 137, 3358

\bibitem[\protect\citeauthoryear{{Masunaga} \& {Inutsuka}}{{Masunaga} \&
  {Inutsuka}}{2000}]{MasInu2000}
{Masunaga} H.,  {Inutsuka} S.-I.,  2000, \apj, 531, 350

\bibitem[\protect\citeauthoryear{{Matzner} \& {McKee}}{{Matzner} \&
  {McKee}}{2000}]{MatMcK2000}
{Matzner} C.~D.,  {McKee} C.~F.,  2000, \apj, 545, 364

\bibitem[\protect\citeauthoryear{{Maxted} \& {Jeffries}}{{Maxted} \&
  {Jeffries}}{2005}]{MaxJef2005}
{Maxted} P.~F.~L.,  {Jeffries} R.~D.,  2005, \mnras, 362, L45

\bibitem[\protect\citeauthoryear{{Maxted}, {Jeffries}, {Oliveira}, {Naylor} \&
  {Jackson}}{{Maxted} et~al.}{2008}]{Maxtedetal2008}
{Maxted} P.~F.~L.,  {Jeffries} R.~D.,  {Oliveira} J.~M.,  {Naylor} T.,
  {Jackson} R.~J.,  2008, \mnras, 385, 2210

\bibitem[\protect\citeauthoryear{{Mazeh}, {Simon}, {Prato}, {Markus} \&
  {Zucker}}{{Mazeh} et~al.}{2003}]{Mazehetal2003}
{Mazeh} T.,  {Simon} M.,  {Prato} L.,  {Markus} B.,    {Zucker} S.,  2003,
  \apj, 599, 1344

\bibitem[\protect\citeauthoryear{{McCarthy} \& {Zuckerman}}{{McCarthy} \&
  {Zuckerman}}{2004}]{McCZuc2004}
{McCarthy} C.,  {Zuckerman} B.,  2004, \aj, 127, 2871

\bibitem[\protect\citeauthoryear{{McDonald} \& {Clarke}}{{McDonald} \&
  {Clarke}}{1995}]{McDCla1995}
{McDonald} J.~M.,  {Clarke} C.~J.,  1995, \mnras, 275, 671

\bibitem[\protect\citeauthoryear{{McKee} \& {Offner}}{{McKee} \&
  {Offner}}{2010}]{McKOff2010}
{McKee} C.~F.,  {Offner} S.~S.~R.,  2010, \apj, 716, 167

\bibitem[\protect\citeauthoryear{{Moeckel} \& {Bate}}{{Moeckel} \&
  {Bate}}{2010}]{MoeBat2010}
{Moeckel} N.,  {Bate} M.~R.,  2010, \mnras, 404, 721

\bibitem[\protect\citeauthoryear{{Moeckel} \& {Clarke}}{{Moeckel} \&
  {Clarke}}{2011}]{MoeCla2011}
{Moeckel} N.,  {Clarke} C.~J.,  2011, ArXiv e-prints

\bibitem[\protect\citeauthoryear{{Monin}, {Guieu}, {Pinte}, {Rebull},
  {Goldsmith}, {Fukagawa}, {M{\'e}nard}, {Padgett}, {Stappelfeld}, {McCabe},
  {Carey}, {Noriega-Crespo}, {Brooke}, {Huard}, {Terebey}, {Hillenbrand} \&
  {Guedel}}{{Monin} et~al.}{2010}]{Moninetal2010}
{Monin} J.-L.,  {Guieu} S.,  {Pinte} C.,  {Rebull} L.,  {Goldsmith} P.,
  {Fukagawa} M.,  {M{\'e}nard} F.,  {Padgett} D.,  {Stappelfeld} K.,  {McCabe}
  C.,  {Carey} S.,  {Noriega-Crespo} A.,  {Brooke} T.,  {Huard} T.,  {Terebey}
  S.,  {Hillenbrand} L.,    {Guedel} M.,  2010, \aap, 515, A91+

\bibitem[\protect\citeauthoryear{{Monin}, {Menard} \& {Duchene}}{{Monin}
  et~al.}{1998}]{MonMenDuc1998}
{Monin} J.-L.,  {Menard} F.,    {Duchene} G.,  1998, \aap, 339, 113

\bibitem[\protect\citeauthoryear{{Monin}, {M{\'e}nard} \& {Peretto}}{{Monin}
  et~al.}{2006}]{MonMenPer2006}
{Monin} J.-L.,  {M{\'e}nard} F.,    {Peretto} N.,  2006, \aap, 446, 201

\bibitem[\protect\citeauthoryear{{Morris} \& {Monaghan}}{{Morris} \&
  {Monaghan}}{1997}]{MorMon1997}
{Morris} J.~P.,  {Monaghan} J.~J.,  1997, J.\ Comp.\ Phys., 136, 41

\bibitem[\protect\citeauthoryear{{Muench}, {Alves}, {Lada} \& {Lada}}{{Muench}
  et~al.}{2001}]{Muenchetal2001}
{Muench} A.~A.,  {Alves} J.,  {Lada} C.~J.,    {Lada} E.~A.,  2001, \apjl, 558,
  L51

\bibitem[\protect\citeauthoryear{{Nakamura} \& {Li}}{{Nakamura} \&
  {Li}}{2007}]{NakLi2007}
{Nakamura} F.,  {Li} Z.-Y.,  2007, \apj, 662, 395

\bibitem[\protect\citeauthoryear{{Natta} \& {Testi}}{{Natta} \&
  {Testi}}{2001}]{NatTes2001}
{Natta} A.,  {Testi} L.,  2001, \aap, 376, L22

\bibitem[\protect\citeauthoryear{{Natta}, {Testi}, {Comer{\'o}n}, {Oliva},
  {D'Antona}, {Baffa}, {Comoretto} \& {Gennari}}{{Natta}
  et~al.}{2002}]{Nattaetal2002}
{Natta} A.,  {Testi} L.,  {Comer{\'o}n} F.,  {Oliva} E.,  {D'Antona} F.,
  {Baffa} C.,  {Comoretto} G.,    {Gennari} S.,  2002, \aap, 393, 597

\bibitem[\protect\citeauthoryear{{Offner}, {Klein} \& {McKee}}{{Offner}
  et~al.}{2008}]{OffKleMcK2008}
{Offner} S.~S.~R.,  {Klein} R.~I.,    {McKee} C.~F.,  2008, \apj, 686, 1174

\bibitem[\protect\citeauthoryear{{Offner}, {Klein}, {McKee} \&
  {Krumholz}}{{Offner} et~al.}{2009}]{Offneretal2009}
{Offner} S.~S.~R.,  {Klein} R.~I.,  {McKee} C.~F.,    {Krumholz} M.~R.,  2009,
  \apj, 703, 131

\bibitem[\protect\citeauthoryear{{Offner} \& {McKee}}{{Offner} \&
  {McKee}}{2011}]{OffMcK2011}
{Offner} S.~S.~R.,  {McKee} C.~F.,  2011, ArXiv e-prints

\bibitem[\protect\citeauthoryear{{Ostriker}, {Stone} \& {Gammie}}{{Ostriker}
  et~al.}{2001}]{OstStoGam2001}
{Ostriker} E.~C.,  {Stone} J.~M.,    {Gammie} C.~F.,  2001, \apj, 546, 980

\bibitem[\protect\citeauthoryear{{Patience}, {Ghez}, {Reid} \&
  {Matthews}}{{Patience} et~al.}{2002}]{Patienceetal2002}
{Patience} J.,  {Ghez} A.~M.,  {Reid} I.~N.,    {Matthews} K.,  2002, \aj, 123,
  1570

\bibitem[\protect\citeauthoryear{{Pinfield}, {Dobbie}, {Jameson}, {Steele},
  {Jones} \& {Katsiyannis}}{{Pinfield} et~al.}{2003}]{Pinfieldetal2003}
{Pinfield} D.~J.,  {Dobbie} P.~D.,  {Jameson} R.~F.,  {Steele} I.~A.,  {Jones}
  H.~R.~A.,    {Katsiyannis} A.~C.,  2003, \mnras, 342, 1241

\bibitem[\protect\citeauthoryear{{Pollack}, {McKay} \&
  {Christofferson}}{{Pollack} et~al.}{1985}]{PolMcKChr1985}
{Pollack} J.~B.,  {McKay} C.~P.,    {Christofferson} B.~M.,  1985, Icarus, 64,
  471

\bibitem[\protect\citeauthoryear{{Preibisch}, {Balega}, {Hofmann}, {Weigelt} \&
  {Zinnecker}}{{Preibisch} et~al.}{1999}]{Preibischetal1999}
{Preibisch} T.,  {Balega} Y.,  {Hofmann} K.-H.,  {Weigelt} G.,    {Zinnecker}
  H.,  1999, New Astronomy, 4, 531

\bibitem[\protect\citeauthoryear{{Price} \& {Bate}}{{Price} \&
  {Bate}}{2007}]{PriBat2007}
{Price} D.~J.,  {Bate} M.~R.,  2007, \mnras, 377, 77

\bibitem[\protect\citeauthoryear{{Price} \& {Bate}}{{Price} \&
  {Bate}}{2008}]{PriBat2008}
{Price} D.~J.,  {Bate} M.~R.,  2008, \mnras, 385, 1820

\bibitem[\protect\citeauthoryear{{Price} \& {Bate}}{{Price} \&
  {Bate}}{2009}]{PriBat2009}
{Price} D.~J.,  {Bate} M.~R.,  2009, \mnras ~submitted, 0

\bibitem[\protect\citeauthoryear{{Price} \& {Monaghan}}{{Price} \&
  {Monaghan}}{2005}]{PriMon2005}
{Price} D.~J.,  {Monaghan} J.~J.,  2005, \mnras, 364, 384

\bibitem[\protect\citeauthoryear{{Price} \& {Monaghan}}{{Price} \&
  {Monaghan}}{2007}]{PriMon2007}
{Price} D.~J.,  {Monaghan} J.~J.,  2007, \mnras, 374, 1347

\bibitem[\protect\citeauthoryear{{Pringle}}{{Pringle}}{1991}]{Pringle1991}
{Pringle} J.~E.,  1991, \mnras, 248, 754

\bibitem[\protect\citeauthoryear{{Quanz}, {Goldman}, {Henning}, {Brandner},
  {Burrows} \& {Hofstetter}}{{Quanz} et~al.}{2010}]{Quanzetal2010}
{Quanz} S.~P.,  {Goldman} B.,  {Henning} T.,  {Brandner} W.,  {Burrows} A.,
  {Hofstetter} L.~W.,  2010, \apj, 708, 770

\bibitem[\protect\citeauthoryear{{Radigan}, {Lafreni{\`e}re}, {Jayawardhana} \&
  {Doyon}}{{Radigan} et~al.}{2009}]{Radiganetal2009}
{Radigan} J.,  {Lafreni{\`e}re} D.,  {Jayawardhana} R.,    {Doyon} R.,  2009,
  \apj, 698, 405

\bibitem[\protect\citeauthoryear{{Raghavan}, {McAlister}, {Henry}, {Latham},
  {Marcy}, {Mason}, {Gies}, {White} \& {ten Brummelaar}}{{Raghavan}
  et~al.}{2010}]{Raghavanetal2010}
{Raghavan} D.,  {McAlister} H.~A.,  {Henry} T.~J.,  {Latham} D.~W.,  {Marcy}
  G.~W.,  {Mason} B.~D.,  {Gies} D.~R.,  {White} R.~J.,    {ten Brummelaar}
  T.~A.,  2010, \apjs, 190, 1

\bibitem[\protect\citeauthoryear{{Rees}}{{Rees}}{1976}]{Rees1976}
{Rees} M.~J.,  1976, \mnras, 176, 483

\bibitem[\protect\citeauthoryear{{Reid}, {Cruz}, {Burgasser} \& {Liu}}{{Reid}
  et~al.}{2008}]{Reidetal2008}
{Reid} I.~N.,  {Cruz} K.~L.,  {Burgasser} A.~J.,    {Liu} M.~C.,  2008, \aj,
  135, 580

\bibitem[\protect\citeauthoryear{{Reid}, {Lewitus}, {Allen}, {Cruz} \&
  {Burgasser}}{{Reid} et~al.}{2006}]{Reidetal2006}
{Reid} I.~N.,  {Lewitus} E.,  {Allen} P.~R.,  {Cruz} K.~L.,    {Burgasser}
  A.~J.,  2006, \aj, 132, 891

\bibitem[\protect\citeauthoryear{{Reipurth} \& {Clarke}}{{Reipurth} \&
  {Clarke}}{2001}]{ReiCla2001}
{Reipurth} B.,  {Clarke} C.,  2001, AJ, 122, 432

\bibitem[\protect\citeauthoryear{{Reipurth}, {Guimar{\~a}es}, {Connelley} \&
  {Bally}}{{Reipurth} et~al.}{2007}]{Reipurthetal2007}
{Reipurth} B.,  {Guimar{\~a}es} M.~M.,  {Connelley} M.~S.,    {Bally} J.,
  2007, \aj, 134, 2272

\bibitem[\protect\citeauthoryear{{Saigo} \& {Tomisaka}}{{Saigo} \&
  {Tomisaka}}{2006}]{SaiTom2006}
{Saigo} K.,  {Tomisaka} K.,  2006, \apj, 645, 381

\bibitem[\protect\citeauthoryear{{Saigo} \& {Tomisaka}}{{Saigo} \&
  {Tomisaka}}{2011}]{SaiTom2011}
{Saigo} K.,  {Tomisaka} K.,  2011, \apj, 728, 78

\bibitem[\protect\citeauthoryear{{Saigo}, {Tomisaka} \& {Matsumoto}}{{Saigo}
  et~al.}{2008}]{SaiTomMat2008}
{Saigo} K.,  {Tomisaka} K.,    {Matsumoto} T.,  2008, \apj, 674, 997

\bibitem[\protect\citeauthoryear{{Salpeter}}{{Salpeter}}{1955}]{Salpeter1955}
{Salpeter} E.~E.,  1955, \apj, 121, 161

\bibitem[\protect\citeauthoryear{{Scally}, {Clarke} \& {McCaughrean}}{{Scally}
  et~al.}{1999}]{ScaClaMcC1999}
{Scally} A.,  {Clarke} C.,    {McCaughrean} M.~J.,  1999, \mnras, 306, 253

\bibitem[\protect\citeauthoryear{{Scholz}, {Jayawardhana} \& {Wood}}{{Scholz}
  et~al.}{2006}]{SchJayWoo2006}
{Scholz} A.,  {Jayawardhana} R.,    {Wood} K.,  2006, \apj, 645, 1498

\bibitem[\protect\citeauthoryear{{Shatsky} \& {Tokovinin}}{{Shatsky} \&
  {Tokovinin}}{2002}]{ShaTok2002}
{Shatsky} N.,  {Tokovinin} A.,  2002, \aap, 382, 92

\bibitem[\protect\citeauthoryear{{Shu}}{{Shu}}{1977}]{Shu1977}
{Shu} F.~H.,  1977, \apj, 214, 488

\bibitem[\protect\citeauthoryear{{Siegler}, {Close}, {Cruz}, {Mart{\'{\i}}n} \&
  {Reid}}{{Siegler} et~al.}{2005}]{Siegleretal2005}
{Siegler} N.,  {Close} L.~M.,  {Cruz} K.~L.,  {Mart{\'{\i}}n} E.~L.,    {Reid}
  I.~N.,  2005, \apj, 621, 1023

\bibitem[\protect\citeauthoryear{{Siegler}, {Close}, {Mamajek} \&
  {Freed}}{{Siegler} et~al.}{2003}]{Siegleretal2003}
{Siegler} N.,  {Close} L.~M.,  {Mamajek} E.~E.,    {Freed} M.,  2003, \apj,
  598, 1265

\bibitem[\protect\citeauthoryear{{Silk}}{{Silk}}{1977a}]{Silk1977a}
{Silk} J.,  1977a, \apj, 214, 152

\bibitem[\protect\citeauthoryear{{Silk}}{{Silk}}{1977b}]{Silk1977b}
{Silk} J.,  1977b, \apj, 214, 718

\bibitem[\protect\citeauthoryear{{Simon}, {Ghez}, {Leinert}, {Cassar}, {Chen},
  {Howell}, {Jameson}, {Matthews}, {Neugebauer} \& {Richichi}}{{Simon}
  et~al.}{1995}]{Simonetal1995}
{Simon} M.,  {Ghez} A.~M.,  {Leinert} C.,  {Cassar} L.,  {Chen} W.~P.,
  {Howell} R.~R.,  {Jameson} R.~F.,  {Matthews} K.,  {Neugebauer} G.,
  {Richichi} A.,  1995, \apj, 443, 625

\bibitem[\protect\citeauthoryear{{Smith}, {Bonnell} \& {Bate}}{{Smith}
  et~al.}{1997}]{SmiBonBat1997}
{Smith} K.~W.,  {Bonnell} I.~A.,    {Bate} M.~R.,  1997, \mnras, 288, 1041

\bibitem[\protect\citeauthoryear{{Soderhjelm}}{{Soderhjelm}}{1997}]{Soderhjelm%
1997}
{Soderhjelm} S.,  1997, in {Docobo} J.~A.,  {Elipe} A.,   {McAlister} H.,  eds,
  Visual Double Stars : Formation, Dynamics and Evolutionary Tracks Vol.~223 of
  Astrophysics and Space Science Library, {Mass-Ratio Distribution from Deltam
  Statistics for Nearby HIPPARCOS Binaries}.
pp 497--+

\bibitem[\protect\citeauthoryear{{Stamatellos}, {Hubber} \&
  {Whitworth}}{{Stamatellos} et~al.}{2007}]{StaHubWhi2007}
{Stamatellos} D.,  {Hubber} D.~A.,    {Whitworth} A.~P.,  2007, \mnras, 382,
  L30

\bibitem[\protect\citeauthoryear{{Stamatellos} \& {Whitworth}}{{Stamatellos} \&
  {Whitworth}}{2009}]{StaWhi2009}
{Stamatellos} D.,  {Whitworth} A.~P.,  2009, \mnras, 392, 413

\bibitem[\protect\citeauthoryear{{Stamatellos}, {Whitworth} \&
  {Hubber}}{{Stamatellos} et~al.}{2011}]{StaWhiHub2011}
{Stamatellos} D.,  {Whitworth} A.~P.,    {Hubber} D.~A.,  2011, \apj, 730, 32

\bibitem[\protect\citeauthoryear{{Stapelfeldt}, {Krist}, {Menard}, {Bouvier},
  {Padgett} \& {Burrows}}{{Stapelfeldt} et~al.}{1998}]{Stapelfeldtetal1998}
{Stapelfeldt} K.~R.,  {Krist} J.~E.,  {Menard} F.,  {Bouvier} J.,  {Padgett}
  D.~L.,    {Burrows} C.~J.,  1998, \apjl, 502, L65+

\bibitem[\protect\citeauthoryear{{Sterzik} \& {Durisen}}{{Sterzik} \&
  {Durisen}}{1998}]{SteDur1998}
{Sterzik} M.~F.,  {Durisen} R.~H.,  1998, \aap, 339, 95

\bibitem[\protect\citeauthoryear{{Sterzik} \& {Tokovinin}}{{Sterzik} \&
  {Tokovinin}}{2002}]{SteTok2002}
{Sterzik} M.~F.,  {Tokovinin} A.~A.,  2002, \aap, 384, 1030

\bibitem[\protect\citeauthoryear{{Tian}, {van Leeuwen}, {Zhao} \& {Su}}{{Tian}
  et~al.}{1996}]{Tianetal1996}
{Tian} K.~P.,  {van Leeuwen} F.,  {Zhao} J.~L.,    {Su} C.~G.,  1996, \aaps,
  118, 503

\bibitem[\protect\citeauthoryear{{Tokovinin}}{{Tokovinin}}{2000a}]{Tokovinin20%
00a}
{Tokovinin} A.,  2000a, in IAU Symposium Vol.~200 of IAU Symposium, {Statistics
  of multiple stars: some clues to formation mechanisms.}.
pp 84--92

\bibitem[\protect\citeauthoryear{{Tokovinin}}{{Tokovinin}}{2008}]{Tokovinin200%
8}
{Tokovinin} A.,  2008, \mnras, 389, 925

\bibitem[\protect\citeauthoryear{{Tokovinin}}{{Tokovinin}}{1993}]{Tokovinin199%
3}
{Tokovinin} A.~A.,  1993, Astronomy Letters, 19, 383

\bibitem[\protect\citeauthoryear{{Tokovinin}}{{Tokovinin}}{2000b}]{Tokovinin20%
00b}
{Tokovinin} A.~A.,  2000b, \aap, 360, 997

\bibitem[\protect\citeauthoryear{{Tomida}, {Machida}, {Saigo}, {Tomisaka} \&
  {Matsumoto}}{{Tomida} et~al.}{2010}]{Tomidaetal2010b}
{Tomida} K.,  {Machida} M.~N.,  {Saigo} K.,  {Tomisaka} K.,    {Matsumoto} T.,
  2010, \apjl, 725, L239

\bibitem[\protect\citeauthoryear{{Tomida}, {Tomisaka}, {Matsumoto}, {Ohsuga},
  {Machida} \& {Saigo}}{{Tomida} et~al.}{2010}]{Tomidaetal2010a}
{Tomida} K.,  {Tomisaka} K.,  {Matsumoto} T.,  {Ohsuga} K.,  {Machida} M.~N.,
   {Saigo} K.,  2010, ArXiv e-prints

\bibitem[\protect\citeauthoryear{{Truelove}, {Klein}, {McKee}, {Holliman} II,
  {Howell} \& {Greenough}}{{Truelove} et~al.}{1997}]{Trueloveetal1997}
{Truelove} J.~K.,  {Klein} R.~I.,  {McKee} C.~F.,  {Holliman} II J.~H.,
  {Howell} L.~H.,    {Greenough} J.~A.,  1997, \apjl, 489, L179

\bibitem[\protect\citeauthoryear{{Urban}, {Martel} \& {Evans}}{{Urban}
  et~al.}{2010}]{UrbMarEva2010}
{Urban} A.,  {Martel} H.,    {Evans} N.~J.,  2010, \apj, 710, 1343

\bibitem[\protect\citeauthoryear{{van Albada}}{{van
  Albada}}{1968}]{vanAlbada1968b}
{van Albada} T.~S.,  1968, \bain, 20, 73

\bibitem[\protect\citeauthoryear{{Vorobyov} \& {Basu}}{{Vorobyov} \&
  {Basu}}{2005}]{VorBas2005}
{Vorobyov} E.~I.,  {Basu} S.,  2005, \apjl, 633, L137

\bibitem[\protect\citeauthoryear{{Wang}, {Li}, {Abel} \& {Nakamura}}{{Wang}
  et~al.}{2010}]{Wangetal2010}
{Wang} P.,  {Li} Z.-Y.,  {Abel} T.,    {Nakamura} F.,  2010, \apj, 709, 27

\bibitem[\protect\citeauthoryear{{Weidner} \& {Kroupa}}{{Weidner} \&
  {Kroupa}}{2006}]{WeiKro2006}
{Weidner} C.,  {Kroupa} P.,  2006, \mnras, 365, 1333

\bibitem[\protect\citeauthoryear{{Weidner}, {Kroupa} \& {Bonnell}}{{Weidner}
  et~al.}{2010}]{WeiKroBon2010}
{Weidner} C.,  {Kroupa} P.,    {Bonnell} I.~A.~D.,  2010, \mnras, 401, 275

\bibitem[\protect\citeauthoryear{{Weights}, {Lucas}, {Roche}, {Pinfield} \&
  {Riddick}}{{Weights} et~al.}{2009}]{Weightsetal2009}
{Weights} D.~J.,  {Lucas} P.~W.,  {Roche} P.~F.,  {Pinfield} D.~J.,
  {Riddick} F.,  2009, \mnras, 392, 817

\bibitem[\protect\citeauthoryear{{Weis}}{{Weis}}{1974}]{Weis1974}
{Weis} E.~W.,  1974, \apj, 190, 331

\bibitem[\protect\citeauthoryear{{Wheelwright}, {Vink}, {Oudmaijer} \&
  {Drew}}{{Wheelwright} et~al.}{2011}]{Wheelwrightetal2011}
{Wheelwright} H.~E.,  {Vink} J.~S.,  {Oudmaijer} R.~D.,    {Drew} J.~E.,  2011,
  ArXiv e-prints

\bibitem[\protect\citeauthoryear{{Whitehouse} \& {Bate}}{{Whitehouse} \&
  {Bate}}{2006}]{WhiBat2006}
{Whitehouse} S.~C.,  {Bate} M.~R.,  2006, \mnras, 367, 32

\bibitem[\protect\citeauthoryear{{Whitehouse}, {Bate} \&
  {Monaghan}}{{Whitehouse} et~al.}{2005}]{WhiBatMon2005}
{Whitehouse} S.~C.,  {Bate} M.~R.,    {Monaghan} J.~J.,  2005, \mnras, 364,
  1367

\bibitem[\protect\citeauthoryear{{Whitworth}}{{Whitworth}}{1998}]{Whitworth199%
8}
{Whitworth} A.~P.,  1998, \mnras, 296, 442

\bibitem[\protect\citeauthoryear{{Wolf}, {Stecklum} \& {Henning}}{{Wolf}
  et~al.}{2001}]{WolSteHen2001}
{Wolf} S.,  {Stecklum} B.,    {Henning} T.,  2001, in {Zinnecker} H.,
  {Mathieu} R.,  eds, The Formation of Binary Stars Vol.~200 of IAU Symposium,
  {Pre-Main Sequence Binaries with Aligned Disks?}.
p.~295

\bibitem[\protect\citeauthoryear{{Worley}}{{Worley}}{1967}]{Worley1967}
{Worley} C.~E.,  1967, in {Dommanget} J.,  ed., On the Evolution of Double
  Stars {Multiple star systems as related to double stars: Multiple stars among
  the visual binaries}.
pp 221--+

\bibitem[\protect\citeauthoryear{{Zapatero Osorio}, {B{\'e}jar},
  {Mart{\'{\i}}n}, {Rebolo}, {Barrado y Navascu{\'e}s}, {Bailer-Jones} \&
  {Mundt}}{{Zapatero Osorio} et~al.}{2000}]{ZapateroOsorioetal2000}
{Zapatero Osorio} M.~R.,  {B{\'e}jar} V.~J.~S.,  {Mart{\'{\i}}n} E.~L.,
  {Rebolo} R.,  {Barrado y Navascu{\'e}s} D.,  {Bailer-Jones} C.~A.~L.,
  {Mundt} R.,  2000, Science, 290, 103

\bibitem[\protect\citeauthoryear{{Zapatero Osorio}, {B{\'e}jar},
  {Mart{\'{\i}}n}, {Rebolo}, {Barrado y Navascu{\'e}s}, {Mundt},
  {Eisl{\"o}ffel} \& {Caballero}}{{Zapatero Osorio}
  et~al.}{2002}]{ZapateroOsorioetal2002}
{Zapatero Osorio} M.~R.,  {B{\'e}jar} V.~J.~S.,  {Mart{\'{\i}}n} E.~L.,
  {Rebolo} R.,  {Barrado y Navascu{\'e}s} D.,  {Mundt} R.,  {Eisl{\"o}ffel} J.,
     {Caballero} J.~A.,  2002, \apj, 578, 536

\bibitem[\protect\citeauthoryear{{Zhu}, {Hartmann} \& {Gammie}}{{Zhu}
  et~al.}{2009}]{ZhuHarGam2009}
{Zhu} Z.,  {Hartmann} L.,    {Gammie} C.,  2009, \apj, 694, 1045

\end{thebibliography}

\end{document}